\documentclass[twoside,fontsize=12pt,BCOR=22.5mm,DIV=15,headsepline=true,headlines=1.5,headinclude=true,listof=totocnumbered,bibliography=totocnumbered,unicode,pdfa]{scrbook}  %
\pdfoutput=1
\immediate\pdfobj stream attr{/N 3}  file{PDFA/sRGBIEC1966-2.1.icm}
\pdfcatalog{%
/OutputIntents [ <<
/Type /OutputIntent
/S/GTS_PDFA1
/DestOutputProfile \the\pdflastobj\space 0 R
/OutputConditionIdentifier (sRGB IEC61966-2.1)
/Info(sRGB IEC61966-2.1)
>> ]
}

\usepackage[utf8]{inputenc}
\usepackage{lmodern}
\usepackage[T1]{fontenc}
\usepackage{csquotes}
\usepackage[ngerman,english]{babel}
\usepackage{setspace}
\AfterTOCHead{\singlespacing}
\KOMAoptions{DIV=last}

\usepackage{scrlayer-scrpage}

\usepackage{threeparttable}
\usepackage{dpfloat}
\usepackage[section]{placeins}

\usepackage{xcolor}
\usepackage{etoolbox}
\usepackage{amsfonts}
\usepackage{amssymb}
\usepackage[intlimits]{amsmath}
\usepackage{amsthm}
\usepackage{mathtools}
\usepackage[h]{esvect}
\usepackage{tensor}
\usepackage{booktabs}
\usepackage{bbm}
\pdfpkmode={supre}
\usepackage{nicefrac}
\usepackage{units}
\usepackage{nextpage}

\usepackage{graphicx}
\usepackage[export]{adjustbox}
\graphicspath{{graphics/}{plot/}{plot/omega_tau_1_transposed.make/}}

\usepackage{siunitx}

\usepackage{wrapfig}
\usepackage{hyperxmp}
\input glyphtounicode.tex
\input glyphtounicode-cmr.tex
\usepackage{bookmark}

\usepackage{calc}
\usepackage{suffix}
\usepackage{ifthen}
\usepackage{titling}
\usepackage{needspace}

\usepackage{makeidx}
\makeindex
\WithSuffix\newcommand\index*[2][]{%
  \ifthenelse{\equal{#1}{}}%
    {#2\index{#2}}%
    {#2\index{#1}}%
}

\usepackage{subfigure}

\usepackage[sorting=none,giveninits=true,url=false,isbn=false,date=year,style=numeric-comp,bibencoding=ascii,backend=bibtex]{biblatex}
\renewbibmacro{in:}{%
  \ifentrytype{article}{}{\printtext{\bibstring{in}\intitlepunct}}}
\DeclareFieldFormat{pages}{\mkfirstpage[{\mkpageprefix[bookpagination]}]{#1}} 
\bibliography{References}

\usepackage[]{hyperref}
\usepackage[all]{hypcap}

\usepackage{dsfont}

\usepackage{tikz}

\usepackage{chngcntr}		
\counterwithout*{footnote}{chapter}

\usepackage{tabularx}
\newcolumntype{Y}{>{\centering\arraybackslash}X}
\usepackage{hhline}
\usepackage{booktabs}
\usepackage{multirow}

\usepackage[format=plain,labelfont=bf]{caption}

\makeatletter
\newcommand{\maketitlefront}{%
  \setparsizes{\z@}{\z@}{\z@\@plus 1fil}\par@updaterelative
  \begin{center}
    {\usekomafont{title}{\huge \@title\par}}%
    \vskip 2em
    {\ifx\@subtitle\@empty\else\usekomafont{subtitle}{\@subtitle\par}\fi}%
    \vskip 3em
    \ifx\@subject\@empty \else
      {\usekomafont{subject}{\@subject\par}}%
      \vskip 3em
    \fi
    {\usekomafont{date}{\@date \par}}%
    \vskip \z@ \@plus3fill
  \end{center}\par
  \vfill
  \begin{minipage}[b]{\linewidth}
  \end{minipage}\par
}

\newcommand{\maketitleback}{%
  \next@tpage
  \begin{minipage}[t]{\linewidth}
    \@uppertitleback
  \end{minipage}\par
  \vfill
  \begin{minipage}[b]{\linewidth}
    \@lowertitleback
  \end{minipage}\par
  \@thanks\let\@thanks\@empty%
}

\newcommand{\makededication}{%
  \ifx\@dedication\@empty
  \else
    \next@tdpage\null\vfill
    {\centering\usekomafont{dedication}{\@dedication \par}}%
    \vskip \z@ \@plus3fill
    \@thanks\let\@thanks\@empty
    \cleardoubleemptypage
  \fi%
}

\newcommand{\mytitlepage}{%
  \begin{titlepage}%
      \begin{spacing}{1}
    \maketitlefront%
    \maketitleback%
  \end{spacing}
  \end{titlepage}%
}
\makeatother

\title{
Numerical~construction~and critical~behavior~of Kaluza-Klein~black~holes \\ \ \\ \Large{Michael Kalisch}
}
\subject{%
  \vspace{3cm}
  \includegraphics[width=0.8\textwidth]{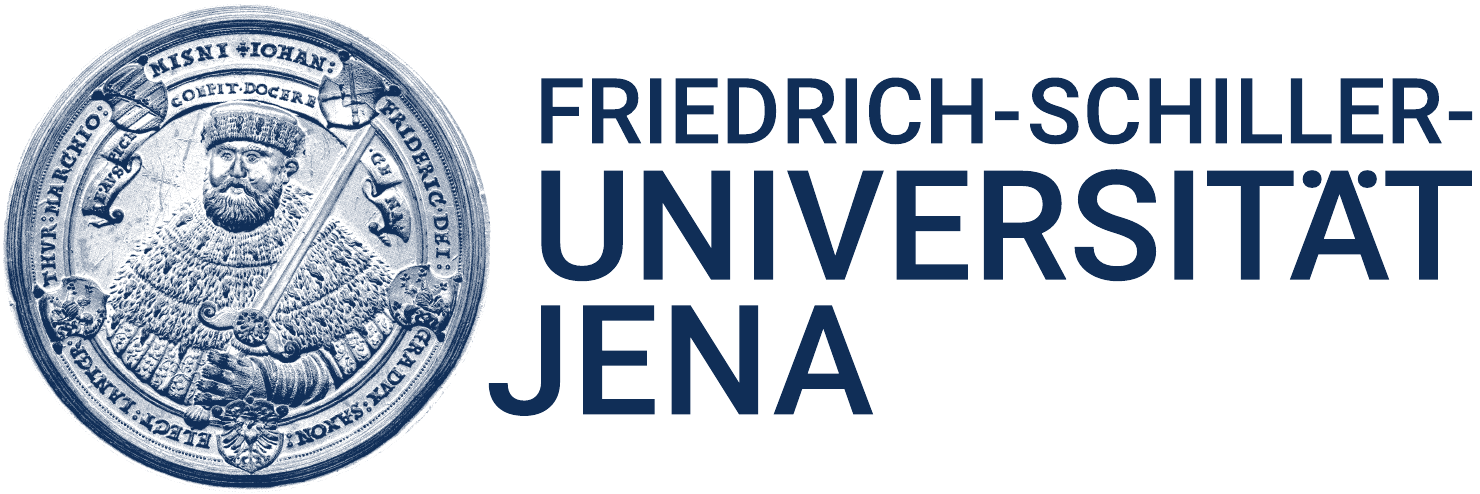} 
  \vspace{5cm}\\
  \Huge{\textbf{--Dissertation--}}
}
\author{Michael Kalisch}
\date{}
\publishers{M.\,Sc.\ Michael Kalisch}
\uppertitleback{}
\lowertitleback{}

\definecolor{rgbcyan}{rgb}{0,1,1}
\definecolor{rgbpurple}{rgb}{1,0,1}
\definecolor{rgbred}{rgb}{1,0,0}
\definecolor{rgbblue}{rgb}{0,0,1}

\hypersetup{
    pdflang = {en},
    unicode = {true},
    pdftitle = {\thetitle},
    pdfsubject = {Dissertation},
    pdfauthor = {\theauthor},
    pdfauthortitle = {Master of Science},
    pdfkeywords = {black strings, black holes, pseudo-spectral method, critical behavior},
    pdfstartview = {FitV},
    pdfview = {FitH},
    pdfpagelayout = {OneColumn},
    colorlinks = {true},
    urlcolor = {rgbcyan},
    citecolor = {rgbpurple},
    filecolor = {rgbred},
    linkcolor = {black}
}

\def\beq{\begin{equation}}
\def\eeq{\end{equation}}
\def\Rmax{R_{\mathrm{max}}}
\def\Rmin{R_{\mathrm{min}}}
\def\D{\mathrm d}
\def\E{\mathrm e}
\def\I{\mathrm i}
\def\GL{\text{GL}}
\def\UBS{\text{UBS}}
\def\LBH{\text{LBH}}
\def\NBS{\text{NBS}}
\def\ST{\text{ST}}
\def\expterm{\E ^{-2\pi \, r/L}}
\newcommand{\lowa}[  1]{#1 _{\mathrm a}}
\newcommand{\lowh}[  1]{#1 _{\mathrm h}}

\addtokomafont{title}{\Huge \bfseries}
\addtokomafont{publishers}{\bfseries \large}

\newcounter{framenum}
\DeclareLayer[{
    clone=scrheadings.head.even,
    area={3mm}{3mm}{\paperwidth-6mm}{\dimexpr\voffset+2.54cm-\headsep\relax},
    align=lt,
    contents={\setcounter{framenum}{(\thepage-1)/2}\hfill\includegraphics[scale=0.8]{thumb/f\theframenum}}
}]{scrheadings.head.bg.even}

\newpairofpagestyles[scrheadings]{headingsthumb}{}
\AddLayersAtBeginOfPageStyle{headingsthumb}{scrheadings.head.bg.even}
\AddLayersAtBeginOfPageStyle{plain.headingsthumb}{scrheadings.head.bg.even}

\begin{document}
  \setlength\footskip{26.83pt}
  \pagenumbering{alph}
  \mytitlepage

  \frontmatter
  
  \setcounter{page}{1}

  {
    \renewcommand*{\chapterheadstartvskip}{\vspace*{-1cm}}
    \begin{onehalfspacing}
\cleardoublepage %
\phantomsection
\pdfbookmark[chapter]{Abstract}{abstract}
\addchap*{Abstract}%

The idea of extra dimensions provides a promising approach to overcome various problems in modern physics. This includes theoretical as well as phenomenological aspects, such as the unification of the fundamental interactions or the hierarchy problem. Based on the seminal works by Kaluza and Klein that were published nearly 100 years ago, we denote theories with at least one compact periodic dimension as Kaluza-Klein theories. 

From a gravitational point of view the question arises, what are the fundamental solutions to Einstein's field equations of general relativity under these assumptions. In particular, in this work we are concerned with black hole solutions in Kaluza-Klein theory. Considering only the static case without electric charge, it turns out that there is a much richer phase space than in the usual four-dimensional theory, where only the Schwarzschild solution exists. There are at least two types of solutions with a completely different horizon topology: localized black holes with an ordinary spherical horizon and black strings with a horizon that wraps the compact dimension. 

Several arguments favor the conjecture that the solution branches of both types are connected via a singular topology changing solution that is controlled by the so-called double-cone metric. We study the regime close to this singular transit solution in five and six spacetime dimensions with the help of a highly accurate numerical scheme that we describe in detail. Consequently, for the first time we are able to show that in this regime the black objects exhibit a critical behavior, indicating that physical quantities are governed by universal critical exponents. Interestingly, such exponents were already derived from the double-cone metric. We show that our data confirms these values extremely well. This provides compelling evidence in favor of the double-cone metric as the local model of the transit solution.

\end{onehalfspacing}

  }

  \mainmatter
  \pagestyle{scrheadings}
    \phantomsection
    \renewcommand{\contentsname}{Table of Contents}
    \pdfbookmark[chapter]{\contentsname}{toc}
    \tableofcontents

  \chapter{Introduction}
\label{chap:Introduction}

One of the most fundamental questions of physics concerns the nature of space and time. About 100 years ago Einstein worked out the theory of general relativity (GR)~\cite{Einstein:1915aa}, which completely changed our understanding in regard of this question. GR unifies space and time to a single entity called spacetime. Moreover, according to this theory the presence of matter causes the spacetime to be curved, while the curvature of spacetime dictates the motion of matter. In fact, this is the origin of gravitational interaction in Einstein's theory. 

While at first glance the concepts of GR seem to be rather bizarre, it turned out that the predictions of GR give a very accurate description of nature on astronomical scales. With the help of this theory Einstein was able to fully explain the perihelion advance of Mercury's orbit, a problem that was unresolved for decades at that time. The first observation of a novel prediction of Einstein's theory happened in 1919 when Eddington led the famous expedition to measure the deflection of light by the sun during a total solar eclipse~\cite{Dyson:1920aa}. Another so-called classical test of GR concerns the redshift of light that is emitted from sources with a huge gravitational potential. In the vicinity of such sources time evolves slower thus leading to a measurable shift in the frequency of the light. Several sophisticated experiments successfully confirmed the gravitational redshift, e.g.\ reference~\cite{Pound:1960aa}. Moreover, we benefit from the knowledge of this effect in everyday life as the accurate positioning via GPS only works if the redshift caused by the earth is taken into account. A comprehensive list of tests of GR is too long to be reviewed here. For more details see for example reference~\cite{Will:2005va}. Nevertheless, we emphasize the most recent developments: the long awaited first direct detection of gravitational waves by the LIGO scientific collaboration~\cite{Abbott:2016blz}. Gravitational waves are tiny perturbations of spacetime caused by the collision of two of the most mysterious objects of the universe: black holes. Nothing from the inside of a black hole can ever escape to the outside, not even light, and we shall come back to these fascinating objects later.

Despite the great success of GR to describe the universe and the macroscopic objects therein, it fails to give a reliable description of nature on microscopic scales. Instead, in the first quarter of the 20th century physicists developed quantum theories that completely changed our understanding of the structure of matter on small scales. Soon after, the emergence of quantum field theory provided a framework that was capable to describe the fundamental electromagnetic, weak and strong interactions and to unify them into a theory called the Standard Model of particle physics. However, a comprehensive and testable theory to unify the Standard Model and GR is still absent and its formulation is one of today's greatest challenges for modern theoretical physics.  

The unification of GR and other fundamental interactions is a problem that already arose when Einstein's theory was still incomplete. In 1914, Nordström had the idea to automatically build in electromagnetism to Einsteins theory with the help of an additional, fourth spatial dimension~\cite{Nordstroem:1914}. A few years later, with the final theory at hand, Kaluza was able to make this idea more explicit in showing that a certain ansatz of five-dimensional GR reproduces the original four-dimensional theory together with electromagnetism~\cite{Kaluza:1921}. The key to this reasoning relies on the reinterpretation of the additional degrees of freedom given by the extra dimension as components of the electromagnetic vector potential. At first, Kaluza imposed a somehow artificial cylinder condition neglecting any dependence on the extra coordinate. Later, in 1926, Klein abandoned the cylinder condition but instead found an argument based on quantum theory~\cite{Klein:1926,Klein:1926fj}: If the extra dimension has finite size, say $L$, and is of periodic nature, we expand all fields into a Fourier series with respect to the extra coordinate. According to quantum mechanics we can assign a momentum to each Fourier mode that is proportional to $k/L$, where $k$ is an integer. Consequently, if $L$ is conveniently small, we are only able to observe the trivial mode $k=0$, since the accessible energy scales are not high enough to probe the $1/L$ modes. In other words, Klein's argument implies that we do not see any evidence in favor of an additional dimension since its size is too small to be detected by today's experiments. 

In the following years, many researchers took the ideas of Kaluza and Klein very seriously, for example Einstein and Pauli wrote in 1943~\cite{Einstein:1943}: ``When one tries to find a unified theory of the gravitational and electromagnetic theory, he cannot help feeling that there is some truth in Kaluza's five-dimensional theory.'' Unfortunately, there is still a drawback in Kaluza's ansatz as it contains an undesirable scalar field with no physical significance. Nevertheless, Kaluza-Klein (KK) theory motivated a great deal of subsequent work that added more and more extra dimensions to the theory of GR and thereby tried to incorporate the weak and strong interaction into the theory. But eventually, all of these attempts turned out to have unresolvable conceptual problems. These developments are nicely reviewed in reference~\cite{Overduin:1998pn}.

By now, a promising approach to answer the question about a unified theory of the Standard Model and GR lies in string theory. While a description of string theory goes far beyond the scope of this work and is the subject of many textbooks, e.g.\ reference~\cite{Zwiebach:2004tj}, we emphasize that one of the most astonishing consequences of this theory is the prediction of six additional spatial dimensions leading to a ten-dimensional spacetime.\footnote{In fact, it is believed that different versions of ten-dimensional string theory are certain limits of an ultimate eleven-dimensional theory, called M-theory.} Again, Klein's argument, the tiny scale of the extra dimensions, is utilized to explain the lack of observational evidence. However, the extremely complicated nature of string theory as well as its many different versions makes it hard to deal with it. Therefore, as usual in physics, to consider a concrete problem one tries to simplify the situation by finding an appropriate approximation of the underlying theory. If we consider the low energy limit of string theory, we recover GR in higher dimensions but with additional matter fields. If, moreover, we neglect these additional matter fields, we obtain GR in higher dimensions. In particular, in case of compact additional dimensions we are led to Kaluza-Klein theory. We arrive at the original four-dimensional theory of GR in the limit of vanishing size of the extra dimensions.

Also based on string theory, there is a remarkable correspondence between gravity and quantum theory. The correspondence involves theories in anti-deSitter (AdS) spacetime, a solution of Einstein's field equations of GR. These are conjectured to be dual to certain conformal quantum field theories (CFTs). Therefore, it is commonly denoted as the AdS/CFT correspondence or more generally as the gauge/gravity duality, e.g.\ see reference~\cite{Ammon:2015wua}. The crucial point is that this duality allows us to relate solutions of higher dimensional GR to quantum systems. Most importantly, in certain parameter regimes, i.e.\ for strongly coupled systems, the calculations on the gravity side are easier to perform than on the quantum side and we thus get new insights into quantum systems by understanding higher dimensional GR.  

Another motivation in favor of higher dimensional GR arises from the hierarchy problem. It concerns the vast discrepancy between the strength of the gravitational interaction and the other fundamental interactions, i.e.\ the weakness of gravity on atomic scales. Surprisingly, one can solve this problem by assuming the existence of extra dimensions, since gravity becomes considerably weakened when spreading over additional dimensions. In contrast, the other fundamental interactions only act on the common three spatial dimensions and are therefore not affected by the extra dimensions. 

As a last point, we emphasize that although Einstein originally formulated GR in four spacetime dimensions, it is straightforward to write down the theory in an arbitrary number of dimensions $D$. If we treat $D$ as a parameter of the theory, we are able to explore the parameter space of the theory rather than restricting ourselves to a certain value. Consequently, we obtain a deeper understanding of the theory. Indeed, in the course of this thesis we will see that GR in higher dimensions yields a lot of surprising results.

\section*{Subject of this work}

Black holes are the most fascinating and mysterious objects arising from GR regardless of the number of spacetime dimensions considered. They are fundamental solutions to Einstein's field equations of GR, since no particular type of matter has to be assumed to describe black holes. Therefore, Einstein's field equations substantially simplify in this case. Indeed, time-independent four-dimensional black hole solutions are well-known analytically, e.g.\ the Schwarzschild or Kerr black hole, and they stand out due to their uniqueness. This changes in higher dimensions $D>4$, where many different types of black holes exist. Often numerical methods are necessary to obtain these solutions. Seeking for a general understanding the study of black holes in higher dimensions has become a continuously growing topic over the last decades~\cite{Horowitz:2012nnc}.

Here, we concentrate on black holes in spacetimes with one compact periodic dimension of size $L$, i.e.\ black holes in KK theory, therefore called KK black holes. To make things simple we restrict ourselves to the \textit{static} case of time-independent and non-moving solutions. Even in this simplified situation at least two different types of solutions exist: black strings and localized black holes.\footnote{We note that the black strings described here have nothing to do with the fundamental strings of string theory.} These solutions are distinguished by their respective shapes, because black strings wrap the compact dimension in contrast to localized black holes.

Fortunately, there is an analytic solution of black strings in $D$ dimensions that are uniform along the compact dimension, thus called \textit{uniform black strings} (UBSs). In the seminal papers of Gregory and Laflamme from the early 1990s~\cite{Gregory:1993vy,Gregory:1994bj} it was shown that small perturbations of the UBS spacetime will rapidly grow in time, if the mass is smaller than a certain value and $L$ is fixed. This Gregory-Laflamme (GL) instability breaks the translation invariance along the compact dimension, which may give rise to another type of static black hole solutions. 

At the beginning of this millennium, Gubser explicitly showed for $D=5$ that, indeed, a new type of solutions emanates from the GL instability~\cite{Gubser:2001ac}. Accordingly, these objects are called \textit{non-uniform black strings} (NBSs). Since there is little hope to find an analytic NBS solution, Gubser developed an iterative perturbative scheme around the UBS, which was adapted later to more than five dimensions~\cite{Wiseman:2002zc,Sorkin:2004qq}. Beyond the perturbative regime, one has to solve Einstein's field equations with the help of a full numerical simulation. This was done in a series of works~\cite{Wiseman:2002zc,Kudoh:2004hs,Kleihaus:2006ee,Sorkin:2006wp,Headrick:2009pv,Figueras:2012xj,Dias:2017uyv}, covering the dimensions $D=5$ up to $D=15$. There are also results available coming from a large $D$ expansion of the field equations~\cite{Emparan:2015hwa,Suzuki:2015axa}.

\textit{Localized black holes} (LBHs) were first discussed in reference~\cite{Myers:1986rx}. Again, there are perturbative techniques to construct these kind of solutions as shown in references~\cite{Harmark:2003yz,Gorbonos:2004uc,Gorbonos:2005px}. Full numerical LBH solutions were obtained in $D=5,6$~\cite{Wiseman:2002ti,Sorkin:2003ka,Kudoh:2003ki,Kudoh:2004hs,Headrick:2009pv} and very recently in $D=10$~\cite{Dias:2017uyv}.

Already in 2002, when numerical data for NBSs and LBHs was rare, several authors conjectured that there is a parametric transition between both branches~\cite{Kol:2002xz,Harmark:2003yz}. In other words, if we move along each branch by changing a certain parameter of the solution, we will find that the two branches eventually merge. On the one hand, this implies for the NBS branch that there is a certain point on the compact dimension where the black string becomes thinner and thinner and finally pinches off at the transition. On the other hand, moving along the LBH branch would reveal that the compact dimension is more and more covered by the black hole until it is completely wrapped. The numerical results mentioned above are in accordance with a common endpoint of the NBS and LBH branch, but break down way before the transition is reached. Moreover, Kol proposed a local model for the singular transit solution, the so-called double-cone metric~\cite{Kol:2002xz}, for which some numerical evidence in $D=6$ is present as well~\cite{Kol:2003ja,Sorkin:2006wp}. Furthermore, in subsequent work Kol derived some interesting implications from perturbations of the double-cone metric~\cite{Kol:2005vy,Asnin:2006ip}. Most importantly, he predicted a critical scaling of physical quantities when the transition is approached. He further specified the corresponding critical exponents. Still, it is not quite clear whether the double-cone metric is indeed the appropriate local model of the transit solution and whether the proposed implications apply.  

The present work aims to close the gap between the NBS and LBH branch, at least in $D=5$ and $D=6$. For this purpose, we develop a sophisticated numerical implementation to find solutions to Einstein's vacuum field equations that describe NBSs and LBHs, respectively. Our numerical method of choice is a pseudo-spectral scheme, which relies on the spectral expansion of any function into a given set of appropriate basis function that in our case are Chebyshev polynomials of the first kind. This method is renowned for its nice convergence properties and hence its ability to provide highly accurate results. However, for the problem at hand it is not straightforward to obtain an accurate implementation, because the functions that we want to solve for are rather involved and require special care particularly in the critical regime close to the transition. Therefore, well-suited adaptions of the method are needed in order to guarantee accurate results obtained in a reasonable computing time. Our adaptions comprise an appropriate decomposition of the domain of integration into several subdomains, the choice of convenient coordinates in each subdomain and, if necessary, a redefinition of the metric functions we solve for. 

\section*{Outline}

The thesis is structured as follows. First, in chapter~\ref{chap:Theoretical_foundations} we review the fundamental properties of black holes in four and higher dimensions. In particular, we provide a detailed description of static KK black holes and the corresponding state of the art. Furthermore, at the end of chapter~\ref{chap:Theoretical_foundations} we discuss the basic concepts of our numerical scheme. Chapters~\ref{chap:Numerical_construction_of_non-uniform_black_string_solutions} and~\ref{chap:Numerical_construction_of_localized_black_hole_solutions} are dedicated to the discussion of our approach to solve Einstein's equations in the given contexts, starting with NBSs in chapter~\ref{chap:Numerical_construction_of_non-uniform_black_string_solutions} followed by LBHs in chapter~\ref{chap:Numerical_construction_of_localized_black_hole_solutions}. The main results of this thesis are presented in chapter~\ref{chap:Complete_phase_diagram_of_static_Kaluza-Klein_black_holes}. Finally, in chapter~\ref{chap:Conclusions} we conclude with an emphasis on the physical relevance of our findings. Moreover, we provide supplementary material in appendix~\ref{chap:Appendix} that describes the concepts of the pseudo-spectral method in more detail.

We note that this thesis relies on results that are already published in references~\cite{Kalisch:2015via,Kalisch:2016fkm,Kalisch:2017bin}.

  \chapter{Theoretical foundations}
\label{chap:Theoretical_foundations}

In this chapter we review the theoretical framework of the work at hand. First, we discuss Einstein's field equations in section~\ref{sec:Einsteins_field_equations_of_general_relativiy}. Then, the remarkable properties of black holes in four and higher dimensions are outlined in section~\ref{sec:General_aspects_of_black_holes}. We focus on static Kaluza-Klein black holes in section~\ref{sec:Static_Kaluza_Klein_black_holes}. Finally, we outline the numerical method used here in section~\ref{sec:Numerical_method}.

\section{Einstein's field equations of general relativity}
\label{sec:Einsteins_field_equations_of_general_relativiy}

The central object in Einstein's theory of general relativity (GR) is the metric tensor $g_{\mu\nu}$, with indices $\mu$ and $\nu$ running from 0 to 3. It encodes the geometry of spacetime, which becomes clear from the line element
\beq
	\D s^2 = g_{\mu\nu} \, \D x^\mu \, \D x^\nu \, ,
	\label{eq:lineelement}
\eeq
as it describes local distances in spacetime. Here, $x^\mu$ denotes a set of coordinates that parametrize the spacetime and we reserve the zeroth entry for the time coordinate, i.e.\ $x^0 = t$. Accordingly, $\D x^\mu$ are the coordinate's differentials. Note that we sum over indices that appear twice. An import fact is that a change of coordinates changes the metric tensor's components but leaves the line element~\eqref{eq:lineelement} invariant.

In GR the metric tensor is determined by Einstein's field equations~\cite{Einstein:1915aa}
\beq
	R_{\mu\nu} - \frac{1}{2} \, R \, g_{\mu\nu} = 8\pi \, G_4 T_{\mu\nu} \, , 
	\label{eq:EFE}
\eeq
where $G_4$ is the usual four-dimensional gravitational constant. We have chosen units in which the speed of light reads $c=1$, which we utilize throughout this work. The Ricci tensor $R_{\mu\nu}$ is derived from the metric $g_{\mu\nu}$ and contains derivatives of the metric with respect to the coordinates $x^\mu$ up to second order. We refer to any standard text book about GR or differential geometry for the definition of $R_{\mu\nu}$, for example see reference~\cite{Wald:1984aa}. From $R_{\mu\nu}$ we get the Ricci scalar via $R = g^{\mu\nu} R_{\mu\nu}$, where $g^{\mu\nu}$ is the inverse of the metric $g_{\mu\nu}$. The last ingredient of Einstein's field equations~\eqref{eq:EFE} is the stress-energy tensor $T_{\mu\nu}$ containing information about the matter. All in all, Einstein's field equations~\eqref{eq:EFE} form a set of partial differential equations for the components of the metric tensor. The left hand side of the field equation~\eqref{eq:EFE} is often summarized to $G_{\mu\nu} := R_{\mu\nu} - R \, g_{\mu\nu} /2$ and is called the Einstein tensor.

We stress that the basic principles of GR allow us to add an expression of the form $\Lambda g_{\mu\nu}$ to the left hand side of Einstein's field equations~\eqref{eq:EFE}, where $\Lambda$ is known as the cosmological constant. Indeed, for positive $\Lambda$ this turns out to be of substantial importance for cosmology. Negative values of $\Lambda$ give rise to anti-deSitter (AdS) solutions, which are conjectured to be dual to certain conformal field theories (CFTs) leading to the famous AdS/CFT correspondence. Nevertheless, in the remainder of this work we focus on $\Lambda =0$. 

It is now straightforward to generalize these concepts to $D>4$ dimensions: Let all indices run from 0 to $D-1$ and replace the tensors by their higher-dimensional counterparts, which actually have the same structure. However, we have to take care about the gravitational constant $G_4$. Let's assume that each of the extra dimensions has a different size $L_i$. Then the $D$-dimensional gravitational constant reads 
\beq
	G_D = G_4 \prod _{i=1}^{D-4} L_i \, ,
	\label{eq:GravConstD}
\eeq
see for instance reference~\cite{Zwiebach:2004tj} for a derivation.

In particular, we are interested in vacuum solutions to Einstein's field equations~\eqref{eq:EFE}. Since there is no matter in vacuum, the stress-energy tensor vanishes $T_{\mu\nu} =0$ and we obtain Einstein's field equations in vacuum
\beq
	R_{\mu\nu} = 0 \, .
	\label{eq:EFEvac}
\eeq
The fundamental solution to this equation is the Minkowski spacetime given by
\beq	
	\D s^2_{\text{Mink}} = -\D t^2 + \delta _{mn} \, \D x^m \, \D x^n \, .
	\label{eq:metricMinkowski}
\eeq
Here, $x^m$ are Cartesian coordinates defined on the $D-1$ spatial dimensions, $\delta _{mn}$ is the Kronecker delta and the indices $m$ and $n$ only run from 1 to $D-1$. This spacetime is entirely flat and we refer to it as $\mathbb M^{D}$. Below, we mainly discuss non-trivial solutions to Einstein's vacuum field equations~\eqref{eq:EFEvac}.

\section{General aspects of black holes}
\label{sec:General_aspects_of_black_holes}       

The most remarkable solutions to Einstein's vacuum field equations~\eqref{eq:EFEvac} describe black holes. In a black hole spacetime there exists a surface called the \textit{event horizon}, which indicates the boundary of the black hole. No particle, not even light, inside a black hole can ever cross the event horizon to escape to the outside. 
 
In the remainder of this work we will mostly denote the event horizon simply as the horizon. At this point one has to be aware of the fact that there are also different notions of horizons that differ from an event horizon in general, such as the apparent and the Killing horizon. However, in the static case all of these notions coincide. A static solutions is time-independent and does not change under time reversal. In contrast, a time-independent solution that does change under time reversal is called stationary and describes rotating configurations. 

In subsection~\ref{subsec:Black_holes_in_four_dimensions} we will discuss black hole solutions in four dimensions and highlight their remarkable properties. Then, we review a surprising connection between black holes and thermodynamics in subsection~\ref{subsec:Black_hole_thermodynamics}. Finally, subsection~\ref{subsec:Black_holes_in_higher_dimensions} provides a discussion of the situation in higher dimensions.

\subsection{Black holes in four dimensions}
\label{subsec:Black_holes_in_four_dimensions}

In 1916, only a few months after Einstein wrote down the field equations of GR, Schwarzschild found one of the most important solutions to the vacuum equations~\eqref{eq:EFEvac}~\cite{Schwarzschild:1916aa}. The Schwarzschild solution describes the exterior of a spherically symmetric source in $D=4$ reading
\beq
	\D s^2_\text{Schw} = - f_4(r) \, \D t^2 + \frac{\D r^2}{f_4(r)} + r^2 \, \D \Omega^2_2 \, ,
	\label{eq:metricSchw}
\eeq
where the function $f_4$ stands for
\beq
	f_4(r) = 1-\frac{r_0}{r} \, .
	\label{eq:f4}
\eeq
The term $\D \Omega ^2_2$ denotes the line element of a unit 2-sphere $\D \Omega ^2_2 = \D \theta ^2 + \sin ^2 \theta \, \D \phi ^2$ with the commonly used angles of spherical coordinates $\theta\in [0,\pi ]$ and $\phi\in [0,2\pi ]$. Here, $r$ has the meaning of a radial coordinate. Additionally, in the asymptotic limit $r\to \infty$ we approach Minkowski spacetime~\eqref{eq:metricMinkowski} expressed in spherical coordinates. 

Obviously, there are two critical values of the coordinate $r$, at which the line element~\eqref{eq:metricSchw} degenerates, $r=0$ and $r=r_0$. It turns out that the former, $r=0$, is a singularity of the spacetime, while the latter, $r=r_0$, is only a coordinate singularity and can be removed by an appropriate coordinate transformation, see for example reference~\cite{Wald:1984aa}. The parameter $r_0$ is called the Schwarzschild radius and is proportional to the mass $M$ of the source of the Schwarzschild spacetime, $r_0 = 2 \, G_4 M$. For ordinary astrophysical objects the radius of the source exceeds the Schwarzschild radius by far.\footnote{For example the Schwarzschild radius of the Earth is about \SI{9}{\mm}.} In such a case the two singularities of the Schwarzschild metric do not play any role, since the Schwarzschild metric is not suitable to describe the object's interior. However, the gravitational collapse of a massive spherical star may lead to an object with radius $r_0$. Then, the surface $r=r_0$ represents the horizon of a Schwarzschild black hole. 

It took nearly 50 years until Kerr found a generalization of the Schwarzschild metric that describes a rotating object and is therefore axisymmetric~\cite{Kerr:1963aa}. Not only is the Kerr solution mathematically much more complicated than the Schwarzschild solution, but it also comes with some surprising physical properties. For example, due to the rotation of the Kerr black hole there is a finite region outside the event horizon, called the ergosphere, where all observers are forced to move. Another interesting fact is that for a given mass there is a maximal angular momentum of the Kerr black hole. In the limit of vanishing angular momentum the Kerr metric reproduces the Schwarzschild spacetime~\eqref{eq:metricSchw}.

The Kerr family of black hole solutions is of particular importance for several reasons. First, all solutions of this family are stationary (or even static in case of Schwarzschild) and hence may serve as possible end states of astrophysical processes, e.g.\ the collapse of a star. Nevertheless, for this to happen a necessary condition on the solution is stability. Whether these solutions are stable against small but finite perturbations remains an open question, but results from numerical relativity and the gravitational wave events detected by LIGO~\cite{Abbott:2016blz,Abbott:2016nmj,Abbott:2017vtc} feature the Kerr black hole as the end state of black hole mergers.  

Moreover, members of the Kerr family are characterized by two asymptotically measured and conserved quantities: mass and angular momentum. The famous \textit{no hair theorem} states that any stationary black hole in vacuum \textit{only} has these two degrees of freedom~\cite{Carter:1971zc}. As a matter of course, this leads to the question if there are further black hole solutions that do not belong to the Kerr family. According to the \textit{black hole uniqueness theorem} the answer is no~\cite{Robinson:1975aa}: If we choose allowed values for mass and angular momentum then there is only one black hole solution to Einstein's equations and this solution belongs to the Kerr family. Obviously, the uniqueness theorem also constrains the possible event horizon topology of a black hole as all Kerr solutions have a spherical horizon topology. The fact that there are only asymptotically flat stationary black hole solutions with spherical event horizon topology was separately proven by Hawking some years before~\cite{Hawking:1973uf}.

We conclude that the phase space of stationary four dimensional black holes in vacuum is rather simple. There is the Kerr family and nothing more.\footnote{We note that when taking electric charge into account, which requires a non-zero right hand side of Einstein's equations~\eqref{eq:EFEvac}, the Kerr solution can even be generalized to the Kerr-Newman solution~\cite{Newman:1965aa,Newman:1965ab} that is described by its mass, angular momentum and electric charge. The no hair theorem~\cite{Robinson:1974nf} as well as the uniqueness theorem~\cite{Mazur:1982db,Blunting:1983aa} can be expanded to hold in this situation as well.} Later we will see that things change dramatically when going to higher dimensions.

\subsection{Black hole thermodynamics}
\label{subsec:Black_hole_thermodynamics}

There are more physical quantities besides the mass $M$ and the angular momentum $J$ that play an important role in black hole physics, in particular, the surface area $A_\mathcal H$ and the surface gravity $\kappa$ of the event horizon. The mathematical definition of the latter is rather technical, thus we do not state it here but refer to any standard textbook of GR, for example reference~\cite{Wald:1984aa}. In simple but not necessarily accurate terms, the surface gravity is the gravitational acceleration at the horizon. Moreover, if the black hole is rotating and thus has a finite angular momentum, one can associate an angular velocity $\Omega _\mathcal H$ with the horizon.    

The physical quantities discussed above allow us to formulate the four laws of black hole mechanics, which concern stationary black hole spacetimes. Bardeen, Carter and Hawking were the first to write down these laws in 1973~\cite{Bardeen:1973gs}:\footnote{However, law \S 3 was only proven later by Israel~\cite{Israel:1986aa}.}
\begin{enumerate}
	\renewcommand{\theenumi}{\S \arabic{enumi}}
  	\setcounter{enumi}{-1}
	\item The surface gravity $\kappa$ is constant over the event horizon.
	\item Consider two slightly different stationary black hole solutions, one with mass $M$, angular momentum $J$ and surface area $A_\mathcal H$, and one with parameters $M+\delta M$, $J+\delta J$ and $A_\mathcal H + \delta A _\mathcal H$. Then, the differences of mass, angular momentum and surface area satisfy
			\beq
			 	\delta M = \frac{\kappa}{8\pi\, G_4} \, \delta A_\mathcal H + \Omega _\mathcal H \, \delta J \, .
				\label{eq:FirstLaw}
			\eeq
	\item The surface area of a black hole can never decrease, i.e.\ $\delta A_\mathcal H \geq 0 $.
	\item No procedure can reduce the surface gravity $\kappa$ to zero in finite time.
\end{enumerate}
The most remarkable feature of these laws is their formal analogy to the four laws of thermodynamics. In this sense the surface gravity $\kappa$ corresponds to the temperature $T$, the surface area $A_\mathcal H$ to the entropy $S$ and the mass $M$ to the internal energy $U$. This analogy turned out to be a physical phenomenon when Hawking showed that a black hole indeed emits thermal radiation if it is coupled to quantum matter fields~\cite{Hawking:1974rv}. He calculated the temperature of a black hole to be\footnote{Note that we use units in which Planck's constant reads $\hbar =1$. Recall that we also have $c=1$.}
\beq
	T = \frac{\kappa}{2\pi} \, .
	\label{eq:TemperatureBH}
\eeq
Moreover, the comparison of the first law of black hole dynamics~\eqref{eq:FirstLaw} with the fundamental thermodynamic equation $\delta U = T\,\delta S - P\,\delta V$ yields an expression for the entropy of a black hole
\beq
	S = \frac{A_\mathcal H}{4\, G_4} \, .
	\label{eq:EntropyBH}
\eeq
Therefore, the above laws are usually referred to as the laws of \textit{black hole thermodynamics}. However, the process of a black hole emitting energy implies that it is shrinking and thus violating the second law. Thus, the second law is rewritten to take into account the total entropy, i.e.\ the sum of the entropies of the black hole and the radiation.

We find another surprising property of black holes by virtue of the thermodynamic interpretation. The temperature of a Schwarzschild black hole reads $T=1/(8\pi\, G_4M)$. Consequently, the temperature of the black hole increases when its mass decreases, i.e.\ the specific heat $\partial M/\partial T$ is negative. In other words: The smaller the black hole is the more energy it radiates away in a given time.

Finally, we conclude by noting that the thermodynamic interpretation of black holes gives rise to a deep connection between GR, quantum field theory and statistical mechanics. Though this connection is not fully understood yet, it manifests for example in the holographic principle or more specifically in the AdS/CFT correspondence, see for instance reference~\cite{Ammon:2015wua}. In this sense, and for the discussion below, we emphasize that the laws of black hole thermodynamics naturally adapt to higher dimensions. 

\subsection{Black holes in higher dimensions}
\label{subsec:Black_holes_in_higher_dimensions}

Now we consider black hole solutions in higher dimensions, but for a moment we restrict ourselves to the situation where all of the additional spatial dimensions are infinitely extended. Therefore, all objects discussed in this subsection approach Minkowski spacetime~\eqref{eq:metricMinkowski} in the asymptotic limit. 

The $D$-dimensional generalization of the Schwarzschild spacetime was found 1963 by Tangherlini~\cite{Tangherlini:1963aa} and reads
\beq
	\D s^2_\text{ST} = - f_D(r) \, \D t^2 + \frac{\D r^2}{f_D(r)} + r^2 \, \D \Omega^2_{D-2} \, ,
	\label{eq:metricST}
\eeq
where the function $f_D$ generalizes $f_4$ to
\beq
	f_D(r) = 1- \left( \frac{r_0}{r} \right) ^{D-3} \, ,
	\label{eq:fD}
\eeq
and  $\D \Omega ^2_{D-2} = \D \theta ^2 + \sin ^2 \theta  \, \D \Omega ^2_{D-3} $ is the line element of a ($D-2$)-sphere. Again, if the matter distribution is compact enough, we find a horizon at $r=r_0$, which represents a coordinate singularity. 

The mass of the Schwarzschild-Tangherlini (ST) black hole is given by 
\beq
	M_\ST = \frac{(D-2)\Omega _{D-2} \, r_0^{D-3}}{16\pi\, G_D} \, , 
	\label{eq:massSTS}
\eeq
where $\Omega _{D-2}$ denotes the surface area of a unit ($D-2$)-sphere and $G_D$ denotes the $D$-dimensional gravitational constant. 

Much like the Kerr solution there is a rotating black hole solution in $D$ dimensions, which was derived by Myers and Perry in 1986~\cite{Myers:1986un}. Since there can be more than only one rotation axis in $D>4$ things become highly involved here. In fact, for every pair of spatial coordinates one can introduce a polar coordinate chart that defines an axis of rotation. Consequently, in $D$ dimensions there are $\lfloor (D-1)/2 \rfloor$ independent rotations possible, each described by a separate angular momentum. 

The Myers-Perry solution contains the Kerr solution for $D=4$. In $D=5$ dimensions both angular momenta can not exceed finite values. Remarkably, this changes for $D\geq 6$, where one of the angular momenta can, in principle, become arbitrarily large as long as some of the others vanish. Such solutions are called ultra-spinning Myers-Perry black holes. In the ultra-spinning regime, where at least one angular momentum is much larger than the others, the event horizon extremely flattens out. This gives rise to an instability due to the tendency of gravity to bind matter in a small region.\footnote{This instability is related to the Gregory-Laflamme instability, which we will explain in section~\ref{subsec:Gregory-Laflamme_instability}.} 

\begin{figure}[ht]
	\centering
	\includegraphics[scale=1]{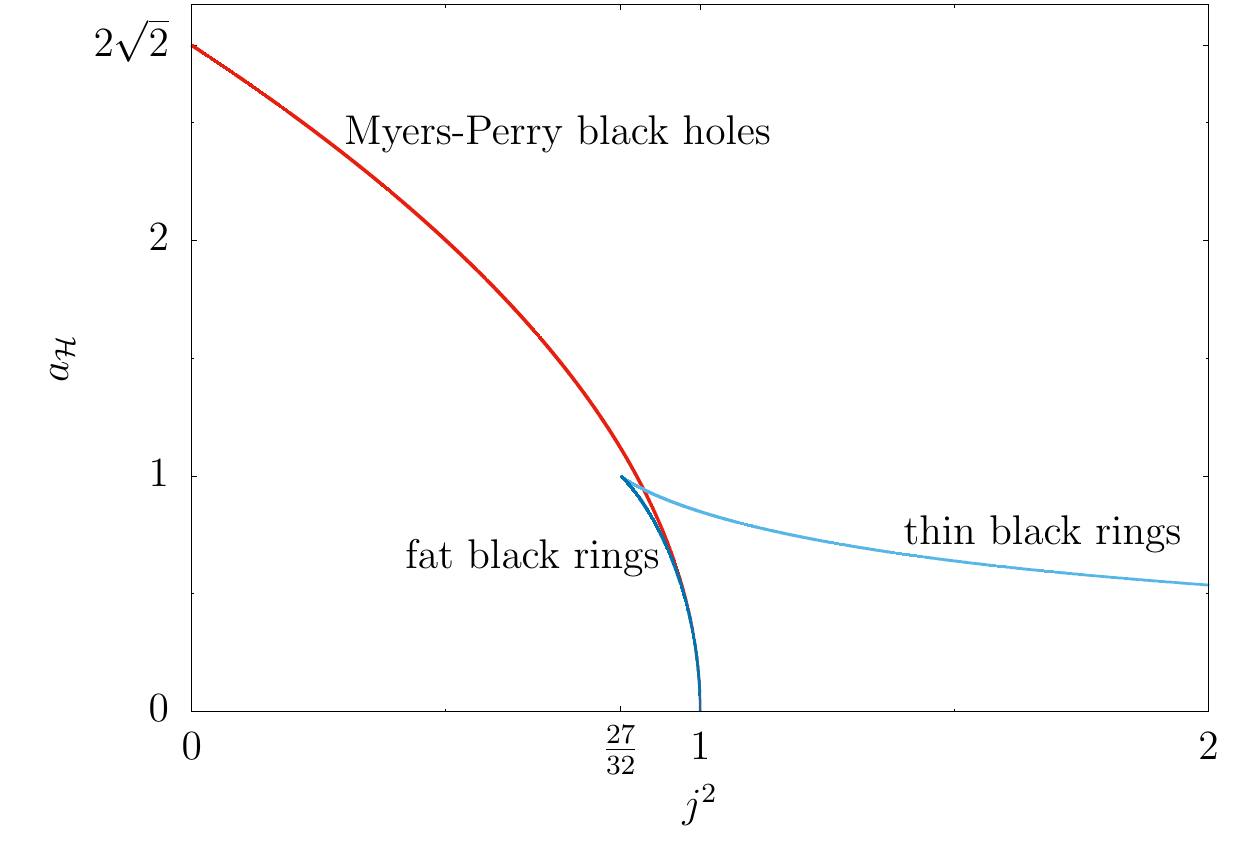}
	\caption{Phase diagram of black rings and Myers-Perry black holes with one vanishing angular momentum in $D=5$. We plot the so-called reduced area of the horizon $a_\mathcal{H} = 3^{3/2} \, A_\mathcal{H} / ( 16 \pi ^{1/2} \, G_5^{3/2} M^{3/2} )$ over the so-called reduced spin squared $j^2 = 27\pi \, J^2 / ( 32\, G_5 M^3 )$. In the range $27/32<j^2<1$ there are three different kinds of solutions: fat (dark blue line) and thin (light blue line) black rings and Myers-Perry black holes (red line). The fat black ring and Myers-Perry black hole branches meet at $(j^2,a_\mathcal{H}) = (1,0)$. }
	\label{fig:BlackRingPhaseDiagram}
\end{figure}

From the discussion in four dimensions one could assume that also for $D>4$ there are no black hole solutions other than the Myers-Perry ones. But about 15 years ago Emparan and Reall explicitly proofed the contrary by constructing a rotating black ring solution in five dimensions~\cite{Emparan:2001wn}. The event horizon of the black ring has the topology $\mathbb S^2\times\mathbb S^1$, which obviously differs from the $\mathbb S^3$ topology of the five-dimensional Myers-Perry solutions. Therefore, Hawking's theorem about the event horizon topology in four dimensions does not apply to higher dimensions. Moreover, if we put black rings and Myers-Perry black holes with one vanishing angular momentum together in a phase diagram, then there is a small range, where three different solutions coexist, fat and thin black rings and Myers-Perry black holes, see figure~\ref{fig:BlackRingPhaseDiagram}. Hence we do have an explicit counter example for black hole uniqueness in higher dimensions! However, Reference~\cite{Santos:2015iua} provides strong evidence that the whole black ring branch is unstable.\footnote{Again, for thin black rings the instability is of Gregory-Laflamme type, see section~\ref{subsec:Gregory-Laflamme_instability}. In contrast, fat black rings are unstable against axisymmetric perturbations.} Black ring solutions were also constructed in six and seven dimensions by using numerical techniques~\cite{Kleihaus:2012xh,Dias:2014cia}.

In 2007 Elvang and Figueras found another remarkable solution to the five-dimensional vacuum Einstein equations that describes a central black hole surrounded by a black ring, thus called a black saturn~\cite{Elvang:2007rd}. The black saturn gives an explicit counter example of the no hair theorem in higher dimensions in its original four-dimensional form, since it is described by a set of four parameters where only two of them are conserved. Nevertheless, one could reformulate the no hair condition without the assumption of conserved quantities. Then, the no hair theorem is expected to hold in higher dimensions as well, i.e.\ all black hole solutions could be described by a finite set of parameters, but a rigorous proof is still pending.

There are a lot more asymptotically flat black hole solutions in higher dimensions, for example multi black ring solutions~\cite{Iguchi:2007is,Evslin:2007fv,Elvang:2007hs,Izumi:2007qx} or so-called black ringoids~\cite{Kleihaus:2014pha}, but this work does not aim to give a comprehensive review of the zoo of higher dimensional black objects and their properties. For this purpose we refer to references~\cite{Emparan:2008eg,Horowitz:2012nnc,Kleihaus:2014pha} but we remark that the zoo is still growing. We gave several examples in order to highlight that neither the no hair theorem nor the uniqueness theorem for stationary black holes adapt straightforwardly to $D>4$ and that there exist solutions with different event horizon topologies and exotic properties. In turn, there is a modified uniqueness theorem in higher dimensions, which states that the ST black hole~\eqref{eq:metricST} is the only \textit{static} asymptotically flat black hole solution~\cite{Gibbons:2002bh}. However, if one allows at least one dimension to be compact, even this theorem does not apply anymore. Static black holes in such a situation will concern us for the remainder of this work.

\section{Static Kaluza-Klein black holes}
\label{sec:Static_Kaluza_Klein_black_holes}

We now turn our attention to black holes in spacetimes with one compact periodic dimension. Since the idea of compact extra dimensions originates from Kaluza and Klein we term such objects Kaluza-Klein (KK) black holes. However, our intention is different from the original one of Kaluza and Klein. We will not interpret the additional degrees of freedom given by the compact dimension as matter fields but we will rather consider the problem in a geometrical way as motivated in the introduction. 

In particular, in this section we review \textit{static} black holes in KK theory. Note that there are already several excellent and much more detailed reviews of this topic~\cite{Kol:2004ww,Harmark:2005pp,Horowitz:2012nnc}, on which this section mainly relies. Here, we start with a discussion of the background metric in subsection~\ref{subsec:Background_metric}. We define the most relevant physical quantities in subsection~\ref{subsec:Physical_quantities} before we discuss the solutions that we are interested in and their properties in subsections~\ref{subsec:Uniform_black_strings} to~\ref{subsec:End_state_of_the_Gregory-Laflamme_instability}. Thereafter, in subsections~\ref{subsec:Phase_diagram_of_static_Kaluza-Klein_black_holes} and~\ref{subsec:Double-cone_metric} we summarize the state of the art by discussing the phase diagram and its conjectured completion. For completeness, we mention some more exotic black hole solutions in KK theory in subsection~\ref{subsec:Copies_and_bubbles}.  

\subsection{Background metric}
\label{subsec:Background_metric}

The Minkowski spacetime $\mathbb M^D$, see equation~\eqref{eq:metricMinkowski}, serves as a background metric in $D$ dimensional asymptotically flat space. If one of the spatial dimensions is of finite size $L$ and of periodic nature, it has the topology of a circle $\mathbb S^1$. Therefore, in KK theory we consider the direct product $\mathbb M^{D-1}\times \mathbb S^1$ as the background metric reading
\beq
	\D s^2_{\text{BG}} = -\D t^2 + \D r^2 + r^2\, \D \Omega ^2_{D-3} + \D z^2 \, .
	\label{eq:metricBG}
\eeq  
The coordinate $z$ denotes the compact dimension, thus we have $z\in [-L/2,L/2]$. For later convenience, we have expressed the $D-2$ spatially extended dimensions in (hyper-)spherical coordinates with the radial coordinate $r\in [0,\infty ]$.\footnote{We use ``hyper-spherical'' as a synonym for the higher dimensional meaning of spherical without restricting ourselves to a certain dimension. Accordingly, ``(hyper-)spherical'' indicates that the usual three-dimensional case is included.} We emphasize that any spacetime in KK theory with only one compact dimension shall approach the background metric~\eqref{eq:metricBG} in the limit $r\to \infty$.

\subsection{Physical quantities}
\label{subsec:Physical_quantities} 

Before we explicitly discuss black hole solutions in KK theory, we define the relevant physical quantities that any KK black hole can be associated with. In particular, we concentrate on two asymptotically measured charges and the thermodynamic quantities that make up the first law of black hole thermodynamics in KK theory. Some more specific quantities will be defined at later stages of this thesis. 

\subsubsection{Asymptotic charges}

The presence of a black hole causes the following leading order corrections to the background metric~\eqref{eq:metricBG} at infinity $r\to\infty$~\cite{Kol:2003if,Harmark:2003dg} 
\beq
	 g_{tt} \simeq - 1 + \frac{c_t}{r^{D-4}} \, , \quad  g_{zz} \simeq 1 + \frac{c_z}{r^{D-4}} \, .
	\label{eq:asymptotic_corrections}
\eeq
Using two different linear combinations of the coefficients $c_t$ and $c_z$ we obtain two physical quantities~\cite{Kol:2003if,Harmark:2003dg} 
\begin{align}
	M &=  \frac{L \Omega _{D-3}}{16 \pi G_D} \left[ (D-3) c_t - c_z  \right] \, , \label{eq:Mass} \\
	\mathcal T &= \frac{\Omega _{D-3}}{16 \pi G_D} \left[ c_t - (D-3) c_z \right] \, . \label{eq:Tension}
\end{align}
As usual $M$ denotes the total mass while $\mathcal T$ is referred to as the tension. We get an intuition of the physical meaning of the tension by inverting equations~\eqref{eq:Mass} and~\eqref{eq:Tension}:
\begin{align}
	c_t & = \frac{16 \pi G_D}{(D-2)(D-4) L \Omega _{D-3}} \left[ (D-3) M -       L\mathcal T \right]	 \, , \label{eq:ct} \\
	c_z & = \frac{16 \pi G_D}{(D-2)(D-4) L \Omega _{D-3}} \left[       M - (D-3) L\mathcal T \right]	 \, . \label{eq:cz} 
\end{align} 
From equations~\eqref{eq:asymptotic_corrections} and~\eqref{eq:cz} we see that increasing the mass increases the $g_{zz}$ component of the metric near infinity and thus corresponds to a leading order expansion of the compact dimension. In contrast, the tension can be seen as a counter force that compresses the size of the compact dimension. 

A convenient normalization of the tension reads
\beq
	n = \frac{L\mathcal T}{M} = \frac{c_t - (D-3) c_z}{(D-3) c_t - c_z} \, 
	\label{eq:RelativeTension}
\eeq
where $n$ is called the relative tension. There are two bounds on $n$ namely~\cite{Traschen:2003jm,Shiromizu:2003gc,Harmark:2003dg}
\beq
	0\leq n\leq D-3 \, .
	\label{eq:RelativeTensionsBounds}
\eeq
Later we will use the relative tension $n$ rather than the tension $\mathcal T$ for characterizing black hole solutions.

\subsubsection{Thermodynamics}

The surface gravity $\kappa$ and the horizon area $A_\mathcal H$ play an important role for KK black holes as well, since they are interpreted as the temperature $T=\kappa /(2\pi )$ and the entropy $S=A_\mathcal H/(4G_D)$ of the black hole. With the new asymptotic charge, the relative tension $n$, the first law of black hole thermodynamics modifies to~\cite{Kol:2003if,Harmark:2003dg}
\beq
	\delta M = T \, \delta S + \mathcal T \, \delta L = T \, \delta S + \frac{nM}{L} \, \delta L \, .
	\label{eq:FirstLawKK}
\eeq
If we compare equation~\eqref{eq:FirstLawKK} with the fundamental thermodynamic equation $\delta U = T\,\delta S - P\,\delta V$ we see again that the tension has the meaning of a force by which the black object tries to compress the length $L$ of the compact dimension. The situation is different in thermodynamics, where the pressure $P$ tries to expand a volume $V$, therefore explaining the opposite sign in the first law. Note that if we fix $L$, the first law reduces to $\delta M = T\, \delta S$.

Furthermore, in the given context Smarr's relation reads~\cite{Kol:2003if,Harmark:2003dg}
\beq
	(D-2) TS = (D-3-n) M  \, .
	\label{eq:SmarrRelation}
\eeq
It represents an integrated version of the first law of black hole thermodynamics.

We note that both the first law and Smarr's relation can serve as non-trivial consistency tests for a numerically obtained solution that describes a KK black hole, as they relate horizon quantities with the asymptotic coefficients $c_t$ and $c_z$.\footnote{In fact, Smarr's relation only contains $c_t$, since $c_z$ drops out of the right hand side of equation~\eqref{eq:SmarrRelation}.}

\subsection{Uniform black strings}
\label{subsec:Uniform_black_strings}

The simplest black hole solution in KK theory describes a \textit{uniform black string} (UBS) given by the metric
\beq	
	\D s^2_\UBS = - f_{D-1}(r) \, \D t^2 + \frac{\D r^2}{f_{D-1}(r)} + r^2 \, \D \Omega ^2_{D-3} + \D z^2 \, ,
	\label{eq:metricUBS}
\eeq
where the function $f_{D-1}$ is given by equation~\eqref{eq:fD}.\footnote{Sometimes UBSs are also referred to as homogenous black strings.} It is apparent from equation~\eqref{eq:metricST} that this spacetime is a direct product of a ($D-1$)-dimensional ST solution and a circle. Since both components separately solve Einstein's vacuum field equations, their direct product~\eqref{eq:metricUBS} is a solution as well. Therefore, any $z=\text{const.}$\ slice of the UBS spacetime resembles exactly an ST solution.

\begin{figure}[ht]
	\centering
		\hfill
		\includegraphics[scale=1]{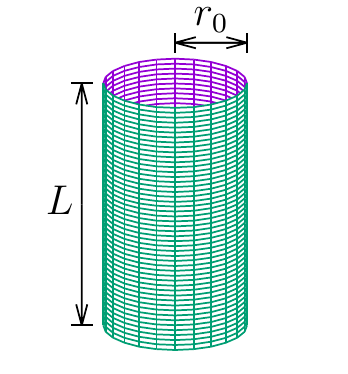} \hfill
		\includegraphics[scale=1]{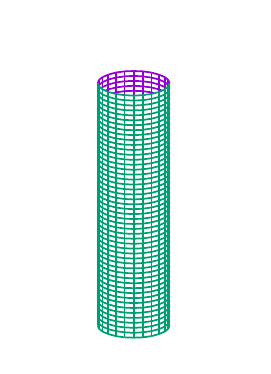}	\hfill
		\includegraphics[scale=1]{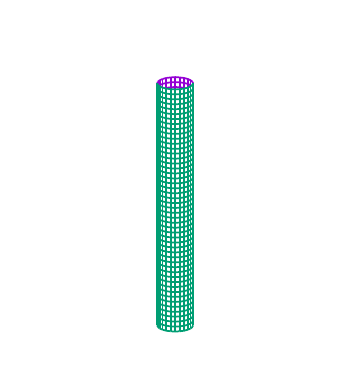} \hfill
	\caption{Spatial embeddings of UBS horizons with different ratios $L/r_0$. The vertical direction corresponds to the compact coordinate $z$ with length $L$. The $z=\text{const.}$\ slices of the horizon are (hyper-)spheres with radius $r_0$, here illustrated as circles. Due to the periodic nature of the compact dimension the end points of the string correspond to the same points in spacetime.}
	\label{fig:horizonsUBS}
\end{figure}

The horizon of a UBS resides at $r=r_0$, where the radius $r_0$ is defined within the function $f_{D-1}$, see equation~\eqref{eq:fD}. Obviously, the horizon radius is uniform along the circle and the horizon wraps around the compact dimension like a string. The topology of the horizon is consequently $\mathbb S^{D-3}\times \mathbb S^1$. Figure~\ref{fig:horizonsUBS} illustrates different UBS horizons. 

Using the close connection of ST black hole and UBS we immediately get an expression for the mass of the latter from equation~\eqref{eq:massSTS}
\beq
	M_\UBS = \frac{(D-3)\Omega _{D-3}\, r_0^{D-4}L}{16\pi\, G_D} = \frac{(D-3)\Omega _{D-3}\, r_0^{D-4}}{16\pi\, G_{D-1}} \, ,
	\label{eq:massUBS}
\eeq
where we have made use of the fact that the circle size $L$ relates the gravitational constants by $G_D = G_{D-1} L$. In fact, the mass of a $D$-dimensional UBS equals the mass of the corresponding ($D-1$)-dimensional ST black hole, cf. equation~\eqref{eq:massSTS}. Interestingly, the same is true for the entropy of the UBS since
\beq
	S_\UBS = \frac{A_{\mathcal H,\UBS}}{4\,G_D} = \frac{\Omega _{D-3}\, r_0^{D-3} L}{4\,G_{D-1}L} = \frac{\Omega _{D-3}\, r_0^{D-3} }{4\,G_{D-1}} \, ,
	\label{eq:entropyUBS}
\eeq	
where the numerator on the right hand side is obviously the horizon area of a $(D-1)$-dimensional ST black hole and hence the whole expression gives the entropy of an ST black hole. 

For completeness, we give the UBS values of relative tension and temperature:
\beq
	n_\UBS = \frac{1}{D-3} \, ,
	\label{eq:tensionUBS}
\eeq
\beq
	T_\UBS = \frac{D-4}{4\pi\, r_0} \, .
	\label{eq:temperatureUBS}
\eeq

There are two free parameters in the UBS solution: the size of the circle $L$ and the radius of the black string horizon $r_0$. Due to the scale invariance of GR it only makes sense to distinguish between solutions that have different ratios $L/r_0=:K$. 

\subsection{Localized black holes}
\label{subsec:Localized_black_holes}

\begin{figure}[ht]
	\centering
		\includegraphics[scale=1]{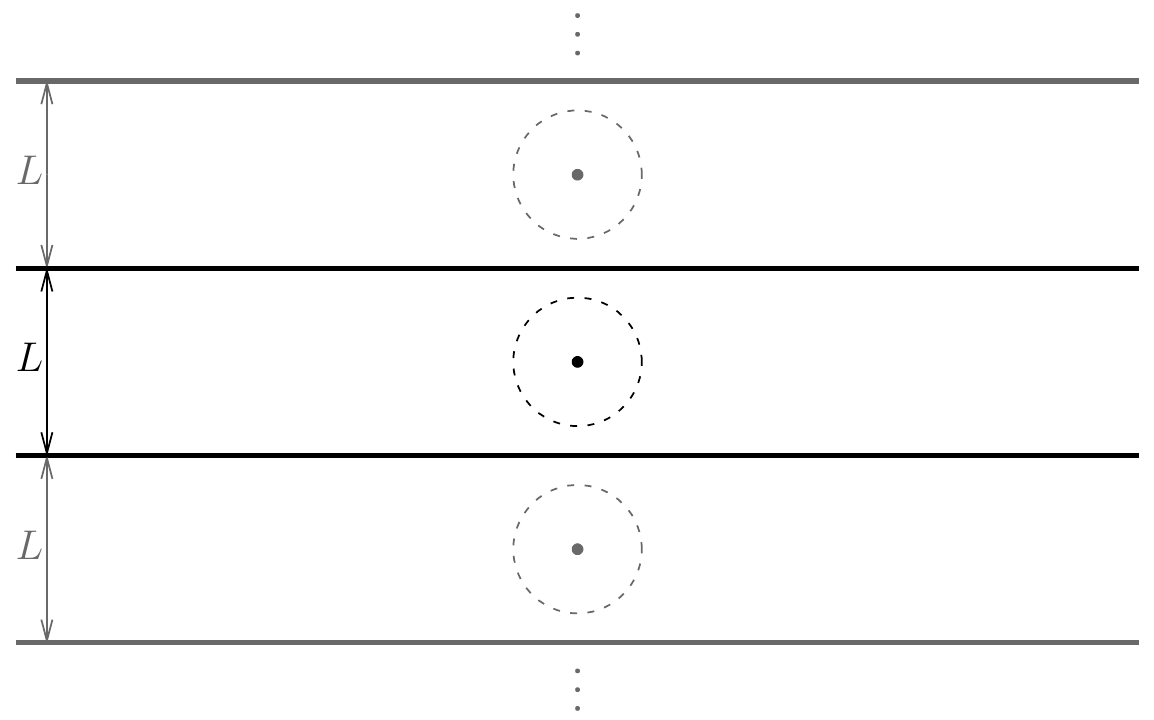}
	\caption{Sketch of a small black hole in a spacetime with one compact periodic dimension of size $L$ and its mirror images. The small black filled circle represents the black hole with horizon topology $\mathbb S^{D-2}$. The dashed circle indicates that the region in vicinity of the black hole is well approximated by the $D$ dimensional ST metric~\eqref{eq:metricST}. Close to the edges of the compact dimension (indicated by the solid horizontal lines) the ST solution is not appropriate to describe the spacetime due to its lack of periodicity.}
	\label{fig:horizonsLBH2Dsketch}
\end{figure}

Consider now a hyper-spherical black object in $D$ dimensions, i.e.\ one with horizon topology $\mathbb S^{D-2}$, where one dimension is of finite size $L$. Clearly, the size $l$ of this object has to be smaller than $L$. We call such objects \textit{localized black holes} (LBHs) as they are localized on the compact dimension.\footnote{Sometimes they were also referred to as caged black holes.}  In the limit $L\to \infty$ an LBH represents a $D$-dimensional ST black hole. Even if $L$ is finite but $L\gg l$, the spacetime in the vicinity of the black hole is locally well approximated by the ST metric~\eqref{eq:metricST}, see figure~\ref{fig:horizonsLBH2Dsketch} for an illustration. However, close to the periodic boundaries the ST metric does not satisfy the periodic boundary conditions. 

\begin{figure}[ht]
	\centering
		\hfill
		\includegraphics[scale=1]{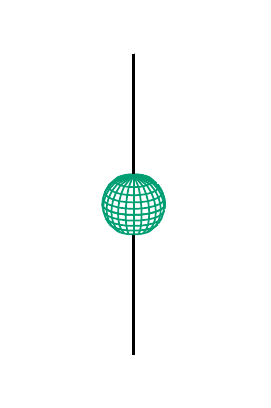} \hfill
		\includegraphics[scale=1]{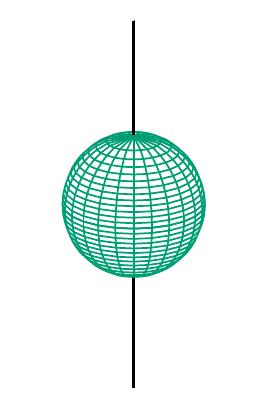} \hfill
		\includegraphics[scale=1]{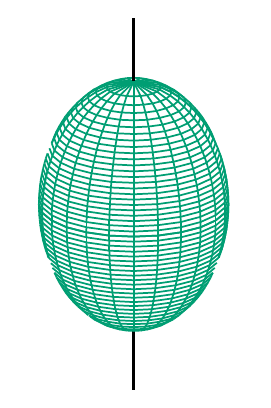} \hfill
		\includegraphics[scale=1]{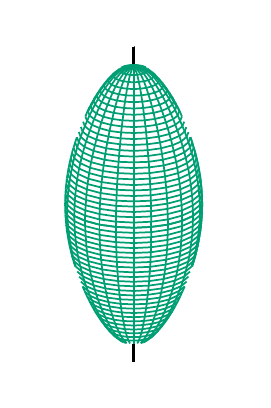} \hfill
		\hfill
	\caption{Spatial embeddings of LBH horizons with different size in $D=5$. The vertical direction corresponds to the compact coordinate $z$. The $z=\text{const}$\ slices of the horizon are (hyper-)spheres, here illustrated as circles. In addition, the axis of (hyper-)spherical symmetry is displayed, which indicates the finite length of the compact dimension. The local length of the compact dimension varies depending on the size of the LBH.}
	\label{fig:horizonsLBH3D}
\end{figure}

Due to the periodicity along the compact dimension one can consider the LBH to have infinitely many mirror images, cf. figure~\ref{fig:horizonsLBH2Dsketch}. Therefore, the horizon shape of an LBH will not be exactly spherical but rather a little bit stretched along the compact dimension, since it is subject to the gravitational field of its mirror images. Again, for small LBHs the effect is nearly negligible, but it becomes significant for larger LBHs, see figure~\ref{fig:horizonsLBH3D}. For the study of small LBHs it is possible to develop a perturbative ansatz that matches the different contributions of the ST spacetime and the appropriate asymptotic behavior at each order of the expansion~\cite{Harmark:2003yz,Gorbonos:2004uc,Gorbonos:2005px}. To get beyond the perturbative regime numerical techniques have to be applied, as was done in a number of works, mainly for $D=5$ or $D=6$~\cite{Wiseman:2002ti,Sorkin:2003ka,Kudoh:2003ki,Kudoh:2004hs,Headrick:2009pv} and very recently for $D=10$~\cite{Dias:2017uyv}. These results suggest that the mass of the LBH is not unbounded, since at some point it will no longer fit into the compact dimension. Most of the numerical implementations mentioned above break down way before this point is reached. In chapter~\ref{chap:Numerical_construction_of_localized_black_hole_solutions} we will present a highly accurate numerical scheme that is capable to construct LBH solutions even in this regime. 

Now that we know already two types of KK black holes, UBSs and LBHs, we can confirm explicitly that the uniqueness theorem for static black holes does not hold in KK theory. Moreover, these two types of black holes have different horizon topologies. But which of these solutions is physically preferred for instance after gravitational collapse? Obviously, there is no problem for high masses, since LBH solutions only exist up to a finite mass. For small masses we can make use of the thermodynamic interpretation of black holes explained in subsection~\ref{subsec:Black_hole_thermodynamics}. In this sense it is the solution with highest entropy which is thermodynamically preferred. We estimate the entropy of a small LBH from the corresponding ST black hole and compare it with the entropy of a UBS, cf. equation~\eqref{eq:entropyUBS}. Writing the entropies in terms of mass $M$ and circle size $L$ and omitting unessential constants we get
\beq
	S_\LBH \sim L^{\frac{1}{D-3}} M^{\frac{D-2}{D-3}} \, , \quad S_\UBS \sim M^{\frac{D-3}{D-4}} \, .
	\label{eq:entropiesLBHUBS}
\eeq
We observe that for fixed $L$ and sufficiently small $M$ the entropy of an LBH will be greater than the entropy of a UBS. Therefore, one may expect the UBS solution to be unstable for small masses.

\subsection{Gregory-Laflamme instability}
\label{subsec:Gregory-Laflamme_instability}

\begin{figure}[ht]
	\centering
		\includegraphics[scale=1]{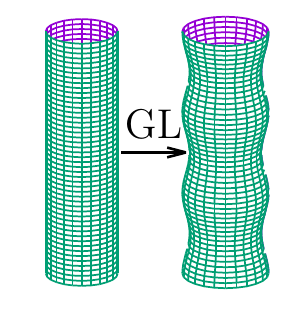}
	\caption{Schematic sketch of the deformation of the horizon of an unstable UBS perturbed by a GL mode. The perturbations that trigger the GL instability are non-uniform along the $z$-direction, thus the shape of the horizon changes.}
	\label{fig:horizonsUBSperturb}
\end{figure}

The thermodynamic argument given above led Gregory and Laflamme to study linear perturbations around the UBS in the early 1990s~\cite{Gregory:1993vy,Gregory:1994bj}. They have found that these perturbations are exponentially decaying in time only if $K=L/r_0$ is small enough. Consequently, there is a threshold value $K_\GL$ where the UBS is marginally stable. Solutions with $K>K_\GL$ are subject to the Gregory-Laflamme (GL) instability meaning that small perturbations around these objects will lead to large deformations and a redistribution of mass, see figure~\ref{fig:horizonsUBSperturb}, until the object settles down to a new, stable configuration. In contrast, if $K<K_\GL$, the energetic costs of the deformation are too high to destabilize the UBS. We list the values of $K_\GL$ in different dimensions in table~\ref{tab:KGL}.
\begin{table}[ht]
	\centering
	\caption{Critical value $K_\GL$ at which the UBS is marginally stable for different dimensions $D$. For $L/r_0>K_\GL$ the UBS is subject to the GL instability, while for $L/r_0<K_\GL$ it is dynamically stable against small perturbations. The values are taken from reference~\cite{Kol:2004pn} and they arise from from numerical calculations.}
	\label{tab:KGL}
 	\begin{tabularx}{\textwidth}{*{12}{Y}}		
 		\toprule
		 $D$		& 5 	& 6		& 7		& 8 	& 9 	& 10	& 11	& 12	& 13	& 14	& 15	\\
		\midrule
		 $K_\GL$ 	& 7.17	& 4.95	& 3.98	& 3.40	& 3.01	& 2.73	& 2.51	& 2.34	& 2.19	& 2.07	& 1.97	\\
 		\bottomrule
	\end{tabularx}	
\end{table} 

One can view the GL instability as an aspect of the generic feature of gravity to form compact objects rather than widely spread structures. Another famous instance of this effect is the instability of an interstellar gas cloud, which was first recognized by Jeans in 1902~\cite{Jeans:1902aa}. In the context of Newtonian gravity Jeans showed that a spherical matter distribution with uniform density and pressure is unstable to gravitational collapse if its size exceeds a critical value. See reference~\cite{Harmark:2007md} for a nice comparison of GL and Jeans instability.

Moreover, instabilities of GL type appear in several other configurations of higher dimensional black objects. The original work by Gregory and Laflamme itself~\cite{Gregory:1993vy} showed that even black branes are subject to the GL instability. Black branes are generalizations of UBS where the horizon is uniform on more than one dimensions. Furthermore, as mentioned earlier in subsection~\ref{subsec:Black_holes_in_higher_dimensions}, GL type instabilities occur for ultra-spinning Myers-Perry black holes and thin black rings.

\subsection{Non-uniform black strings}
\label{subsec:Non-uniform_black_strings}

The GL instability of black strings leads to the question what happens to a perturbed unstable UBS, i.e.\ to which kind of configuration will it evolve. Of particular interest is the question whether the horizon of a black string will finally pinch-off and form an LBH (or a sequence of LBHs). At the beginning of this millennium Horowitz and Maeda showed that this does not happen after finite horizon time~\cite{Horowitz:2001cz}, i.e.\ for finite values of the affine parameter along the horizon generators. They thus ruled out the LBH as the end state of the GL instability. Instead they conjectured a new type of static black string solutions to be the end state. This stimulated the search for black string solutions that are non-uniform along the circle, thus called \textit{non-uniform black strings} (NBSs).\footnote{Sometimes they were also referred to as inhomogenous black strings.} Gubser was first to construct NBS solutions in $D=5$ by developing a perturbation theory around the UBS~\cite{Gubser:2001ac}. Thereafter, Wiseman applied this perturbative scheme to $D=6$ and, moreover, he developed a numerical algorithm to obtain solutions beyond the perturbative regime~\cite{Wiseman:2002zc}. Sorkin provided a generalization of the perturbation theory to arbitrary dimensions~\cite{Sorkin:2004qq}. Later, a number of works numerically solved equations~\eqref{eq:NBSfield_eqns} directly in order to leave the perturbative regime. Results are presented in the dimensions $D=5$ up to $D=15$~\cite{Wiseman:2002zc,Kudoh:2004hs,Kleihaus:2006ee,Sorkin:2006wp,Headrick:2009pv,Figueras:2012xj,Dias:2017uyv}. 

\begin{figure}
	\centering
		\hfill
		\includegraphics[scale=1]{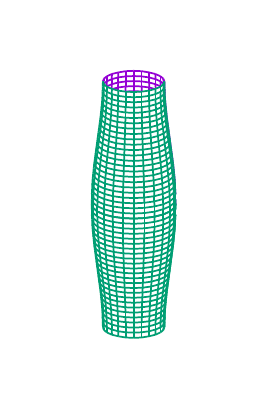} \hfill
		\includegraphics[scale=1]{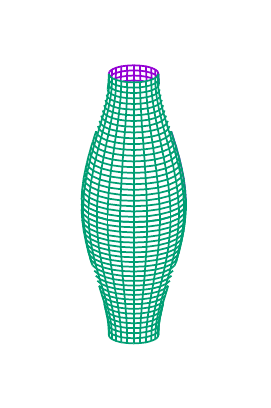} \hfill
		\includegraphics[scale=1]{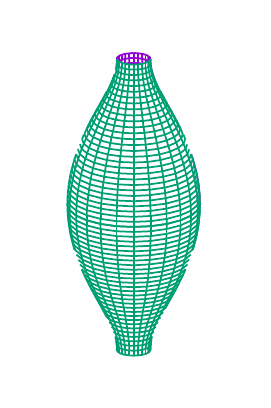} \hfill
		\includegraphics[scale=1]{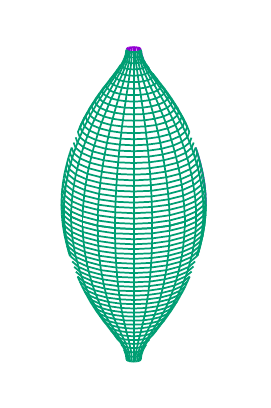} \hfill
		\hfill
	\caption{Spatial embeddings of NBS horizons with different shapes in $D=5$. The vertical direction corresponds to the compact coordinate $z$. The $z=\text{const}$\ slices of the horizon are (hyper-)spheres, here illustrated as circles. Depending on the shape of the NBS the local length of the compact dimension varies. Moving along the NBS branch the waist of the string is more and more shrinking. We refer to NBSs close to the UBS as slightly deformed NBSs (leftmost embedding) and to NBS with a nearly pinching horizon as strongly deformed NBSs (rightmost embedding).}
	\label{fig:horizonsNBS3D}	
\end{figure}

We illustrate the shape of different NBS horizons in figure~\ref{fig:horizonsNBS3D}. The non-uniformity results in the formation of a bulge region, where the string radius increases, and a waist region, where the string radius decreases. Moving along the NBS branch one observes that the string's waist is more and more shrinking. For a numerical implementation it is highly demanding to attain this critical regime where the waist of the NBS becomes extremely thin. In chapter~\ref{chap:Numerical_construction_of_non-uniform_black_string_solutions} we will present a highly accurate numerical scheme that is capable to tackle this regime for the first time. 

Coming back to the question of stability, the references mentioned above have drawn the following picture: Slightly deformed NBSs, i.e.\ those solutions close to the UBS, have lower entropy than the corresponding UBS with equal mass for $D\leq 13$, while for $D\geq 14$ the entropy is greater~\cite{Sorkin:2004qq}. Furthermore, the numerical results suggest that the \textit{whole} NBS branch has smaller entropy than the UBS at least for $D\leq 11$. In contrast, it appears that the \textit{whole} branch has greater entropy for $D\geq 14$, while this is the case only for a part of the branch in $D=12$ and $D=13$~\cite{Figueras:2012xj}. Consequently, at least for $D\leq 11$ NBSs can not serve as the final configuration of a perturbed UBS that is subject to the GL instability. This calls the conjecture of Horowitz and Maeda into question but we proceed in the next paragraph resolving this issue.

\subsection{End state of the Gregory-Laflamme instability}
\label{subsec:End_state_of_the_Gregory-Laflamme_instability}

The best way to identify the end state of the GL instability is to follow the time evolution of a perturbed unstable UBS. Therefore, we now leave the scope of static solutions to Einstein's vacuum equations for a moment. Already in 2003 Choptuik et~al.\ tackled the problem in $D=5$ numerically and observed that under time evolution the shape of the unstable UBS horizon becomes hyper-spherical but with its poles connected by a thin string along the compact dimension~\cite{Choptuik:2003qd}. Unfortunately, their code was not able to approach an equilibrium configuration. However, later studies~\cite{Garfinkle:2004em} provided an argument that an LBH can indeed be the end state of the GL instability, though it would take infinite horizon time to get there. They showed that, even if the horizon time diverges towards a pinch-off of the black string, the asymptotic time can stay finite, see also reference~\cite{Marolf:2005vn}. 

Finally, in 2010 Lehner and Pretorius were able to perform an improved simulation~\cite{Lehner:2010pn}, see reference~\cite{Lehner:2011wc} for a more detailed review. They showed that the thin string segment that forms is again subject to a GL instability, thus forming another smaller hyper-spherical object with its poles connected to the bigger one by even thinner black string segments. These new string segments then give rise to yet another transformation of this type and consequently to a cascade that will only terminate when the string segments reach zero size and the horizon pinches off. In this case a naked curvature singularity, i.e.\ a singularity not hidden by a horizon, will eventually form. Therefore, the final configuration is not accessible by numerics, but Lehner and Pretorius were able to extrapolate their data to show that the horizon will pinch off in finite asymptotic time. This is remarkable since it gives an explicit example of the violation of cosmic censorship in higher dimension from generic initial data.\footnote{In four-dimensional gravity the cosmic censorship conjecture states that the time evolution of realistic generic initial data can not lead to a naked singularity. Note that recent work found counter examples of this conjecture also in higher-dimensional asymptotically flat spacetimes, namely in the time evolution of perturbed thin black rings~\cite{Figueras:2015hkb} and ultraspinning Myers-Perry black holes~\cite{Figueras:2017zwa}.} However, the reasoning to avoid such a naked singularity relies on the fact that quantum effects will play a crucial role right before the horizon pinches off. Then, we would be left with an array of LBHs of different size arranged along the compact dimension. But this configuration is of course highly unstable against perturbations of the distance between two black holes. The consequence would be that the black holes will move towards each other and finally merge into a single LBH.  

In reference~\cite{Emparan:2015gva} the time evolution of a perturbed unstable UBS in the large $D$ limit was investigated. There, for thin enough UBSs the system shows a similar behavior as described above. In contrast, for initial black strings around the GL point the final configuration is an NBS. Indeed, these results are in accordance with the fact that for $D\geq 12$ there are NBS solutions with higher entropy than the UBS~\cite{Sorkin:2004qq,Figueras:2012xj}. Besides reference~\cite{Emparan:2015gva} no more time evolution of the unstable UBS for $D>5$ is present, therefore we can only speculate about the end state there. Most likely, for $D<12$ the situation is qualitatively similar to the findings of Lehner and Pretorius~\cite{Lehner:2010pn} described above while for $D\geq 12$ also the NBS can serve as a final state for perturbed unstable UBSs, at least around the GL point.

\subsection{Phase diagram of static Kaluza-Klein black holes}
\label{subsec:Phase_diagram_of_static_Kaluza-Klein_black_holes}

\begin{figure}[ht]
	\centering
		\includegraphics[scale=1]{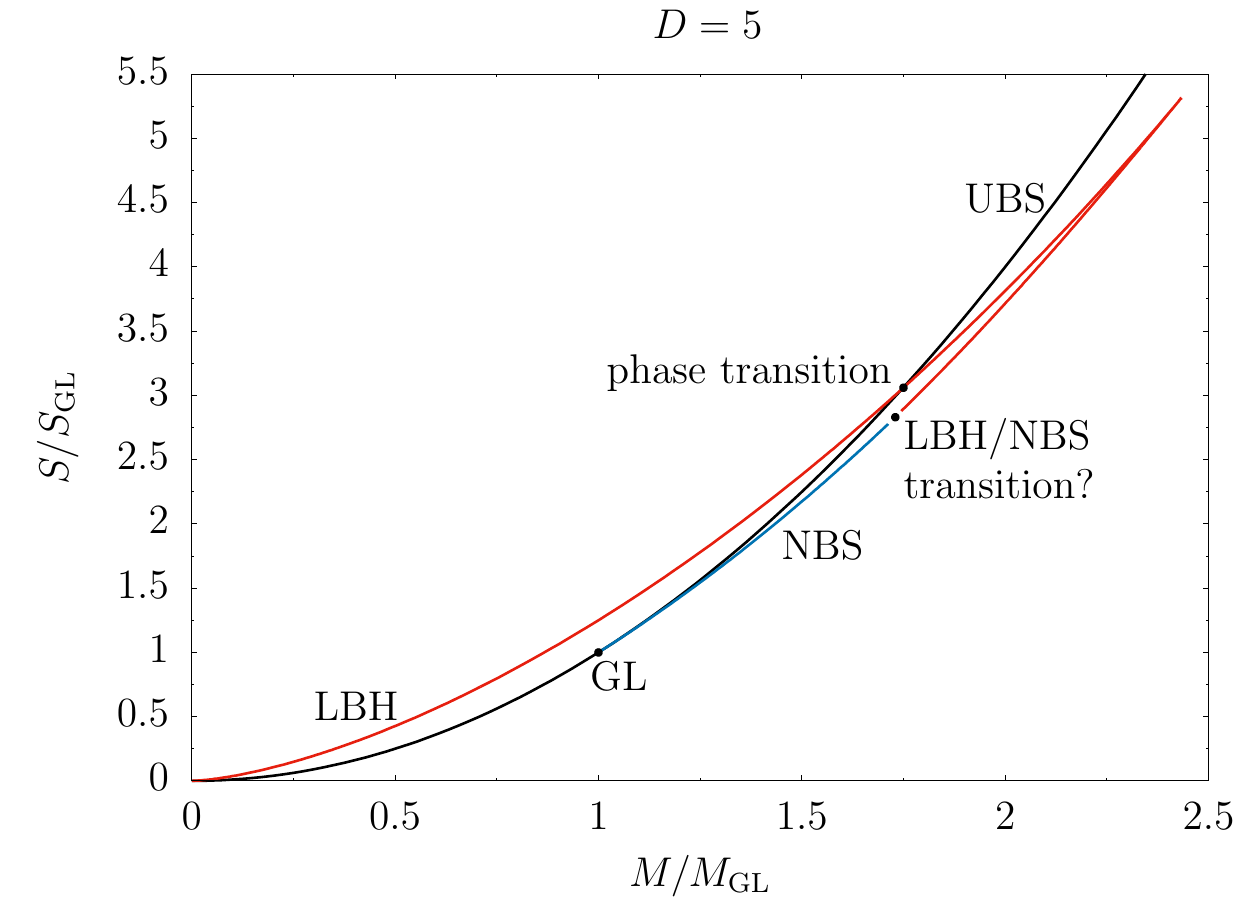} 
	\caption{Phase diagram of KK black holes in the microcanonical ensemble in $D=5$. Mass $M$ and entropy $S$ are normalized by the corresponding value of a UBS at the GL point. For small masses LBHs (red line) have highest entropy and are thermodynamically preferred. This changes when the LBH branch crosses the UBS branch (black line), i.e.\ for high masses the UBSs are thermodynamically preferred. The crossing point of the LBH and UBS branches thus marks a first order phase transition. The NBS branch (blue line) emanates from the GL instability but has lower entropy than the UBS branch. After the LBH branch reaches a maximum of entropy and mass it turns towards the NBS branch. It is expected that both branches will finally meet at a topology changing singular transit solution.}
	\label{fig:PhaseDiagram5Dold}	
\end{figure}

We come back to the static solutions discussed earlier and combine them now into one single phase diagram. The situation in $D=5$ is summarized in figure~\ref{fig:PhaseDiagram5Dold}, for more details see reference~\cite{Horowitz:2011cq}. 

Figure~\ref{fig:PhaseDiagram5Dold} displays the phase diagram in the microcanonical ensemble, where the configuration with highest entropy at given mass is the physically preferred one. We see that UBSs dominate for large masses while LBHs dominate for small masses. Indeed, this is what we have already learned from the estimation of the entropy of small LBHs, see equation~\eqref{eq:entropiesLBHUBS}. But even for moderate masses these two systems dominate the entropy and, consequently, there is a phase transition at the point where the two branches cross each other. Consider a UBS above this point. It has highest entropy and is therefore considered to be globally stable. If its mass is reduced, its stability changes from global to local when its entropy becomes smaller than that of an LBH. Below this point the UBS is locally stable until the GL point is reached and an unstable mode evolves. At the GL point already small perturbations around the UBS will inevitably grow in a manner as described in subsection~\ref{subsec:End_state_of_the_Gregory-Laflamme_instability}. Eventually, an LBH forms.  

On the contrary, consider a small LBH that has higher entropy than a UBS. For increasing mass it is globally stable until the entropy of a UBS becomes greater. Nevertheless, above this point the LBH is locally stable until it reaches the maximum mass, see figure~\ref{fig:PhaseDiagram5Dold}. Another increase of the mass will force the LBH to evolve towards another configuration, since there is simply no LBH solution with this mass. According to the phase diagram~\ref{fig:PhaseDiagram5Dold} the end state of such a deformation is a UBS. Thus, an unstable mode at the maximum LBH mass is expected and, indeed, numerical evidence for such a mode was found~\cite{Headrick:2009pv}.

We note that the phase transition discussed above is of first order. It is accompanied by a considerable release of energy. This may lead to test signatures for extra dimensions in accelerator experiments or astronomical observations as discussed in reference~\cite{Kol:2002hf}.

Let us discuss the remaining parts of the phase diagram. Obviously, the LBH branch does not terminate at the maximum mass solution, it rather continues with solutions of lower mass and entropy. This turning point has a cuspy appearance in the phase diagram, which is completely sound due to the first law of black hole thermodynamics $\delta M = T\,\delta S$, cf. subsection~\ref{subsec:Physical_quantities}. Moreover, the NBS branch emanates from the GL point and reference~\cite{Kleihaus:2006ee} provides numerical results that show such a turning point in this branch as well. However, other authors could not reproduce this feature. Therefore, one of the motivations of the work at hand is to clarify this issue.  

Another remarkable feature of the phase diagram, figure~\ref{fig:PhaseDiagram5Dold}, is that the LBH and NBS branches seem to approach each other. Indeed, already as numerical data was rare, several authors conjectured that both branches meet~\cite{Kol:2002xz,Harmark:2003yz}. Clearly, this would imply a change of the horizon topology, e.g.\ when going from LBHs to NBSs the topology changes from $\mathbb S^{D-2}$ to $\mathbb S^{D-3}\times \mathbb S^1$. Then, if this transition is continuous, there has to be a transit solution where the poles of the LBH touch or the horizon of an NBS pinches off, respectively. At the critical point where this happens the spacetime will exhibit a curvature singularity. Kol strengthened the conjecture by giving a local model of this singular transit solution at the critical point~\cite{Kol:2002xz}. This local model is rather simple but it comes with some interesting implications for the LBH/NBS transition, as we will review in the next subsection.  

At this point we have to clarify our terminology. In contrast to the phase transition from UBSs to LBHs (or the other way round) we denote the merger of the LBH and NBS branch as the LBH/NBS transition or simply as the transition. In the remainder of this work we focus on the LBH/NBS transition as it is the missing part in the phase diagram~\ref{fig:PhaseDiagram5Dold}. Note that one could actually regard the LBH/NBS transition as a second order phase transition. Moreover, in the subsequent chapters we denote the regime close to the LBH/NBS transition as the critical regime.

It is believed that when going to higher dimensions the situation does not change qualitatively until one reaches $D=12$, apart from details about the LBH/NBS transition. As we already mentioned, for $D\geq 12$ parts of the NBS branch (for $D\geq 14$ even the whole branch) have higher entropy than the corresponding UBSs. However, in this work we will eventually concentrate on $D=5$ and $D=6$. For a discussion of the predicted phase diagram for higher dimensions see reference~\cite{Figueras:2012xj}.

\subsection{Double-cone metric}
\label{subsec:Double-cone_metric}

\begin{figure}[ht]
	\centering
		\includegraphics[scale=1]{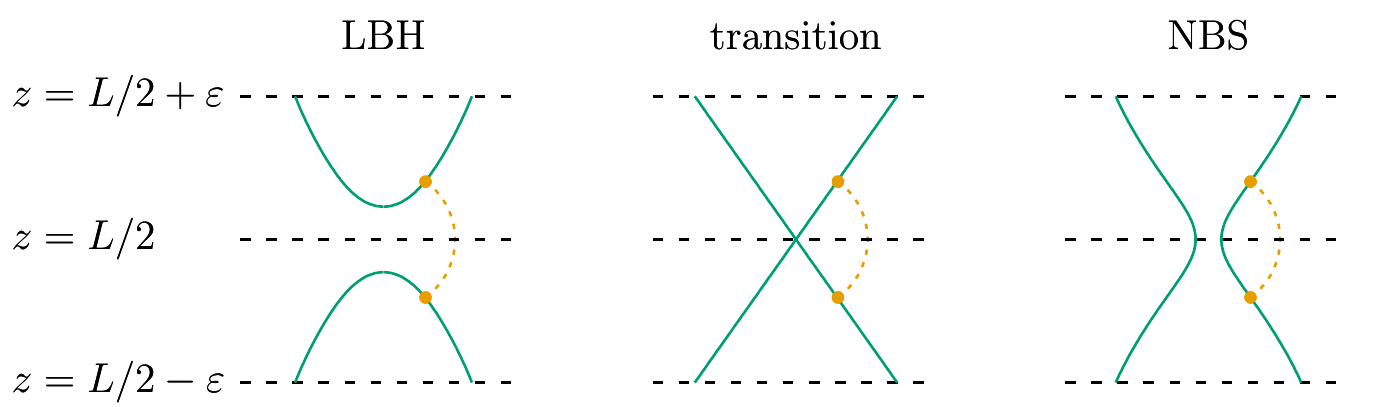} 
	\caption{Sketch of the horizon shape of a nearly merging LBH, the double-cone and a nearly pinching NBS in vicinity of the critical point where the LBH horizon is supposed to merge, the double-cone metric is singular or the NBS is supposed to pinch-off, respectively. At the starting and end point of the indicated paths the Euclidean time circle has zero size, since these points lie on the horizon. }
	\label{fig:SketchDoubleCone}	
\end{figure}

Kol's first step in finding a local model of the LBH/NBS transit solution was to consider the spacetime in the vicinity of the supposed critical point. In this region he identified two important (hyper-)spheres: an $\mathbb S^{D-3}$ and an $\mathbb S^2$. The former obviously represents the inherent (hyper-)spherical symmetry of the setup. To understand where the latter comes from one has to perform a Wick rotation, i.e.\ a coordinate transformation to Euclidean time $\tau = \I\, t$. A static black hole solution has to be periodic in Euclidean time $\tau$ in order to avoid a conical singularity. At the horizon this Euclidean time circle has zero size, since the corresponding metric coefficient vanishes. Consider now a path in the LBH or NBS spacetime that starts and ends at the horizon, like illustrated in figure~\ref{fig:SketchDoubleCone}. Following this path the Euclidean time circle first has zero size, then increases before it shrinks back to zero size. Therefore, the fibration of Euclidean time circles along such a path produces a surface that is topologically a sphere $\mathbb S^2$. 

Furthermore, Kol observed that in the LBH spacetime the $\mathbb S^{D-3}$ is contractible to zero size due to the exposed axis of (hyper-)spherical symmetry, while the $\mathbb S^2$ is not contractible to zero size due to the spatial separation of the horizon's poles. The contrary is true for the NBS spacetime. However, both (hyper-)spheres are contractible in the spacetime of the transit solution, since it may exhibit a critical point on the exposed axis where the horizon is marginally connected, see figure~\ref{fig:SketchDoubleCone}. These considerations led Kol to search for a Ricci-flat metric that describes the two separate cones with their tips at the same place. He found the \textit{double-cone metric}~\cite{Kol:2002xz}
\beq
	\D s^2_\text{DC} = \D \rho ^2 + \frac{\rho ^2}{D-2} \left[ \, \D \Omega^2_2 + (D-4) \, \D \Omega ^2_{D-3} \right] \, .
	\label{eq:metricDoubleCone}
\eeq
The tips of both cones are located at $\rho =0$, which marks a curvature singularity. For a more detailed and pictorial introduction of the double-cone metric see reference~\cite{Kol:2004ww}. 

From a geometric point of view, the respective $D$-dependent prefactors in front of the different (hyper-)spheres in the double-cone metric~\eqref{eq:metricDoubleCone} dictate its shape. Let us embed the double-cone into $(D-1)$-dimensional flat space
\beq
	\D s^2_\text{flat} = \D R^2 + \D Z^2 + R^2 \, \D\Omega ^2_{D-3} \, .
	\label{eq:metricFlatSpace}
\eeq
Then, we find that the double-cone is described by
\beq
	Z - Z_L / 2 = \sqrt{\frac{2}{D-4}} \, |R| \, ,
	\label{eq:doubleconeEmbedding}
\eeq
with an arbitrary constant of integration $Z_L$. If the double-cone indeed controls the geometry of the critical transit solution, then the transit solution should exhibit a similar behavior at the point, where its horizon shrinks to zero size. In other words, the double-cone metric prescribes the angle under which the poles of an LBH merge or the horizon of an NBS pinches off when the transit solution is approached, cf.\ figure~\ref{fig:SketchDoubleCone}.

Furthermore, references~\cite{Kol:2002xz,Asnin:2006ip} analyzed perturbations from the double-cone metric of the form
\beq
	\D s^2_\text{PDC} = \D \rho ^2 + \frac{\rho ^2}{D-2} \left[ \E ^{\epsilon (\rho )} \, \D \Omega^2_2 + (D-4) \E ^{-2\epsilon (\rho ) /(D-3)} \, \D \Omega ^2_{D-3} \right]  \, .
	\label{eq:perturb_cone}
\eeq
Linear perturbations in $\epsilon$ give rise to the solutions~\cite{Kol:2002xz}
\beq
	\epsilon = \rho^{s_\pm} \, , 
	\label{eq:epsilon_cone}
\eeq
with the complex exponents
\beq
	s_\pm   = \frac{D-2}{2} \left( -1 \pm \text{i} \, \sqrt{\frac{8}{D-2} - 1} \right) \, .
	\label{eq:complex_exponents}
\eeq
It is apparent that the exponents $s_\pm$ are purely real for $D\geq 10$, while for $D<10$ their imaginary part produces oscillations in $\epsilon (\rho )$. To give these exponents a physical meaning we follow the arguments of reference~\cite{Kol:2005vy}: Suppose $\epsilon = \Delta p := p-p_\text{c}$, with $p$ denoting a physical quantity, such as the mass, and $p_\text{c}$ denoting the critical value of this physical quantity that is associated with the transit solution. Furthermore, consider a characteristic length scale $\rho _0$ of the perturbed double-cone. For instance, one can think of $\rho _0^{-2}$ being a measure of the maximal curvature of the perturbed double-cone spacetime. According to reference~\cite{Kol:2005vy} we obtain from equations~\eqref{eq:epsilon_cone} and~\eqref{eq:complex_exponents}
\beq
	\Delta p = \tilde a \, \rho _0^{-s_+} + \tilde d \, \rho _0^{-s_-} \, ,
	\label{eq:dp_perturbed_cone}
\eeq
where $\tilde a$ and $\tilde d$ are constants. After a straightforward algebra we obtain for $D<10$
\beq
	\Delta p = a \, \rho _0^b \cos (c\log \rho _0 +d) \, ,
	\label{eq:dp_perturbed_cone_real}
\eeq
with $b=-\mathfrak{Re} (s_+)$, $c=\mathfrak{Im} (s_+)$ and real constants $a$ and $d$. 

The previous analysis may have some interesting implications for the phase diagram of KK black holes. Once we have identified an appropriate length scale $\rho _0$ that parametrizes the LBH and NBS branch and approaches zero for the transit solution, we can express physical quantities in terms of equation~\eqref{eq:dp_perturbed_cone}, at least close to the transition. This implies a scaling of physical quantities in the critical regime where the transit solution is approached. In particular, for $D<10$ this scaling comes with an infinite number of oscillations, cf. equation~\eqref{eq:dp_perturbed_cone_real}.

However, the implications of the double-cone metric for the LBH/NBS transition are still a conjecture that needs evidence from numerical data. Reference~\cite{Kol:2003ja} provided the first comparison of the local geometry of the double-cone metric and NBS solutions in $D=6$. An improvement of these calculations can be found in reference~\cite{Sorkin:2006wp}. Indeed, both results give evidence in favor of the double-cone. In addition, in reference~\cite{Sorkin:2006wp} the proposed scaling of NBS solutions was tested but could not be confirmed convincingly. Therefore, another goal of this work is to investigate the critical regime close to the LBH/NBS transition, in particular with regard to the double-cone metric.

\subsection{Copies and bubbles}
\label{subsec:Copies_and_bubbles}

Finally, we also mention other known static black hole solutions of KK gravity. First, consider one of the solutions we have discussed above, e.g.\ an LBH, which lives in a compact dimension of size $L$. Since the compact dimension is periodic, such a configuration is equivalent to extending this dimension to infinity and place infinitely many LBHs with separation $L$ along this dimension, cf. figure~\ref{fig:horizonsLBH2Dsketch}. Then, one can take $2L$ as the new period, which gives us a solution with two identical LBHs. By proceeding in this manner one can construct solutions with an arbitrary number $k$ of copies of the original solution. This procedure was first mentioned in the black string context in reference~\cite{Horowitz:2002dc}. Depending on $k$, the physical quantities of the new solution will transform as
\beq
	M_k = \frac{M}{k^{D-4}} \, , \quad n_k = n \, , \quad  T_k = kT \, , \quad S_k = \frac{S}{k^{D-3}} \, ,
	\label{eq:QuantitiesCopies}
\eeq
with respect to the original solution. However, these copy solutions are highly unstable, e.g.\ recall that an array of LBHs is unstable against perturbations of their relative distances. We refer to reference~\cite{Harmark:2003eg} for a wider discussion.

For the sake of completeness, we note that there exist also static solutions to Einstein's vacuum equations containing so-called bubbles of nothing. The boundary of such a bubble of nothing is an inner boundary of the spacetime. It is remarkable that one can construct not only an analytic solution of a single bubble but also solutions containing sequences of bubbles and different types of black holes~\cite{Elvang:2004iz}. Nevertheless, any bubble of nothing is unstable against contraction or expansion. 

At this point it should be mentioned that certain (non-static) bubble solutions can have negative energy. Therefore the KK vacuum, the background $\mathbb M^{D-1}\times \mathbb S^1$~\eqref{eq:metricBG}, appears to be unstable, which calls the whole KK approach into question. While pointing this out in reference~\cite{Witten:1981gj}, Witten also showed a way out of this dilemma. If one assumes that a \lq realistic\rq{} theory should support fermions, one can show that the critical bubble solutions are ruled out. We note that all black hole solutions discussed in this thesis are still allowed under these circumstances.

\section{Numerical methods}
\label{sec:Numerical_method}

We aim to solve Einstein's field equations in the context of LBHs and NBSs numerically. As described above, there are already plenty of numerical studies of this system. To improve on previous calculations we need a numerical scheme that is capable to enter the critical regime and to provide sufficiently accurate results. Our method of choice relies on a \textit{pseudo-spectral} scheme. Here, we introduce the main ideas of this method and refer to appendix~\ref{sec:Pseudo-spectral_method} for more details. 

The pseudo-spectral method relies on the following approximation of a real-valued function $f(x)$ defined on a finite interval $x\in [a,b]$:
\beq
	f(x) \approx \sum _{k=0}^{N-1} c_k  \Phi _k(x) \, . 
	\label{eq:MaintextSpectralApproximation}
\eeq
We refer to this approximation as a spectral expansion of order $N$ of the function $f(x)$. The original function $f(x)$ is approximated by a linear combination of a set of basis functions $\Phi _k(x)$ with coefficients $c_k$. For example, trigonometric basis functions yield a truncated Fourier series representation of the function $f(x)$. In contrast, a common choice of basis functions $\Phi _k(x)$ for non-periodic functions $f(x)$ are Chebyshev polynomials of the first kind $T_k(y) = \cos (k \arccos y )$, which are defined on the interval $y\in [-1,1]$. Indeed, in this work we solely utilize the Chebyshev polynomials as a basis and we thus have 
\beq	
	\Phi _k(x) = T_k \left( \frac{2\, x-b-a}{b-a} \right) \, .
	\label{eq:MaintextChebyshevBasis}
\eeq
We consider the function $f(x)$ on so-called Lobatto grid points 
\beq
	x_k = \frac{b+a}{2} - \frac{b-a}{2} \cos \left(  \frac{\pi \, k}{N-1} \right)  \, , \quad k = 0,1, \ldots ,N-1 \, .
	\label{eq:MaintextLobattoPoints}
\eeq
Other choices of grid points are possible but Lobatto grid points include the boundaries $x=a$ and $x=b$ and are thus particularly suitable for boundary value problems. 

We demand that the approximation~\eqref{eq:MaintextSpectralApproximation} is exact at the $N$ Lobatto grid points yielding the pseudo-spectral coefficients $c_k$. If these are known, we are able to interpolate at any point $x\in [a,b]$. Moreover, using various identities of the Chebyshev polynomials we obtain a simple recursion formula for the spectral coefficients of the derivative of $f(x)$.  

In order to solve a differential equation we discretize the problem on a Lobatto grid. We solve the resulting set of algebraic equations with a Newton-Raphson scheme. For non-linear problems the Newton-Raphson scheme needs a good initial guess to converge to the actual solutions. In each of the several iterative steps of this scheme we have to solve a linear system numerically, which is computationally expensive. Therefore, our strategy is to reformulate the problem in such a way that the spectral approximation~\eqref{eq:MaintextSpectralApproximation} is sufficiently accurate for moderate expansion orders $N$. For this purpose we have to understand the properties of the underlying functions and how they affect the convergence of the spectral expansion. 

It is well known that the spectral approximation~\eqref{eq:MaintextSpectralApproximation} converges rapidly with increasing $N$ for analytic functions $f(x)$, i.e.\ functions that have a converging Taylor series expression at every point of the interval $x\in [a,b]$. In this case, the difference to the actual function decreases exponentially for increasing $N$. We denote such a decay as \textit{geometric} rate of convergence. The situation slightly worsens if the function $f(x)$ is not analytic but smooth, i.e.\ all derivatives of $f(x)$ exist on $x\in [a,b]$. The corresponding rate of convergence is called \textit{subgeometric}. If only a finite number $m$ of derivatives of $f(x)$ are bounded, the approximation error of the expansion~\eqref{eq:MaintextSpectralApproximation} decreases as an inverse power of $N$, which we call an \textit{algebraic} rate of convergence. In general, small values of $m$ lead to a slow convergence.

  \chapter{Numerical construction of non-uniform black string solutions}
\label{chap:Numerical_construction_of_non-uniform_black_string_solutions}

\begin{figure}[ht]
	\centering
	\includegraphics[scale=1]{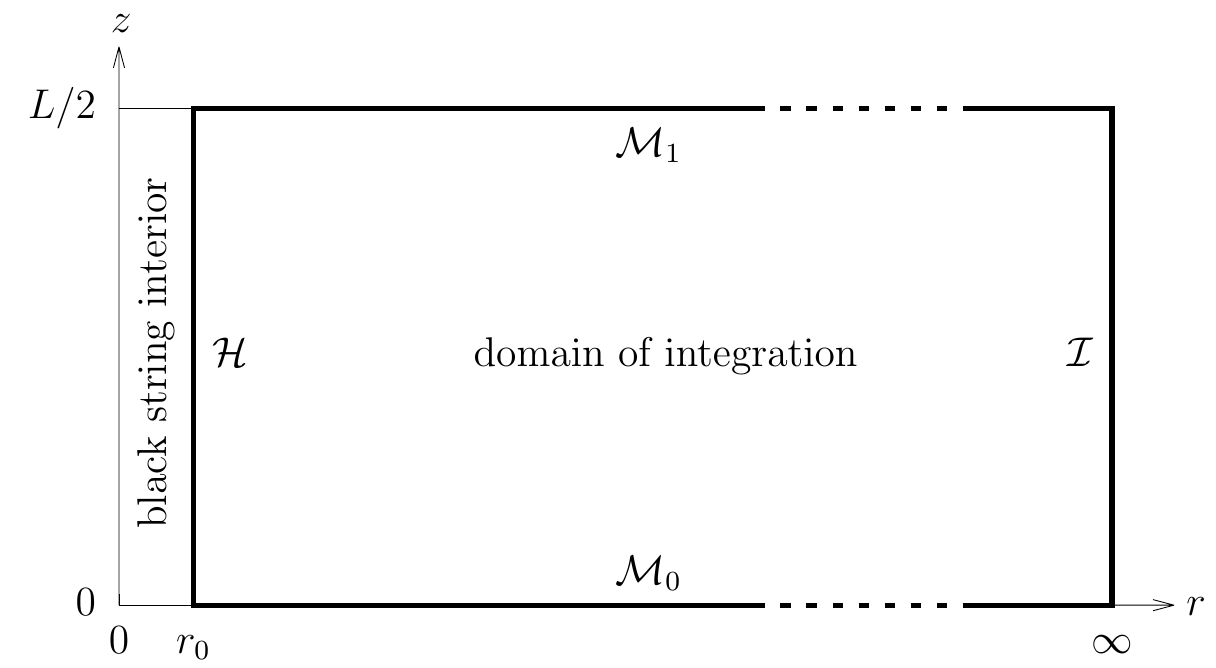}
	\caption{Domain of integration for the construction of NBS solutions. The boundaries are the horizon $\mathcal H$, the lower and upper mirror boundaries $\mathcal M_0$ and $\mathcal M_1$ and the asymptotic boundary $\mathcal I$.}
	\label{fig:NBSdomain_bare}
\end{figure}

The horizon of a $D$-dimensional non-uniform black string (NBS) wraps the compact dimension leading to an $\mathbb S^{D-3}\times \mathbb S^1$ horizon topology. We adopt the $(r,z)$ coordinates introduced in section~\ref{sec:Static_Kaluza_Klein_black_holes} and choose a gauge such that the horizon of the NBS lies at the constant coordinate line $r=r_0$. Then, the coordinate $r\in [r_0,\infty ]$ denotes a radial coordinate defined in the $(D-2)$ spatially extended dimensions in which we assume a (hyper-)spherical symmetry. Though the compact dimension has asymptotic length $L$, we only consider the upper part $z\in [0,L/2]$ since, due to the presupposed reflection symmetry, the lower part is only a duplicate of the former. In addition, we only consider a static spacetime and thus end up with a two-dimensional problem with the following boundaries also depicted in figure~\ref{fig:NBSdomain_bare}:
\begin{itemize}
	\item the asymptotic boundary $\mathcal I = \left\{ (r,z) \colon r\to\infty \, , ~ 0\leq z \leq L/2 \right\}$,
	\item the lower mirror boundary $\mathcal M_0 = \left\{ (r,z) \colon r\geq r_0 \, , ~ z=0 \right\}$,
	\item the upper mirror boundary $\mathcal M_1 = \left\{ (r,z) \colon r\geq r_0 \, , ~ z=L/2 \right\}$,
	\item the horizon $\mathcal H =  \left\{ (r,z) \colon r=r_0 \, , ~ 0\leq z \leq L/2 \right\}$.
\end{itemize}

Our goal is to numerically construct NBS solutions with a highly deformed horizon shape compared to the uniform black string (UBS), i.e\ NBS solutions with an extremely thin waist , cf.\ figure~\ref{fig:horizonsNBS3D}. The numerical techniques rely on the pseudo-spectral method. In the following, we first discuss an appropriate metric ansatz and the corresponding field and boundary equations in section~\ref{sec:Metric_ansatz_and_field_equations}. Thereafter, in section~\ref{sec:Perturbations_around_the_uniform_black_string}, we only solve linearized versions of these equations describing perturbations around the UBS. This will help us to develop an ansatz for the full non-linear equations. In section~\ref{sec:Construction_of_non-perturbative_solutions} we discuss the necessary adaptions of the pseudo-spectral method to solve these equations accurately even in the critical regime, where the horizon is close to pinching. A discussion of physical and unphysical quantities is provided in section~\ref{sec:Parameters_and_physical_quantities_NBS}. Finally, we discuss the accuracy of the numerical results in section~\ref{sec:Accuracy_of_the_numerical_solutions_NBS}.

\section{Metric ansatz and field equations}
\label{sec:Metric_ansatz_and_field_equations}

We consider the static NBS metric in $D$ dimensions and with the background $\mathbb M^{D-1}\times \mathbb S^1 $ in the form 
\beq
	\D s^2_{\text{NBS}} = -\E ^{2A} f_{D-1}(r) \, \D t^2 + \E ^{2B} \left( \frac{\D r^2}{f_{D-1}(r)} + \D z^2 \right) + r^2\E ^{2C} \, \D \Omega ^2_{D-3} \, .
	\label{eq:metricNBS}
\eeq
We have omitted the explicit $r$- and $z$-dependence of the yet unknown functions $A$, $B$ and $C$. In this section, we will develop a scheme to solve for these functions in the NBS context. For convenience, we revisit here the function $f_{D-1}$, which already appeared in the Schwarzschild-Tangherlini metric~\eqref{eq:metricST} 
\beq
	f_{D-1} (r) = 1 - \left( \frac{r_0}{r} \right)^{D-4} \, .
	\label{eq:SchwarzschildTangherliniFunctionDm1}
\eeq
As claimed before, by this construction the horizon of the black string resides at the constant coordinate value $r=r_0$, no matter how deformed the horizon of the NBS actually is. 

It is apparent that if $A\equiv B\equiv C \equiv 0$ we recover the UBS metric~\eqref{eq:metricUBS}. As discussed in section~\ref{subsec:Uniform_black_strings} the UBS solutions are subject to the Gregory-Laflamme instability if $L/r_0>K_\GL$, which breaks the translation invariance along the $z$-direction and leads to the NBS branch with non-vanishing functions $A$, $B$ and $C$. 

From Einstein's vacuum field equations we get the following system of second order partial differential equations~\cite{Kleihaus:2006ee} (we define $':= \partial /\partial r$ and $\dot{} := \partial /\partial z$):
\refstepcounter{equation} \label{eq:NBSfield_eqns}
\begin{align}
	0 =& \, A'' + \frac{\ddot A}{f_{D-1}} + A'^2 + \frac{\dot A^2}{f_{D-1}} + (D-3) \left( A'C' + \frac{\dot A\dot C}{f_{D-1}} + \frac{A'}{r} + \frac{f'_{D-1}C'}{2f_{D-1}} \right) \nonumber \\ 
	   & \, + \frac{3f'_{D-1}A'}{2f_{D-1}} \, ,	
			\tag{\theequation a} \label{eq:NBSfield_eqnsA} \\
	0 =& \, B'' + \frac{\ddot B}{f_{D-1}} - (D-3) \left( A'C' + \frac{\dot A\dot C}{f_{D-1}} + \frac{A'}{r} + \frac{f'_{D-1}C'}{2f_{D-1}} \right) 	 + \frac{f'_{D-1}B'}{2f_{D-1}}								    \nonumber \\
	   & \, - \frac{(D-3)(D-4)}{2r^2} \left( \frac{1 - \E^{2B-2C}}{f_{D-1}}  + r^2C'^2 + 2rC' + \frac{r^2\dot C^2}{f_{D-1}} \right) \, ,		              
	   		\tag{\theequation b} \label{eq:NBSfield_eqnsB}	\\
	0 =& \, C'' + \frac{\ddot C}{f_{D-1}} + A'C' + \frac{\dot A \dot C}{f_{D-1}} + \frac{A'}{r} + \frac{f'_{D-1}C'}{f_{D-1}}  + \frac{(D-4)}{r^2} \frac{ (1-\E ^{2B-2C})}{f_{D-1}}																		    \nonumber \\
	   & \, + (D-3) \left( C'^2 + \frac{2C'}{r} + \frac{\dot C^2}{f_{D-1}} \right) \, .  			  
	   		\tag{\theequation c} \label{eq:NBSfield_eqnsC}	
\end{align}
To be more precise, the above equations follow from the Einstein tensor's components $G^t_t = 0$, $G^r_r + G^z_z = 0$ and $G^\theta _\theta = 0$. Actually, there are two more independent equations arising from $G^r_z=0$ and $G^r_r-G^z_z=0$. These two additional equations are denoted as constraint equations, since a solution of the system~\eqref{eq:NBSfield_eqns} is only a solution of Einstein's vacuum field equations if \textit{all} components of the Einstein tensor vanish. However, a solution of~\eqref{eq:NBSfield_eqns} automatically satisfies the constraints if we choose appropriate boundary conditions~\cite{Wiseman:2002zc}. 

The following boundary conditions, cf.\ figure~\ref{fig:NBSdomain_bare}, arise:
\begin{itemize}
	\item 	Asymptotic boundary $\mathcal I$ ($r\to\infty$): \nopagebreak \\
		  	At infinity the spacetime has to resemble the KK flat space~\eqref{eq:metricBG}. Therefore the metric functions have to vanish
		  	\beq
		  		0 = A = B = C \, .
		  		\label{eq:NBSBCasymp}
		  	\eeq
	\item 	Mirror boundaries $\mathcal M_0$ ($z=0$) and $\mathcal M_1$ ($z=L/2$): \nopagebreak  \\
			Periodicity and reflection symmetry in $z$ require the metric to be symmetric at these boundaries. Consequently we have
			\beq
				0 = \frac{\partial{A}}{\partial z} = \frac{\partial{B}}{\partial z} = \frac{\partial{C}}{\partial z}\, .
				\label{eq:NBSBCmirror}
			\eeq
	\item 	Horizon $\mathcal H$ ($r=r_0$): \nopagebreak   \\ 
			On the horizon the field equations~\eqref{eq:NBSfield_eqns} are singular and automatically provide boundary conditions. The first reads 
			\beq
				0 = \frac{\partial{A}}{\partial z} - \frac{\partial{B}}{\partial z}  \, .
				\label{eq:NBSBCconstT}
			\eeq 
			This condition ensures that the surface gravity (or the temperature) along the horizon is constant. Upon integration we obtain an undetermined constant, which we fix by prescribing the value of the function $B$ at the upper horizon edge $z=L/2$ in such a way that the following relation holds
			\beq
				0 = \E ^{-2B} -\beta _\text{c}  \, .
				\label{eq:NBSBCbetac}
			\eeq
			We are free to specify any value of $\beta _\text{c}\in [0,1]$. To be more specific, it turns out that $\beta _\text{c}$ (having no significant physical meaning) serves as an appropriate control parameter to distinguish between physically inequivalent solutions. Moreover, the field equations~\eqref{eq:NBSfield_eqns} give us two further conditions on the horizon. Instead of using these conditions directly, we follow reference~\cite{Kleihaus:2006ee} and introduce a modified radial coordinate $\tilde r$ via $r/r_0 = \sqrt{\tilde r^2 +1}$. The corresponding boundary conditions are then regularity conditions and imply that the derivatives with respect to $\tilde r$ vanish on the horizon
			\beq
				0 = \frac{\partial A}{\partial \tilde r} = \frac{\partial C}{\partial \tilde r} \, .
				\label{eq:NBSBCregularity}
			\eeq		
			Note that later we will use different coordinate transformations to achieve these regularity conditions. The crucial point is that $r(\tilde r )$ behaves quadratic (to leading order) at the horizon $r=r_0$ and thus we have $\partial r/\partial \tilde r = 0$ there.
\end{itemize}
The full set of equations to solve numerically are the field equations~\eqref{eq:NBSfield_eqns} together with the boundary conditions~\eqref{eq:NBSBCasymp}--\eqref{eq:NBSBCregularity}. We note that the boundary conditions are in accordance with the constraint rule of reference~\cite{Wiseman:2002zc}. As a consequence we do not have to demand the additional regularity condition $\partial B/\partial \tilde r =0$ separately. Instead it will be automatically satisfied by all solutions. Of course, for a numerical solution this will not be exactly true. Therefore, we check the constraints \textit{a posteriori} to verify the consistency of the numerically obtained solution.

Finally, to obtain a unique solution to the equations~\eqref{eq:NBSfield_eqns}--\eqref{eq:NBSBCregularity} we fix the length scale and the value of the parameter $\beta _\text{c}$. Naturally, we achieve the former by setting $L/r_0 = K_\GL$. Then, for $\beta _\text{c} =1$ the solution is a marginal stable UBS at the GL point. For $\beta _\text{c}\lesssim 1$ we enter the NBS branch with only small deformations of the black string horizon. If $\beta _\text{c}$ is even smaller, the horizon becomes considerably pinched at the coordinate value $z=L/2$. In particular, we are interested in the critical limit, where the horizon is about to pinch-off, which is obtained by $\beta _\text{c}\to 0$.

\section{Perturbations around the uniform black string}
\label{sec:Perturbations_around_the_uniform_black_string}

Before we will describe a numerical scheme that is capable to approach the critical limit $\beta _\text{c}\to 0$, we investigate small perturbations around the UBS, where $\beta _\text{c}\lesssim 1$. This will help us to understand the behavior of the metric functions, in particular near the asymptotic boundary, and to design an appropriate ansatz for the non-perturbative regime. Along the way we will get highly accurate values for $K_\GL$. Furthermore, since our numerical implementation relies on a Newton-Raphson scheme, we utilize the solution of the perturbation equations as an initial guess for the full non-linear system.  

First order perturbations around the UBS are governed by the marginal GL mode and read 
\refstepcounter{equation} \label{eq:NBSLinPerturb}
\begin{align}
	A &= \varepsilon \, a(r) \cos \left( \tfrac{2\pi}{L}z \right) \, ,
		\tag{\theequation a}  \label{eq:NBSLinPerturbA} \\
	B &= \varepsilon \, b(r) \cos \left( \tfrac{2\pi}{L}z \right) \, , 
		\tag{\theequation b}  \label{eq:NBSLinPerturbB} \\
	C &= \varepsilon \, c(r) \cos \left( \tfrac{2\pi}{L}z \right) \, .
		\tag{\theequation c}  \label{eq:NBSLinPerturbC}
\end{align}
For a small perturbation parameter $\varepsilon$ we substitute this ansatz into the field equations~\eqref{eq:NBSfield_eqns} and solely take linear orders of $\varepsilon$ into account. This leads to the following set of ordinary differential equations describing only first order perturbations (see also reference~\cite{Kol:2004pn}):
\refstepcounter{equation} \label{eq:NBSLinPerturbEqns}
\begin{align}
	0 =& \, f_{D-1} a'' + \left[ \dfrac{3}{2}  f'_{D-1}  + (D-3) \frac{f_{D-1}}{r}  \right] a' + \frac{1}{2} (D-3) f'_{D-1}  c' - \frac{4\pi ^2}{L^2} a \, ,
		\tag{\theequation a}  \label{eq:NBSLinPerturbEqnA} \\
	0 =& \, f_{D-1} b'' - (D-3) \frac{f_{D-1}}{r} a' + \frac{1}{2} f'_{D-1} b' - \frac{1}{2} (D-3) \left[  f'_{D-1} + 2 (D-4) \frac{f_{D-1}}{r} \right] c' \nonumber \\ 
	   & \, +  (D-3)(D-4) \frac{b-c}{r^2} - \frac{4\pi ^2}{L^2} b   \, , 
		\tag{\theequation b}  \label{eq:NBSLinPerturbEqnB} \\
	0 =& \, f_{D-1} c'' + \frac{f_{D-1}}{r} a' + \left[ f'_{D-1} + 2 (D-3) \frac{f_{D-1}}{r} \right] c' - 2 (D-4) \frac{b-c}{r^2} -   \frac{4\pi ^2}{L^2} c \, .
		\tag{\theequation c}  \label{eq:NBSLinPerturbEqnC}
\end{align}Now the benefit of this approach becomes clear: The field equations are considerably simplified. However, a solution of~\eqref{eq:NBSLinPerturbEqns} is only a rough approximation of a slightly deformed NBS. A better approximation can be obtained by taking into account higher order corrections. Reference~\cite{Gubser:2001ac} provides a scheme for the construction of perturbations to arbitrary order in $\varepsilon$. The advantage of this scheme is that at each level one obtains a system of ordinary differential equations. Since the original scheme was designed for five dimensional NBS solutions, generalizations to six~\cite{Wiseman:2002zc} and higher dimensions~\cite{Sorkin:2004qq} followed. 

Below we analyze the first order perturbations and we describe an approach to obtain highly accurate numerical solutions. We start with the $D=6$ case in subsection~\ref{subsec:Perturbations_in_six_dimensions} and we continue with the $D=5$ case in subsection~\ref{subsec:Perturbations_in_five_dimensions}. 


\subsection{Linear perturbations in six dimensions}
\label{subsec:Perturbations_in_six_dimensions}

An analysis of the linearized field equations~\eqref{eq:NBSLinPerturbEqns} for $D=6$ in the asymptotic limit $r\to \infty$ reveals the leading behavior of the functions $a$, $b$ and $c$ at infinity. We extract the respective behavior from these functions giving rise to the ansatz
\refstepcounter{equation} \label{eq:NBSLinPerturbAnsatz6D}
\begin{align}
	a &= \tilde a(r) \, \expterm \left( \frac{r_0}{r} \right) ^{3/2} \, ,
		\tag{\theequation a} \label{eq:NBSLinPerturbAnsatz6DA} \\
	b &= \tilde b(r) \, \expterm \, , 
		\tag{\theequation b} \label{eq:NBSLinPerturbAnsatz6DB} \\
	c &= \tilde c(r) \, \expterm \left( \frac{r_0}{r} \right) \, .
		\tag{\theequation c} \label{eq:NBSLinPerturbAnsatz6DC} 
\end{align}
Note that the exponential factor $\expterm$ strongly suppresses the $z$-dependent modes at infinity and appears generically in KK theory. 

We now want to solve the linearized field equations~\eqref{eq:NBSLinPerturbEqns} for $D=6$ with the ansatz~\eqref{eq:NBSLinPerturbAnsatz6D} by using a pseudo-spectral scheme. In order to do so, we first have to find a coordinate transformation of $r$ which satisfies the following requirements:
\begin{enumerate}
	\item The coordinate transformation shall \textit{compactify the asymptotic boundary} to a finite coordinate value. Consequently, the whole domain $r\in [r_0,\infty )$ will be numerically attainable. 
	\item After the compactification, half integer powers of $r_0/r$, cf.\ equation~\eqref{eq:NBSLinPerturbAnsatz6DA}, may give rise to non-smooth functions. Such functions have a slowly converging spectral representation, see appendix~\ref{subsec:Basic_ideas_and_concepts} for more details. Therefore, the coordinate transformation shall \textit{regularize the half integer powers of} $r_0/r$.
	\item According to our discussion of the boundary conditions in section~\ref{sec:Metric_ansatz_and_field_equations} the coordinate transformation shall \textit{lead to vanishing derivatives of the functions with respect to the new coordinate on the horizon} $r=r_0$.
\end{enumerate}
A transformation that satisfies all of the three requirements is
\begin{equation}
	\frac{r_0}{r} = \left[ 1 - (1-\xi )^2 \right] ^2 = \xi ^2 (2-\xi )^2 \, .
	\label{eq:NBScoordxi}
\end{equation}
The new coordinate $\xi$ ranges from $\xi =0$, which corresponds to the asymptotic limit $r\to \infty$, to $\xi =1$, which corresponds to the horizon $r =r_0$. 

To solve the set of \textit{homogenous} linear ordinary differential equations~\eqref{eq:NBSLinPerturbEqns} for the functions $\tilde a$, $\tilde b$ and $\tilde c$ with respect to the coordinate $\xi$ numerically, we need to impose an arbitrary scaling condition. We decide to choose $\tilde c=1$ on the horizon $\xi =1$. The remaining boundary conditions follow directly from the differential equations. These are $\tilde a _{,\xi} = \tilde c_{,\xi} =0$ at $\xi =1$ as well as $\tilde a_{,\xi} = \tilde b_{,\xi} = 0$ and $\tilde b + 2\pi \, r_0 /L \,\tilde c= 0$  at the asymptotic boundary $\xi =0$. Finally, we have to prescribe a value for the dimensionless quantity $K=L/r_0$, which is the only physical parameter in the equations. Doing so we are then able to solve the system numerically in a straightforward way described in the appendix section~\ref{sec:Pseudo-spectral_method}. 

Interestingly, the solution function $\tilde b$ does not generically obey the regularity requirement $\tilde b_{,\xi} =0$ at the horizon $\xi =1$. However, as we saw in section~\ref{sec:Metric_ansatz_and_field_equations}, this condition has to be satisfied in order to fulfill the constraint equations. It turns out that if the parameter $K$ is fine tuned, one can obtain a solution where $\tilde b_{,\xi}$ vanishes on the horizon. We obtain this value with high accuracy by considering $K$ as an additional unknown in the system of equations. Accordingly, the additional equation $\tilde b_{,\xi}=0$ enters the system.\footnote{\label{foot:FindKGL}We note that, if the system is solved with a prescribed value of $K$, we do not have to take care of an initial guess for the functions $\tilde a$, $\tilde b$ and $\tilde c$ within the Newton-Raphson scheme, since the equations are linear. If $K$ is treated as an additional unknown, the equations become non-linear but we simply get an initial guess from a numerical solution obtained from the linear system with prescribed $K$.} The value of $K$, which we get by this procedure, is exactly the value where the GL instability causes a UBS to be marginally stable. We give this value with unprecedented accuracy
\beq
	K_\GL = 4.9516154200735(1) \quad \text{for} \quad D=6 \, .
	\label{eq:NBSKGL6D}
\eeq 

At this point we emphasize that the spectral coefficients of the solution functions $\tilde a$, $\tilde b$ and $\tilde c$ show a geometric decay. Obviously, due to the exponential factor $\expterm$ this would not be the case for the functions $a$, $b$ and $c$, cf. appendix section~\ref{sec:Pseudo-spectral_method}. Therefore, it seems to be a good idea to adopt the ansatz~\eqref{eq:NBSLinPerturbAnsatz6D} together with the coordinate transformation~\eqref{eq:NBScoordxi} when solving the full non-linear field equations~\eqref{eq:NBSfield_eqns}.

\subsection{Linear perturbations in five dimensions}
\label{subsec:Perturbations_in_five_dimensions}

We repeat the asymptotic analysis of the linearized field equations~\eqref{eq:NBSLinPerturbEqns} for the case $D=5$. This yields the ansatz
\refstepcounter{equation} \label{eq:NBSLinPerturbAnsatz5D}
\begin{align}
	a &= \tilde a(r) \, \expterm \left( \frac{r_0}{r} \right) ^{1+\pi \, r_0/L} \, , 
		\tag{\theequation a} \label{eq:NBSLinPerturbAnsatz5DA} \\ 
	b &= \tilde b(r) \, \expterm \left( \frac{r_0}{r} \right) ^{  \pi \, r_0/L} \, ,
		\tag{\theequation b} \label{eq:NBSLinPerturbAnsatz5DB} \\
	c &= \tilde c(r) \, \expterm \left( \frac{r_0}{r} \right) ^{1+\pi \, r_0/L} \, . 
		\tag{\theequation c} \label{eq:NBSLinPerturbAnsatz5DC}
\end{align}
Note again the exponential factor $\expterm$, but this time it is accompanied by some odd powers of $r_0/r$. Here, we do not try to regularize these terms with a coordinate transformation like in the $D=6$ case, since it is not straightforward to get rid of them. However, in view of a numerical implementation, we keep the other two requirements on the coordinate transformation, see section~\ref{subsec:Perturbations_in_six_dimensions}. To be more precise, we want to compactify the asymptotic boundary and we want to have vanishing first derivatives on the horizon. We utilize the transformation, cf.\ equation~\eqref{eq:NBScoordxi},
\beq
	\frac{r_0}{r} = 1-(1-\chi )^2 = \chi (2-\chi ) \, ,
	\label{eq:NBScoordchi}
\eeq
where the coordinate $\chi$ ranges from $\chi =0$, which corresponds to the asymptotic limit $r\to \infty$, to $\chi =1$, which corresponds to the horizon $r =r_0$. 

Now, in order to solve the linearized field equations~\eqref{eq:NBSLinPerturbEqns} for $D=5$ we proceed in the same manner as described in the previous section, this time using the ansatz~\eqref{eq:NBSLinPerturbAnsatz5D} and the coordinate $\chi$. Boundary conditions on the horizon $\chi =1$ are given again by the scaling condition $\tilde c=1$ and the regularity conditions $\tilde a_{,\chi}=\tilde c_{,\chi}=0$. As before, conditions at $\chi =0$ follow from the degeneracy of the resulting field equations. Their explicit form is not important here. By enforcing the additional regularity condition $\tilde b_{\chi}=0$ on the horizon $\chi =1$ and treating $K=L/r_0$ as an additional unknown in the solution scheme, we obtain the critical GL value of $K$ with unprecedented accuracy:$^{\ref{foot:FindKGL}}$
\beq
	K_\GL = 7.1712728543704(1) \quad \text{for} \quad D=5 \, .
	\label{eq:NBSKGL5D}
\eeq 


\section{Construction of non-perturbative solutions}
\label{sec:Construction_of_non-perturbative_solutions}

\begin{figure}[ht]
	\centering
	\includegraphics[scale=1]{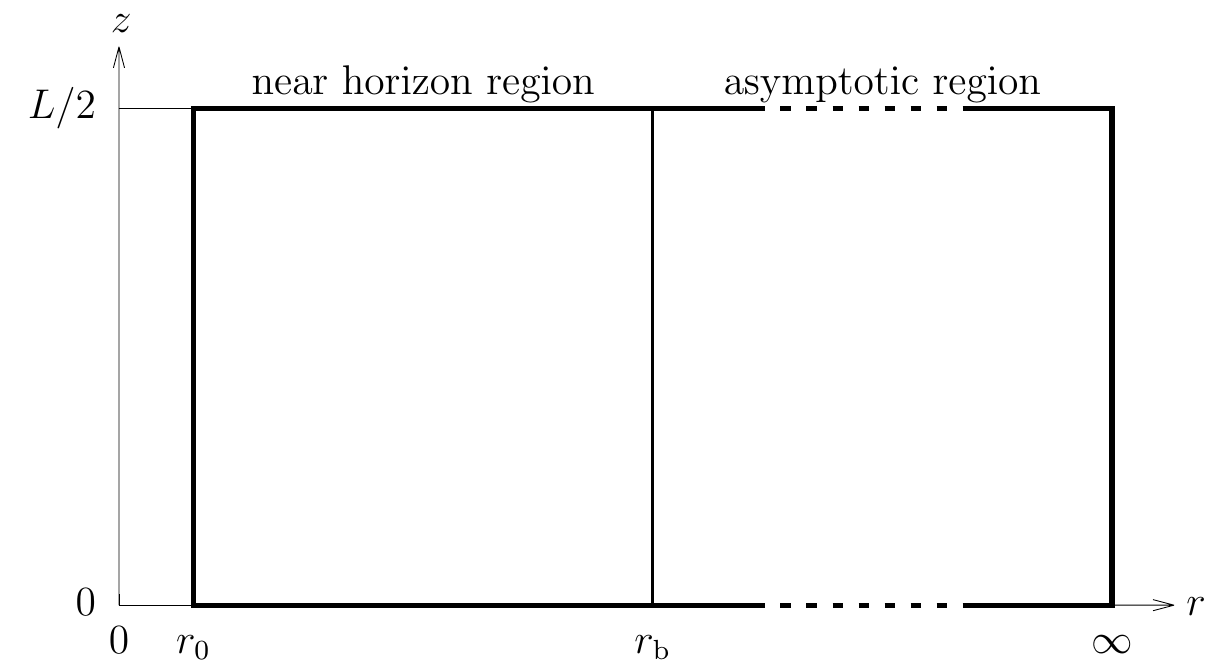}
	\caption{Domain of integration for the construction of NBS solutions with a decomposition into a near horizon region and an asymptotic region. These two regions are separated by the coordinate line $r=r_\text{b}$. For $r_0<r<r_\text{b}$  we find the near horizon region and for $r>r_\text{b}$ we find the asymptotic region.}
	\label{fig:NBSdomain_bare_split}
\end{figure}

The full set of field equations~\eqref{eq:NBSfield_eqns} are a system of partial differential equations. Hence, it is much more demanding to construct solutions to these equations than for the ordinary differential equations arising in perturbation theory. However, the analysis of the first order perturbations indicated the involved behavior of the metric functions near the asymptotic boundary. Additionally, since we are mainly interested in NBS solutions that are nearly pinching, we also have to take care about what happens to the metric functions near the horizon. Therefore, we split the domain of integration into a near horizon and an asymptotic region, as depicted in figure~\ref{fig:NBSdomain_bare_split}. Then, we utilize separate ansätze and coordinate transformations in each region in order to adapt the numerical setup to the respective behavior of the metric functions. This is of particular advantage in the scope of a pseudo-spectral method, which we utilize here. 

First, we discuss the asymptotic region in subsections~\ref{subsec:Treatment_of_the_asymptotics_in_six_dimensions} and~\ref{subsec:Treatment_of_the_asymptotics_in_five_dimensions}, where we distinguish different approaches in $D=6$ and $D=5$, starting with the former. Finally, we present our adaptions in the near horizon region in subsection~\ref{subsec:Treatment_of_the_horizon}. 

\subsection{Treatment of the asymptotics in six dimensions}
\label{subsec:Treatment_of_the_asymptotics_in_six_dimensions}

Recall the linear perturbations in six dimensions, subsection~\ref{subsec:Perturbations_in_six_dimensions}, and, in particular, the ansatz~\eqref{eq:NBSLinPerturbAnsatz6D}. We will rely on this also in the non-perturbative regime. However, additional modifications are needed in order to obtain overall accurate numerical results.

\subsubsection{Ansatz}

The linear perturbations around the UBS only yield the marginal GL mode, see equations~\eqref{eq:NBSLinPerturb}. Obviously, these do not take the leading asymptotics of the metric functions~\eqref{eq:asymptotic_corrections} into account, but they are of particular importance since they carry information about the mass $M$ and the relative tension $n$, cf.\ equations~\eqref{eq:Mass} and~\eqref{eq:Tension}. It turns out that the modes that carry this information are independent of $z$ and they first occur in second order perturbation theory, see for example reference~\cite{Wiseman:2002zc}. In order to take care of the leading asymptotic behavior of both the $z$-dependent and $z$-independent modes we utilize the following ansatz for the metric functions
\refstepcounter{equation} \label{eq:NBSsplitAnsatz6D}
\begin{align}
	A(r,z) & = A_0(r) \left( \frac{r_0}{r} \right) ^2            + \, A_1(r ,z) \, \cos\left( \tfrac{2\pi}{L}z\right) \, \expterm  \left( \frac{r_0}{r} \right) ^{3/2} \, , 	
		\tag{\theequation a} \label{eq:NBSsplitAnsatz6DA}  \\
	B(r,z) & =  B_0(r) \left( \frac{r_0}{r} \right) ^2            + \, B_1(r ,z) \, \cos\left( \tfrac{2\pi}{L}z\right) \, \expterm  \, ,					                           
		\tag{\theequation b} \label{eq:NBSsplitAnsatz6DB} \\ 
	C(r,z) & = C_0(r) \left( \frac{r_0}{r} \right) \hphantom{^2} + \, C_1(r ,z) \, \cos\left( \tfrac{2\pi}{L}z\right) \, \expterm  \left( \frac{r_0}{r} \right) \, . 
		\tag{\theequation c} \label{eq:NBSsplitAnsatz6DC}  					  
\end{align}
Instead of having three metric functions that depend on $(r,z)$, namely $A$, $B$ and $C$, we now end up with three new metric functions that depend on $(r,z)$, namely $A_1$, $B_1$ and $C_1$, \textit{and} three additional metric functions that only depend on $r$, namely $A_0$, $B_0$ and $C_0$. It is important to note that there is the following one-to-one map between $\{A_0, A_1\}$ and $A$:
\begin{align}
	A_0(r) &   =  \left(\dfrac{r_0}{r} \right)^{-2} \left. A(r,z)\right| _{z=L/4} \, ,  
 		\label{eq:NBSA0_of_A} \\
	A_1(r,z) & = \frac{A(r,z)-\left. A(r,z)\right| _{z=L/4}}{\cos\left( \tfrac{2\pi}{L}z\right)} \, \E ^{2\pi r/L} \, \left( \frac{r_0}{r} \right) ^{-3/2} \, ,
		\label{eq:NBSA1_of_A}
\end{align}
and similarly for $B$ and $C$. Moreover, we get equations for the new functions in a similar way by using the original field equations~\eqref{eq:NBSfield_eqns}. As an additional benefit the ansatz~\eqref{eq:NBSsplitAnsatz6D} allows us to calculate the asymptotic charges directly from the asymptotic values $A_\infty := \lim_{r\to\infty} A_0(r)$ and $B_\infty := \lim_{r\to\infty} B_0(r)$. In contrast, one would have to perform two numerical derivatives to extract the asymptotic charges from the original functions $A$ and $B$ diminishing the accuracy. For later convenience we also define $C_\infty := \lim_{r\to\infty} C_0(r)$.


Again, we make use of the coordinate transformation $r_0/r = \xi ^2 (2-\xi )^2$ discussed in subsection~\ref{subsec:Perturbations_in_six_dimensions}. This coordinate transformation compactifies the asymptotic boundary to the coordinate value $\xi =0$, while the horizon is mapped to $\xi =1$. We find another useful coordinate transformation for the transverse direction by exploiting the periodic nature of the coordinate $z$ as well as the mirror symmetry with respect to $z=0$. The transformation reads
\beq
	u = \cos\left( \tfrac{2\pi}{L}z\right) \, ,
	\label{eq:NBScoordu}
\eeq
where $u$ runs from $u=-1$, corresponding to $z=L/2$ , to $u=1$, corresponding to $z=0$.\footnote{\label{footnote:oddN}We note that our ansatz~\eqref{eq:NBSsplitAnsatz6D} privileges the use of an odd number of grid points with respect to the $u$-direction in the numerical implementation. As explained in appendix~\ref{sec:Pseudo-spectral_method} we discretize all functions on a Lobatto grid~\eqref{eq:LobattoPoints}, which only contains the central point $u=0$, corresponding to $z=L/4$, for odd resolutions.} We note that a Chebyshev expansion with respect to the coordinate $u$ is equivalent to an even Fourier expansion with respect to $z$.

\subsubsection{Asymptotic boundary conditions}

Below, we discuss the asymptotic boundary conditions for the auxiliary functions that constitute the metric components through~\eqref{eq:NBSsplitAnsatz6D}. The conditions for the $z$-independent functions $A_0$, $B_0$ and $C_0$, now expressed as functions of $\xi$, follow from the analysis of a power series in terms of the coordinate $\xi$ of the field equations~\eqref{eq:NBSfield_eqns} around $\xi =0$. As said before, the functions $A_0$, $B_0$ and $C_0$ are associated with the field equations at $u=0$, or equivalently $z=L/4$. To zeroth order, we obtain
\refstepcounter{equation} \label{eq:NBSAsymptoticAnalysisX0_0thorder}
\begin{align}
	0 & = A_{0,\xi} \, , \tag{\theequation a} \label{eq:NBSAsymptoticAnalysisX0_0thorderA} \\
	0 & = C_{0,\xi} \, , \tag{\theequation b} \label{eq:NBSAsymptoticAnalysisX0_0thorderB} 	
\end{align}
where the first condition arises from the field equation~\eqref{eq:NBSfield_eqnsA}, while both equation~\eqref{eq:NBSfield_eqnsB} and~\eqref{eq:NBSfield_eqnsC} yield the second condition. Taking the next expansion order into account we get
\refstepcounter{equation} \label{eq:NBSAsymptoticAnalysisX0_1storder}
\begin{align}
	0 & = 3            \, A_{0,\xi\xi} - 24\, (1-2\, A_\infty )C_\infty \, , \tag{\theequation a} \label{eq:NBSAsymptoticAnalysisX0_1storderA} \\
	0 & = \hphantom{3} \, C_{0,\xi\xi} + 4\, ( 2\, A_\infty + 4\, B_\infty + C_\infty ^2 ) \, . \tag{\theequation b} \label{eq:NBSAsymptoticAnalysisX0_1storderB} 	
\end{align}
Again, equation~\eqref{eq:NBSfield_eqnsA} yields the first condition and equations~\eqref{eq:NBSfield_eqnsB} and~\eqref{eq:NBSfield_eqnsC} yield the second condition. It seems as if there is some trivial condition on the function $B_0$ missing, for example a similar condition as in~\eqref{eq:NBSAsymptoticAnalysisX0_0thorder}. Such a condition only arises at the second order and reads
\beq
 	0 = B_{0,\xi}  \, . 
	\label{eq:NBSAsymptoticAnalysisX0_2ndorderB}
\eeq 
It turns out that a pseudo-spectral numerical scheme that incorporates the boundary conditions of vanishing $\xi$-derivatives of $A_0$, $B_0$ and $C_0$ at $\xi =0$ yields solutions with unsatisfactory accuracy, since the corresponding versions of equations~\eqref{eq:NBSfield_eqnsB} and~\eqref{eq:NBSfield_eqnsC} are identical up to second order in $\xi$.
Thus, in order to restore accuracy, we perform another decomposition 
\beq
	C_0(\xi) = C_{\infty} + \xi ^2 \, C_{01}(\xi) \, ,
	\label{eq:NBSC0split} 
\eeq
which already incorporates the condition $C_{0,\xi} =0$ at $\xi = 0$. 
After utilizing the decomposition~\eqref{eq:NBSC0split} we end up with the following set of four boundary conditions at $\xi =0$:
\beq
	0 = A_{0,\xi} = B_{0,\xi} =  C_{01} + C_{01,\xi} = C_{01} + 2\, (2 \, A_\infty + 4 \, B_\infty + C_{\infty}^2) \, .
	\label{eq:NBSasympBC_X0}
\eeq 
Note that we now need four instead of three conditions since, besides the functions $A_0$, $B_0$ and $C_{01}$, the additional unknown parameter $C_\infty$ enters the numerical scheme. Furthermore, we note that the set of equations~\eqref{eq:NBSasympBC_X0} is equivalent to the conditions~\eqref{eq:NBSAsymptoticAnalysisX0_0thorderA},~\eqref{eq:NBSAsymptoticAnalysisX0_0thorderB},~\eqref{eq:NBSAsymptoticAnalysisX0_1storderB} and~\eqref{eq:NBSAsymptoticAnalysisX0_2ndorderB}. We neglect condition~\eqref{eq:NBSAsymptoticAnalysisX0_1storderA}, but it emerges \textit{a posteriori} as a property of the numerical solution.  

We get a condition for the two-dimensional functions $A_1$, $B_1$ and $C_1$ by evaluating the field equations~\eqref{eq:NBSfield_eqns} at $\xi =0$. For convenience, we write $X_1(\xi ,u) = \{ A_1(\xi, u), B_1(\xi ,u), C_1(\xi ,u) \}$ and obtain at $\xi =0$
\beq
	u(1-u^2)X_{1,uu} - (2-3u^2) X_{1,u} - 2 X_{1,u}|_{u=0} = 0 \, .
	\label{eq:NBSasympBC_X1ode}
\eeq
The trivial solution $X_{1,u} = 0$ is the only regular solution of this ordinary differential equation. Consequently, we obtain the following conditions on $A_1$, $B_1$ and $C_1$ at $\xi = 0$
\beq
	0 = A_{1,u} = B_{1,u} = C_{1,u} \, .
	\label{eq:NBSasympBC_X1}
\eeq 
Another integration implies that $A_1$, $B_1$ and $C_1$ take yet unknown constant values at $\xi =0$. In fact, continuation of the perturbation theory around the UBS to higher orders reveals that the functions $X_1$ again split into two parts: one part that is $u$-independent and another part that depends on both $\xi$ and $u$ but is exponentially suppressed at infinity. 
Keeping this in mind, we deduce additional boundary conditions at $\xi =0$ from a power law expansion of the $X_1$'s in terms of $\xi$:
\beq
	0 = A_{1,\xi} = B_{1,\xi} = B_1 + 2\pi \, \frac{r_0}{L} C_1 \, .
	\label{eq:NBSasympBC_X1u0}
\eeq
These conditions are necessary to fix the constants of integration that arise from equation~\eqref{eq:NBSasympBC_X1}. In practice, our numerical scheme uses condition~\eqref{eq:NBSasympBC_X1u0} only at $(\xi ,u) = (0,0)$ but condition~\eqref{eq:NBSasympBC_X1} on all other points along $\xi = 0$. 

\subsubsection{Numerical grid}

\begin{figure}[ht]
	\centering
	\includegraphics[scale=1]{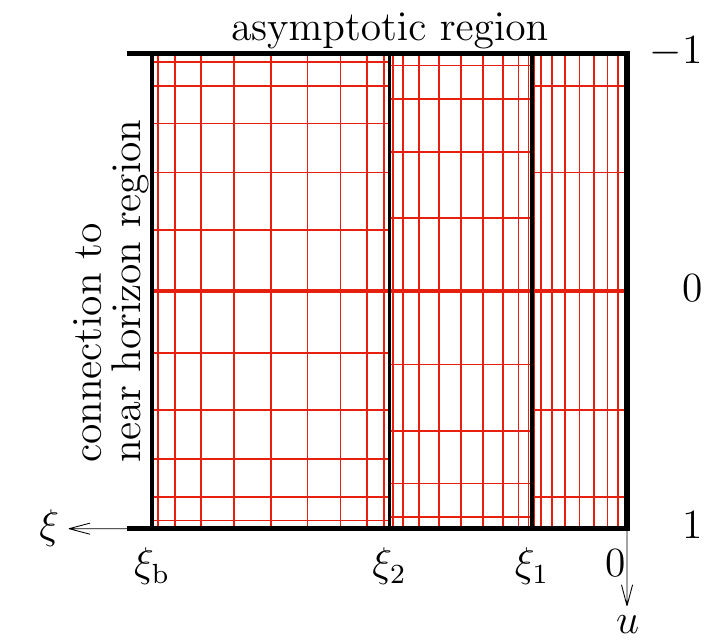}
	\caption{Numerical grid in the asymptotic region for $D=6$. In this part of the domain of integration, we evaluate the functions $A_1(\xi ,u)$, $B_1(\xi ,u)$ and $C_1(\xi ,u)$. Furthermore, we consider the one-dimensional functions $A_0(\xi )$, $B_0(\xi )$ and $C_0(\xi )$ at the coordinate line $u=0$, which is indicated by a bolder line in the plot. This region is connected to the near horizon region at $\xi =\xi _\text{b} =\xi (r_\text{b} )$. 
	}
	\label{fig:NBSdomain_grid_asymptotics_6D}
\end{figure}

We emphasize that we employ the ansatz~\eqref{eq:NBSsplitAnsatz6D} only in the asymptotic region, where $r>r_\text{b}$ or $\xi >\xi _\text{b} = \xi (r_\text{b} )$. Clearly, the numerical grid in this region relies on the coordinates $\xi$ and $u$. Moreover, we exploit the special behavior of the metric functions near the asymptotic boundary $\xi =0$ by decomposing the asymptotic region into several rectangular subdomains, see figure~\ref{fig:NBSdomain_grid_asymptotics_6D}. These subdomains are separated by lines of constant $\xi$, which we denote as $\xi _1$ and $\xi _2$ in figure~\ref{fig:NBSdomain_grid_asymptotics_6D}. Note that $0<\xi _1<\xi _2<\xi _\text{b}$. The benefit of this decomposition is twofold. On the one hand, we obtain a rapid fall-off of the spectral coefficients with respect to $\xi$ in each of the subdomains, if we choose narrow domains in vicinity of $\xi =0$. In particular, this improves the accuracy by taking into account the non-analytic behavior of the metric functions caused by the exponential factor $\expterm$ when considered in terms of $\xi$. On the other hand, we use different resolutions in the $u$-direction in each subdomain. Near infinity $\xi =0$, the $u$-dependency is strongly suppressed by the exponential factor and thus we only need a small resolution. In contrast, far from the $\xi =0$ boundary, a much higher resolution in $u$-direction is necessary, since the non-uniformity of the horizon becomes appreciable. All in all, this heavily reduces the number of grid points to be taken and, speaking in terms of computational costs, we thus save a lot of memory capacity and computing time. 

\subsection{Treatment of the asymptotics in five dimensions}
\label{subsec:Treatment_of_the_asymptotics_in_five_dimensions}

It is tempting to proceed in $D=5$ in a similar way as described in the previous subsection for $D=6$, i.e.\ to decompose the functions $A$, $B$ and $C$ into a $z$-independent and a $z$-dependent part. However, we face an additional challenge in $D=5$: the appearance of logarithmic terms in the asymptotics. Obviously, such terms are absent in the first order perturbations around the UBS, but indeed they occur in higher orders. Especially, when we use a pseudo-spectral scheme, these terms are cumbersome because we need high resolutions for an accurate representation.

\subsubsection{Ansatz}

An appropriate decomposition of the metric functions that takes the logarithmic behavior into account reads
\refstepcounter{equation} \label{eq:NBSsplitAnsatz5D}
\begin{align}
	A(r,z) & = A_0(r) \frac{r_0}{r} \hphantom{\ln \frac{r_0}{r}} + \, \tilde A_1(r ,z) \, \cos\left( \tfrac{2\pi}{L}z\right) \, , 	
		\tag{\theequation a} \label{eq:NBSsplitAnsatz5DA}  \\
	B(r,z) & = B_0(r) \frac{r_0}{r} \hphantom{\ln \frac{r_0}{r}} + \, \tilde B_1(r ,z) \, \cos\left( \tfrac{2\pi}{L}z\right) \,  ,					                           
		\tag{\theequation b} \label{eq:NBSsplitAnsatz5DB} \\ 
	C(r,z) & = C_0(r) \frac{r_0}{r}            \ln \frac{r_0}{r} + \, \tilde C_1(r ,z) \, \cos\left( \tfrac{2\pi}{L}z\right) \, . 
		\tag{\theequation c} \label{eq:NBSsplitAnsatz5DC}  					  
\end{align}
We see the logarithm $\ln (r_0/r)$ appearing in the function $C$, but similar expressions are present in the other functions as well though they are accompanied by higher powers of $r_0/r$. 
Note again that this ansatz allows us to read off the asymptotic charges directly from the function values of $A_\infty = \lim_{r\to\infty} A_0(r)$ and $B_\infty = \lim_{r\to\infty} B_0(r)$. We also set $C_\infty =\lim_{r\to\infty} C_0(r)$.


For the radial direction we consider the coordinate transformation $r_0/r = \chi (2-\chi )$ as defined in equation~\eqref{eq:NBScoordchi}, where $\chi = 0$ corresponds to the asymptotic boundary $r\to \infty$ and  $\chi =1$ to the horizon $r=r_0$. Again, we find that the transverse direction is well described by the coordinate $u=\cos (2\pi\, z/L)$. 

Obviously, the ansatz~\eqref{eq:NBSsplitAnsatz5D} does not extract the leading behavior of the marginal GL mode from the functions $\tilde A_1$, $\tilde B_1$ and $\tilde C_1$, which we worked out in equation~\eqref{eq:NBSLinPerturbAnsatz5D}. The reason for this is that, this time, we want to keep the exponential factor $\expterm$ within these functions, since it suppresses any logarithmic expressions of the kind $\chi ^l\ln ^k\chi$ with $k,l>0$. Therefore, the functions $\tilde A_1$, $\tilde B_1$ and $\tilde C_1$ are guaranteed to be smooth and a spectral representation converges subgeometrically. 

Unfortunately, such a trick is not available for the asymptotically dominant functions $A_0$, $B_0$ and $C_0$. In this case, we perform the coordinate transformation 
\beq
	\chi = \chi _\text{b} \, \E ^{1-\frac{1}{\eta}} \, ,
	\label{eq:NBScoordeta}
\eeq
in order to deal with the logarithmic behavior of these functions. The new coordinate $\eta$ runs from $\eta =0$, which corresponds to $\chi =0$, to $\eta =1$, where $\chi = \chi _\text{b}$ corresponding to $r=r_\text{b}$. Consider now typical expressions $\chi ^l\ln ^k\chi$ that occur in the functions $A_0$, $B_0$ and $C_0$. As already mentioned, a spectral representation of these expressions with respect to $\eta$ converges only algebraically, but by utilizing the coordinate transformation $\chi (\eta )$ we enhance the order of convergence to being subgeometric, because the problematic logarithmic expressions are smooth with respect to $\eta$:
\beq
	\chi ^l \ln ^k \chi = \chi _b^l \, \E ^{l \left(1-\frac{1}{\eta} \right) } \left( \ln \chi _b + 1 - \frac{1}{\eta} \right) ^k \, . 
	\label{eq:NBSlogTermInEta}
\eeq 

Taking the above findings into account, our strategy for the numerical implementation is to consider the functions $\tilde A_1$, $\tilde B_1$ and $\tilde C_1$ on a $(\chi ,u)$-grid and the functions $A_0$, $B_0$ and $C_0$ on an $\eta$-grid at the coordinate line $u=0$. Recall that this only concerns the asymptotic region $r>r_b$, see figure~\ref{fig:NBSdomain_bare_split}. However, another technical difficulty arises from this approach. Consider the field equations for the functions $A_0$, $B_0$ and $C_0$ with respect to $\eta$, which originate from equations~\eqref{eq:NBSfield_eqns} at the specific coordinate value $u=0$ (or equivalently $z=L/4$). We can bring them into the form
\beq
	0 = F_0 \left( X_{0,\eta\eta}, X_{0,\eta} , X_0 ; \eta \right) + \left( \frac{r(\eta )}{r_0} \right) ^4 F_1 \left( \tilde X_{1,u} ; \eta \right) \, ,
	\label{eq:NBSsplitEqns5D}	
\eeq
where $X_0=\{ A_0, B_0, C_0\}$ and $\tilde X_1 = \{ \tilde A_1, \tilde B_1, \tilde C_1 \}$. Then we have a part $F_0$ that depends on the one-dimensional functions $X_0$ and its derivatives, and a part $F_1$ that depends on the $u$-derivatives of the two-dimensional functions $\tilde X_1$. For small values of $\eta$ the factor $(r/r_0)^4$ strongly blows up due to the exponential mapping~\eqref{eq:NBScoordeta}. Nevertheless, the part $F_1$ of equation~\eqref{eq:NBSsplitEqns5D} vanishes for $\eta \to 0$ since the functions $\tilde X_1$ carry the exponential factor $\expterm$, which suppresses the $(r/r_0)^4$ term. However, things are not that clear in a numerical implementation, where we have only finite machine precision. In this situation, we can not guarantee that the values of $\tilde X_1$ are always tiny enough to compensate the $(r/r_0)^4$ behavior. Therefore, we extract a $(r_0/r)^4$ factor out of the two-dimensional functions in order to cancel exactly the problematic factor in equation~\eqref{eq:NBSsplitEqns5D}, i.e.\
\beq
	\tilde A_1 = \left( \frac{r_0}{r} \right)^4 A_1 \, , \quad \tilde B_1 = \left( \frac{r_0}{r} \right)^4 B_1 \, , \quad \tilde C_1 = \left( \frac{r_0}{r} \right) ^4 C_1 \, . 
	\label{eq:NBSX1extract5D}
\eeq
Despite the fact that the fall-off of the spectral coefficients of the functions $X_1$ with respect to $\chi$ is slower than that of $\tilde X_1$, the rate of convergence still stays subgeometric. Again, the reason for this is the asymptotically dominant exponential factor $\expterm$.

\subsubsection{Asymptotic boundary conditions}

We are in place to provide appropriate boundary conditions for the functions $A_0$, $B_0$, $C_0$, $A_1$, $B_1$ and $C_1$ at the asymptotic boundary. The previously discussed equations for the one-dimensional functions yield in the limit $\eta \to 0$:
\beq
	0 = A_{0,\eta} = B_{0,\eta} = A_\infty + 2 \, B_\infty + C_\infty \, .
	\label{eq:NBSasymp_BCX05D}
\eeq
Furthermore, as we repeated frequently, the two-dimensional functions decay very rapidly to zero in the asymptotic limit. Therefore, we require at $\chi =0$
\beq
	0 = A_1 = B_1 = C_1 \, .
	\label{eq:NBSasymp_BCX15D}
\eeq

\subsubsection{Numerical grid}

\begin{figure}[ht]
	\centering
	\includegraphics[scale=1]{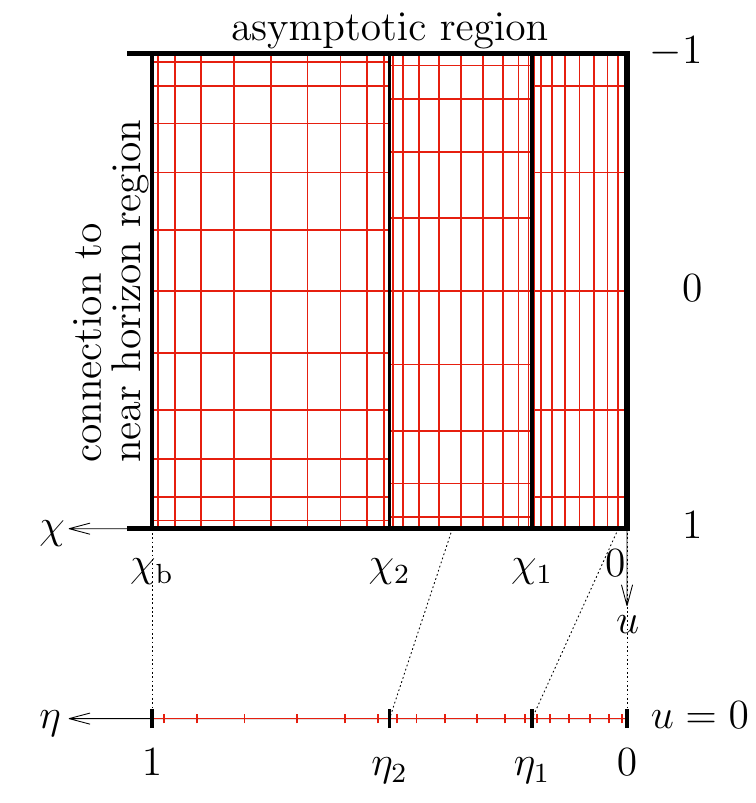}
	\caption{Numerical grid in the asymptotic region for $D=5$. In this part of the domain of integration, we evaluate the functions $A_1(\chi ,u)$, $B_1(\chi ,u)$ and $C_1(\chi ,u)$. Furthermore, we consider the one-dimensional functions $A_0(\eta )$, $B_0(\eta )$ and $C_0(\eta )$ at the coordinate line $u=0$ but on an $\eta$-grid. This region is connected to the near horizon region at $\chi =\chi _\text{b} =\chi (r_\text{b} )$ corresponding to $\eta =0$. }
	\label{fig:NBSdomain_grid_asymptotics_5D}
\end{figure}

We summarize the numerical approach in the asymptotic region for $D=5$ in figure~\ref{fig:NBSdomain_grid_asymptotics_5D}. While the two-dimensional functions $A_1$, $B_1$ and $C_1$ are considered on an $(\chi ,u)$-grid, we regard the one-dimensional functions $A_0$, $B_0$ and $C_0$ as functions of $\eta$. The corresponding $\eta$-grid actually lies at the coordinate line $u=0$, since the one-dimensional functions are evaluated there, see also footnote~\ref{footnote:oddN}. As a consequence of the different grids, an interpolation between both is necessary during the numerical solution of the field equations. Fortunately, as this is done only at one coordinate line, $u=0$, the corresponding computational costs barely come into account.

Again, we utilize the numerical trick presented in the previous subsection~\ref{subsec:Treatment_of_the_asymptotics_in_six_dimensions}: the split of the asymptotic region into several subdomains with inner boundaries, for example at $\chi = \chi _1$ and $\chi = \chi _2$ with $0<\chi _1<\chi _2<\chi _\text{b} = \chi (r_\text{b})$, cf.\ figure~\ref{fig:NBSdomain_grid_asymptotics_5D}. Furthermore, we do the same with the $\eta$-grid by considering several intervals, e.g.\ $0\leq \eta \leq \eta _1$ and $\eta _1\leq \eta \leq \eta _2$ and $\eta_2\leq \eta \leq 1$. The idea and benefit of this trick is similar as discussed above. 

\subsection{Treatment of the horizon}
\label{subsec:Treatment_of_the_horizon}

In the near horizon region, the metric functions $A$, $B$ and $C$ show similar behavior for $D=5$ and $D=6$. Therefore, we implement the same adaptions in both cases and describe them below.

In case of slightly deformed NBS solutions the functions $A$, $B$ and $C$ are well behaved near the horizon. However, when we leave the regime close to the UBS and consider solutions with stronger horizon deformations, it turns out that the functions $A$, $B$ and $C$ develop clearly pronounced peaks at $(r,z)=(r_0,L/2)$. This happens exactly at the waist of the black string horizon, cf.\ figure~\ref{fig:horizonsNBS3D}. In order to avoid exceedingly high values of the metric functions at the waist we consider the redefinition
\beq
	\alpha = \E ^{-2\, A} \, , \quad \beta = \E ^{-2\, B} \, , \quad \gamma = \E ^{2\, C} \, .
	\label{eq:NBSalbega}
\eeq
The new functions $\alpha$, $\beta$ and $\gamma$ are bound and, moreover, they approach zero at the waist when the NBS solutions approach the critical transition. However, this redefinition does not solve another problem that arises at the waist: Even the newly defined functions $\alpha$, $\beta$ and $\gamma$ are still plagued with steep gradients. We realize that the benefits of the coordinate $u$ are lost in the critical regime of high horizon deformations, since derivatives with respect to $u$ rise in vicinity of the waist and in particular at the coordinate line $u=-1$ (or equivalently $z=L/2$). Therefore, we refrain from the coordinate $u$ in the near horizon region and return to the original coordinate $z$.\footnote{In the numerical implementation, we use dimensionless quantities and thus we rather carry out the calculations with respect to $z/L$.} Note that we still utilize the coordinate transformation from $r$ to $\xi$, see equation~\eqref{eq:NBScoordxi}, for $D=6$ and from $r$ to $\chi$, see equation~\eqref{eq:NBScoordchi}, for $D=5$, respectively.

\begin{figure}[ht]
	\centering
	\includegraphics[scale=1]{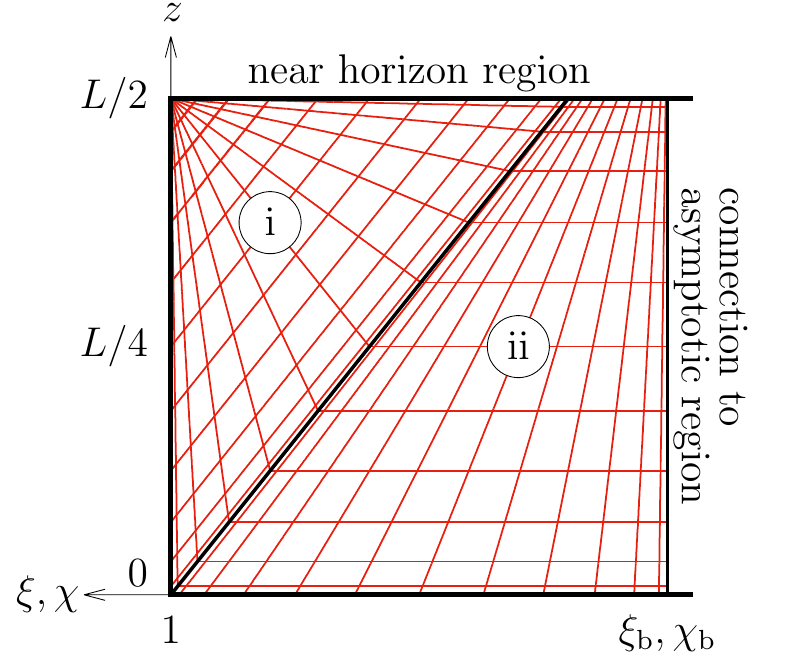}
	\caption{Numerical grid in the near horizon region. In this part of the domain of integration, we evaluate the functions $\alpha$, $\beta$ and $\gamma$. Depending on the dimension under consideration we use the radial coordinate $\xi$ ($D=6$) or $\chi$ ($D=5$). The numerical grid in the triangular subdomain (i) relies on polar-like coordinates centered around the waist where $\xi ,\chi =1$ and $z=L/2$. Likewise, we use appropriate coordinates in the trapezoidal subdomain (ii). This region is connected to the asymptotic region at $\xi =\xi _\text{b} =\xi (r_\text{b} )$ or $\chi = \chi _\text{b} = \chi (r_\text{b} )$, respectively.}
	\label{fig:NBSgrid_horizon}
\end{figure}

Nonetheless, we can not completely avoid steep gradients with respect to the radial direction when going towards the waist. We resolve these peaks by introducing polar-like coordinates that are centered around the waist. To keep the corresponding coordinate transformations simple, we perform a decomposition of the near horizon region into several subdomains. Here, we illustrate the resulting numerical grid in figure~\ref{fig:NBSgrid_horizon}.

Below, we describe the corresponding coordinate transformations to cover the subdomains (i) and (ii) depicted in figure~\ref{fig:NBSgrid_horizon}. For convenience, we only discuss the $D=6$ case and perform the coordinate transformations with respect to the coordinate $\xi$. However, the $D=5$ case works analogously and can be obtained by simply replacing $\xi$ by $\chi$ below.
 
We cover the subdomain (ii) with the coordinates $(v,z)$, where 
\beq
	v = \xi _\text{b} + \frac{(\xi _\text{i} - \xi _\text{b}) (\xi - \xi _\text{b} )}{(\xi _\text{i}-\xi _\text{b}) + (1-\xi _\text{i}) ( 1-4\, z/L )} \, ,
	\label{eq:NBScoordv}
\eeq
and $v\in [\xi _\text{i},\xi _\text{b}]$. For $v=\xi _\text{b}$ we recover $\xi = \xi _\text{b}$, while the coordinate line $v = \xi _\text{i}$ represents the diagonal domain boundary, which is described by $1-4\, z/L = (\xi -\xi _\text{i})/( 1 - \xi _\text{i} )$. Note that $\xi _\text{i}$ denotes the $\xi$ value of the center of this diagonal. 

In the triangular domain we define the coordinates $w\in [\xi _\text{i},1]$ and $p\in [-1,1]$ as 
\begin{align}
	w &= 1 - \frac{1}{2}\left[ (1-\xi ) + (1-\xi _\text{i})(2 -4\, z/L ) \right] \, , \label{eq:NBScoordw} \\
	p &= 1 - 2\frac{1-\xi}{(1-\xi ) + (1-\xi _\text{i})(2 -4\, z/L)} \, .  \label{eq:NBScoordp}
\end{align}
From the inverted form 
\begin{align}	
	\xi         &= 1-(1-w )(1-p ) \, , \label{eq:NBScoordxiofwp} \\
	1 - 4\, z/L &= \frac{(1-w )(1+p )}{1-\xi _\text{i}} -1 \, , \label{eq:NBScoordzofwp}
\end{align}
we see that the waist $(\xi ,z ) = (1,L/2)$ is obtained by $w =1$ and the diagonal by $w=\xi _\text{i}$. Moreover, $p=1$ represents the remaining part of the horizon and $p=-1$ the upper mirror boundary $z=L/2$. By this construction the coordinate $w$ is now similar to a radial coordinate with respect to the waist $(\xi ,z ) = (1,L/2)$, whereas $p$ behaves like an angular coordinate. The bottom line is that the triangular subdomain in $(\xi ,z)$ coordinates is mapped to a rectangular domain in $(w,p)$ coordinates with the original point $(\xi ,z) = (1,L/2)$ blown up to an edge $w=1$ in the new coordinates. Accordingly, the grid points around the waist are densely distributed in order to account for the steep gradients of the metric functions there. 

\begin{figure}[ht]
	\centering
	\includegraphics[scale=1]{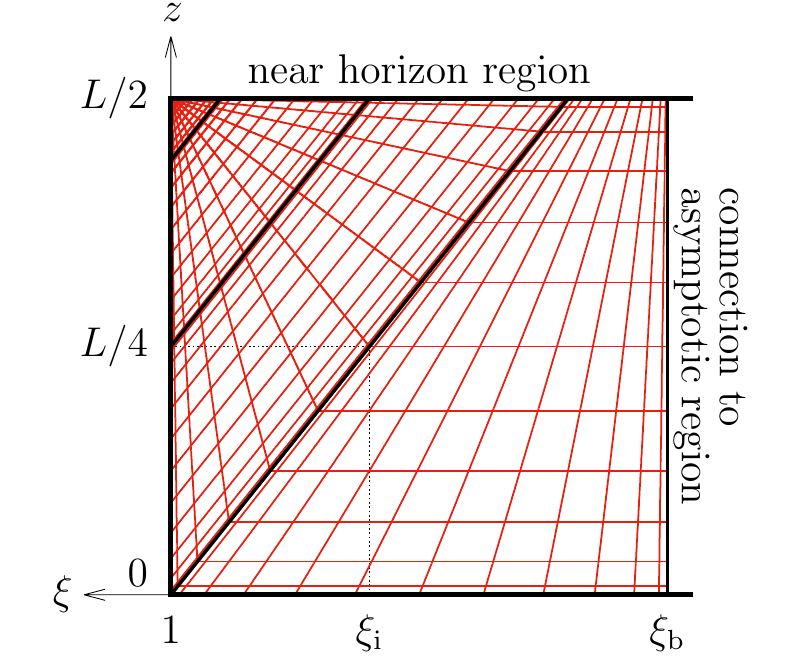}
	\caption{Numerical grid in the near horizon region with a decomposition of the triangular subdomain (i). This setup allows us to separately adapt the numerical resolution in each of the subdomains. Consequently, we are able to highly increase the density of grid points around waist $(\xi ,z)=(1,L/2)$, while we can utilize a moderate resolution where the behavior of the metric functions is not problematic. This region is connected to the asymptotic region at $\xi =\xi _\text{b} =\xi (r_\text{b} )$, respectively.}
	\label{fig:NBSgrid_horizon_full}
\end{figure}

We incorporate the following two additional adaptions to enhance the benefit of this polar-like coordinates. Similar to the domain decomposition in the asymptotic region we split the triangular subdomain into several linearly connected subdomains that are separated along constant lines of $w=w_i$. This allows us to significantly increase the resolution particularly near the waist and to take care of the specific behavior of the metric functions there. We illustrate the resulting grid in figure~\ref{fig:NBSgrid_horizon_full}.

Finally, we utilize an analytic mesh refinement to flatten out the steep gradients at the waist $w=1$, see appendix subsection~\ref{subsec:Further_common_techniques} for more details. In the setup at hand we perform the coordinate transformation
 \beq
	w = 1 - (1-w _{0}) \frac{\sinh \left( \lambda \, \frac{1-\bar w}{1-w _{0}} \right)}{\sinh \lambda} \, ,
	\label{eq:NBSanalytic_mesh_refinement}
\eeq
in the very last domain, which contains the waist, i.e.\ the triangular domain in figure~\ref{fig:NBSgrid_horizon_full}. Here, $\bar w\in [w_0,1]$ is the new coordinate and $w_0$ denotes the boundary of this domain, such that $w\in [w_0,1]$. In addition, $\lambda$ is an arbitrary parameter that has to be chosen appropriately but is normally $\mathcal O(1)$. 

\section{Parameters and physical quantities}
\label{sec:Parameters_and_physical_quantities_NBS}

In the aforedescribed scheme we introduced a lot of unphysical parameters that essentially control the numerical grid. Below, we specify explicit values for these parameters that turned out to work well in a numerical implementation in subsection~\ref{subsec:Parameter_values_NBS}. Moreover, in section~\ref{subsec:Extraction_of_physical_quantities_NBS} we discuss how to extract physically relevant quantities from the numerical data. 

\subsection{Parameter values}
\label{subsec:Parameter_values_NBS}

There are two length scales that enter the numerical scheme: the asymptotic size of the compact dimension $L$ and the coordinate radius of the horizon $r_0$. In fact, we can get rid of both scales with only the ratio $K=L/r_0$ still appearing in the field equations. Then, we fix this ratio to the corresponding value for a marginally stable UBS at the GL point, cf.\ equation~\ref{eq:NBSKGL6D} for $D=6$ and equation~\ref{eq:NBSKGL5D} for $D=5$. In tables~\ref{tab:ParametersNBS5D} and~\ref{tab:ParametersNBS6D} we summarize appropriate values for the remaining parameters that control the numerical grid. 

In section~\ref{sec:Metric_ansatz_and_field_equations} we also introduced the control parameter $\beta _\text{c}$, which is the value of the function $\beta$ at the waist of the black string horizon. Obviously, we obtain slightly deformed NBS solutions for $\beta _\text{c} \lesssim 1$ and approach the critical regime of nearly pinching solutions for $\beta _\text{c}\to 0$. Employing the scheme described above we were able to construct numerical NBS solutions down to $\beta _\text{c} = 10^{-5}$ for $D=6$ and $\beta _\text{c} = 10^{-4}$ for $D=5$. 

Finally, we note that the parameter $\lambda$ of the analytic mesh refinement~\eqref{eq:NBSanalytic_mesh_refinement} is missing in the tables~\ref{tab:ParametersNBS5D} and~\ref{tab:ParametersNBS6D}, since we have to adjust it several times when approaching the critical regime. For our numerical NBS solutions with smallest $\beta _\text{c}$ a value of $\lambda \approx 10$ proved to be optimal, while $\lambda$ has to be smaller for solutions with larger $\beta _\text{c}$.  
\begin{table}
	\centering
	\caption{Parameter values that shape the numerical grid for $D=5$. In particular, these values were used to find NBS solutions in the critical regime of a nearly pinching horizon. See figures~ \ref{fig:NBSdomain_grid_asymptotics_5D}, \ref{fig:NBSgrid_horizon_full} and the text for the meaning of the parameters. We split the triangular subdomain (i) into four subdomains with inner boundaries at $w=w_0$, $w=w_1$ and $w=w_2$. }
	\label{tab:ParametersNBS5D}
	\begin{tabularx}{\textwidth}{*{9}{Y}}	
		\toprule 				
		 $\eta _1$ 	& $\eta _2$	&  	$\chi _1$ 	& $\chi _2$	& $\chi _\text{b}$	& $\chi _\text{i}$	& $w_2$ & $w_1$ & $w_0$		\\
		\midrule
		 0.15		& 0.4		&	0.03		& 0.12		& 0.3	 			& 0.8		 		& 0.9  	& 0.95	& 0.99		\\
		\bottomrule	
	\end{tabularx}
\end{table}
\begin{table}
	\centering
	\caption{Parameter values that shape the numerical grid for $D=6$. In particular, these values were used to find NBS solutions in the critical regime of a nearly pinching horizon. See figures~\ref{fig:NBSdomain_grid_asymptotics_6D}, \ref{fig:NBSgrid_horizon_full} and the text for the meaning of the parameters. We split the triangular subdomain (i) into three subdomains with inner boundaries at $w=w_0$ and $w=w_1$. }
	\label{tab:ParametersNBS6D}
	\begin{tabularx}{\textwidth}{*{6}{Y}}			
		\toprule 		
		 $\xi _1$ 	& $\xi _2$	& $\xi _\text{b}$	& $\xi _\text{i}$	& $w_1$ 	& $w_0$		\\
		\midrule
		 0.1		& 0.25		& 0.5	 			& 0.8		 		& 0.9  			& 0.975		   		\\
		\bottomrule		
	\end{tabularx}
\end{table}

\subsection{Extraction of physical quantities}
\label{subsec:Extraction_of_physical_quantities_NBS}

In this section we give formulas to calculate several physical quantities of interest. Mainly, these are the thermodynamic quantities that we already introduced in subsection~\ref{subsec:Physical_quantities}. In addition, we introduce relevant quantities that describe the geometry of the horizon.

\subsubsection{Thermodynamic quantities}

The complicated decomposition of the metric functions~\eqref{eq:NBSsplitAnsatz6D} and~\eqref{eq:NBSsplitAnsatz5D} allows us to read off the mass $M$ and the relative tension $n$ directly from the asymptotic values of the one-dimensional functions:
\begin{align}
	M/M_\GL & = 1 - 2\, A_\infty - \frac{2}{D-3} B_\infty \, , \label{eq:NBSMass} \\
	n/n_\GL & = \frac{1 - 2\, A_\infty - 2 \, (D-3) \, B_\infty}{1 - 2\, A_\infty - 2/(D-3) \, B_\infty} \label{eq:NBSRelativeTension} \, .
\end{align}
Here, $M_\GL$ and $n_\GL$ are the corresponding values of a UBS, cf.\ equations~\eqref{eq:massUBS} and~\eqref{eq:tensionUBS}, at the GL point where $L/r_0 = K_\GL$.\footnote{Note that all UBSs have the same relative tension, but for the sake of consistency we use here the notation $n_\GL$ rather than $n_\UBS$.} Obviously, the normalization by $M_\GL$ and $n_\GL$ renders the considered quantities dimensionless. 

Temperature and entropy are both evaluated on the horizon $r=r_0$ and read
\begin{align}
	T/T_\GL & =  \E ^{A-B}   \, , \label{eq:NBSTemperature} \\
	S/S_\GL & =  \frac{2}{L} \int _{0}^{L/2} \E ^{B + (D-3) \, C } \, \D z  \, . \label{eq:NBSEntropy}
\end{align}
Again, $T_\GL$ and $S_\GL$ denote the corresponding values of a UBS, cf.\ equations~\eqref{eq:entropyUBS} and~\eqref{eq:temperatureUBS}, at the GL point. We recall that the temperature is constant along the horizon, which is explicitly imposed by the boundary condition~\eqref{eq:NBSBCconstT}. 

\subsubsection{Geometric quantities}

Interesting quantities to describe the geometry of the horizon are: the minimal and maximal horizon areal radius
\refstepcounter{equation} \label{eq:NBSHorizonArealRadii}
\begin{align}
	R_\text{min} &= r_0 \, \E ^{C} \quad \text{at} \quad z = L/2 \, , \tag{\theequation a}  \label{eq:NBSHorizonArealRadiusmin} \\
	R_\text{max} &= r_0 \, \E ^{C} \quad \text{at} \quad z = 0   \, , \tag{\theequation b}  \label{eq:NBSHorizonArealRadiusmax}
\end{align}
and the proper length of the horizon along the compact dimension
\beq
	L_\mathcal{H} = 2 \int _0^{L/2} \E ^B \, \D z \, .
	\label{eq:NBSHorizonLength}
\eeq
All of these quantities are considered on the horizon $r=r_0$. In particular, $\Rmin$ is a useful physical parameter that takes the value $\Rmin = r_0$ for UBSs and gradually decreases along the NBS branch, while the limit $\Rmin \to 0$ describes the critical transit solution with a pinching horizon. 

Finally, we want to embed the NBS horizon into $(D-1)$-dimensional flat space~\eqref{eq:metricFlatSpace}. A comparison with the NBS metric~\eqref{eq:metricNBS} yields 
\refstepcounter{equation} \label{eq:NBSHorizonEmbedding}
\begin{align}
	R ( z ) &= r_0 \, \E ^C \, , \tag{\theequation a}  \label{eq:NBSHorizonEmbeddingR} \\
	Z ( z ) &= \int_{0}^{z} \sqrt{ \E ^{2B} - \left( \D R / \D \tilde z \right) ^2  } \, \D \tilde z \, . \tag{\theequation b}  \label{eq:NBSHorizonEmbeddingZ}
\end{align}
Again, we evaluate these expressions at the horizon $r=r_0$. Note that we have fixed an arbitrary constant of integration in order to have $Z=0$ at $z=0$.

\section{Accuracy of the numerical solutions}
\label{sec:Accuracy_of_the_numerical_solutions_NBS}

\begin{figure}[]
	\centering
	\includegraphics[scale=1]{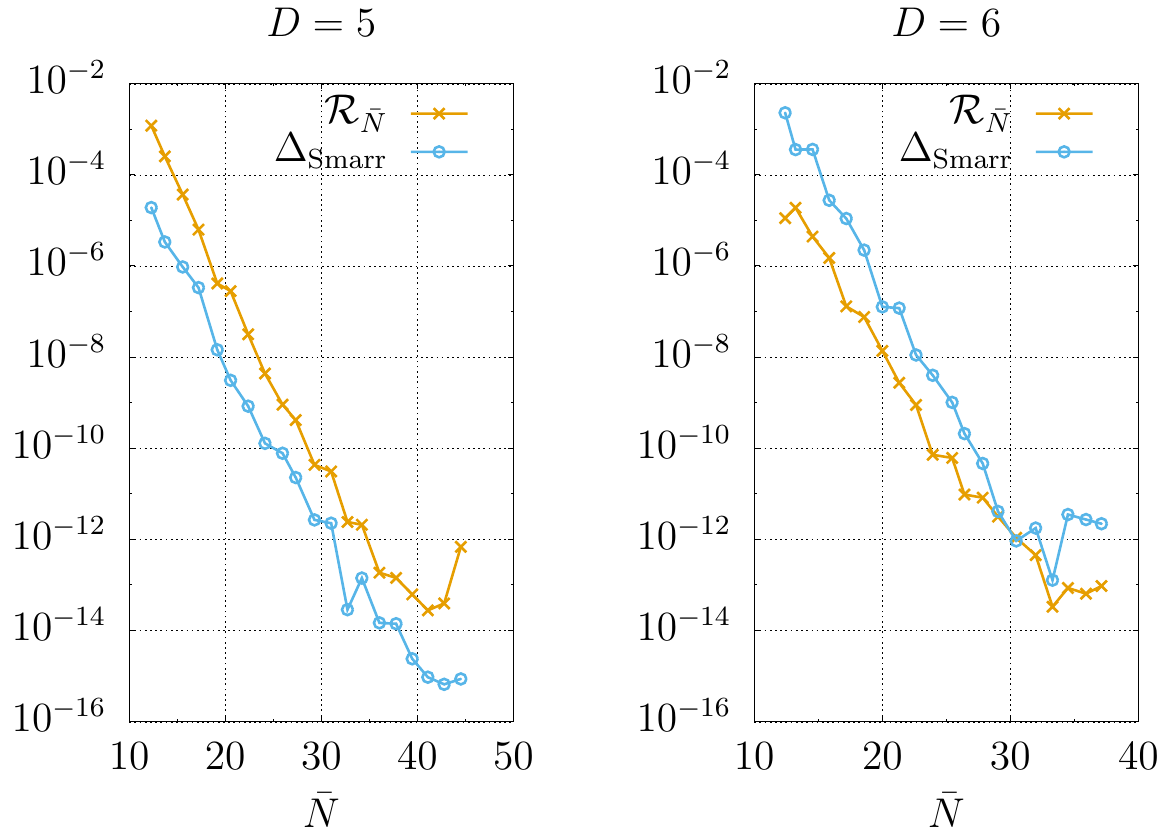}
	\caption{Convergence of the residue $\mathcal R_{\bar N}$ and the deviation from Smarr's relation $\Delta _\text{Smarr}$ as a function of the mean resolution $\bar N$. The respective numerical solutions correspond to NBSs with $\beta _\text{c} = 10^{-4}$ for $D=5$ and $\beta _\text{c} = 10^{-5}$ for $D=6$, or, in more physical terms, $\Rmin /L \approx 0.0040$ for $D=5$ and $\Rmin /L \approx 0.00086$ for $D=6$. }
	\label{fig:NBSconvergence}
\end{figure}

A numerical solution is worth nothing without an estimate of its accuracy. Without an analytic solution at hand, the most reliable and common way to test the accuracy of a pseudo-spectral solution is to compare a reference solution with high resolution with several solutions of lower resolutions that are obtained by the same procedure. In particular, we interpolate each solution on the same fine grid by using spectral interpolation techniques. Then, it is straightforward to calculate the differences of the reference solution to all of the less resolved solutions at each of these grid points. For a solution with resolution $N$ the greatest magnitude of these differences to the reference solution is referred to as the residue $\mathcal R_N$. We determined the residue for several NBS solutions. In figure~\ref{fig:NBSconvergence} we depict the convergence of the residue for our numerical solutions closest to the critical transit solution, i.e.\ those solutions that are highly demanding to obtain due to their steep gradients near the horizon. We see that even in this critical regime the sophisticated numerical scheme is able to produce solutions with residues of up to approximately $10^{-13}$. Moreover, figure~\ref{fig:NBSconvergence} displays the deviation $\Delta _\text{Smarr}$ from Smarr's relation~\eqref{eq:SmarrRelation} for the different resolutions $\bar N$ and we observe a similar fall-off as for the residue when the resolution is increased. Note that $\bar N$ denotes the mean resolution averaged over all subdomains and directions. As can be seen, the finite machine precision and rounding errors lead to a saturation of $\mathcal R_{\bar N}$ and $\Delta _\text{Smarr}$ for high $\bar N$.\footnote{The numerical calculations are performed in long double precision (80-bit extended precision).}  

Besides Smarr's relation we checked our solutions by comparison with the first law of black hole thermodynamics~\eqref{eq:FirstLawKK} on different parts of the NBS branch. In our case we parametrize the thermodynamic quantities with the control parameter $\beta _\text{c}$ and write the first law as
\beq
	\frac{\delta M}{\delta \beta _\text{c}} = T \frac{\delta S}{\delta \beta _\text{c}} \, .
	\label{eq:FirstLawParameterbeta}
\eeq
Now, consider different numerical NBS solutions at values of $\beta _\text{c}$ that are distributed on a Lobatto grid~\eqref{eq:MaintextLobattoPoints}. Accordingly, equation~\ref{eq:FirstLawParameterbeta} is evaluated using pseudo-spectral techniques. Again, we check equality of right and left hand side of equation~\eqref{eq:FirstLawParameterbeta} for different numbers of grid points with respect to $\beta _\text{c}$. At the end, a similar picture as in figure~\ref{fig:NBSconvergence} arises, where the deviations decrease rapidly with increasing resolutions until a saturation at orders of $10^{-10}$ is reached. 

Finally, it is essential to check the constraint equations for numerical NBS solutions since these are not explicitly solved in our scheme, cf.\ section~\ref{sec:Metric_ansatz_and_field_equations}. Moreover, it was pointed out in reference~\cite{Kudoh:2003ki} that a violation of the constraints would not cause any deviation from Smarr's relation or the first law as long as the field equations~\eqref{eq:NBSfield_eqns} are satisfied. For our numerical solutions the highest constraint violation is of the order $10^{-8}$ but only near the waist of the horizon, i.e.\ the critical point where the horizon nearly pinches. Far from this point the constraints are satisfied better by several orders of magnitude.

  \chapter{Numerical construction of localized black hole solutions}
\label{chap:Numerical_construction_of_localized_black_hole_solutions}

\begin{figure}[ht]
	\centering
	\includegraphics[scale=1]{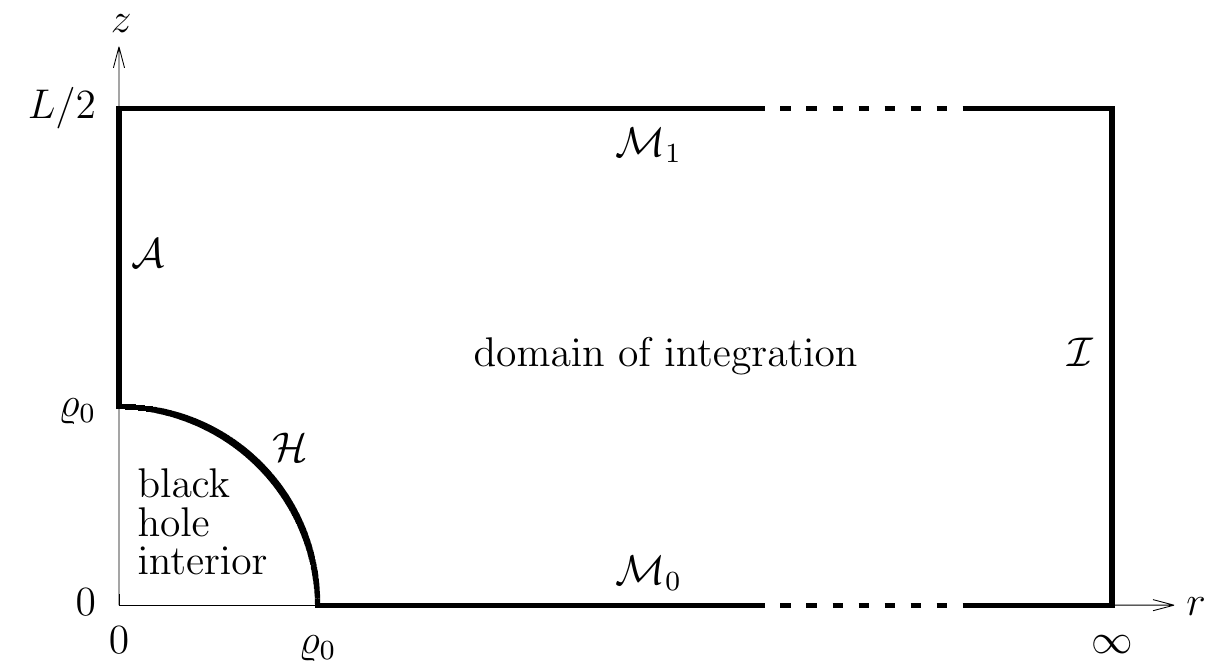}
	\caption{Domain of integration for the construction of LBH solutions.  The boundaries are the horizon $\mathcal H$, the exposed axis of spherical symmetry $\mathcal A$, the lower and upper mirror boundary $\mathcal M_0$ and $\mathcal M_1$, and the asymptotic boundary $\mathcal I$.}
	\label{fig:LBHint_domain_bare}
\end{figure}

Localized black hole (LBH) solutions exhibit a hyper-spherical horizon topology that is $\mathbb S^{D-2}$ in $D$-dimensional spacetime. Therefore, in contrast to non-uniform black strings (NBSs), the horizon does not wrap the entire compact dimension. We consider the LBH to be centered at the origin of the $r$-$z$ plane, where $r\in [0,\infty ]$ denotes the radial coordinate in the $D-2$ spatially extended dimensions and $z$ goes along the compact dimension with asymptotic length $L$, cf.\ section~\ref{sec:Static_Kaluza_Klein_black_holes}. Again, we assume (hyper-)spherical symmetry and reflection symmetry with respect to $z=0$, thus we focus on $z\in [0,L/2]$. Considering static solutions makes the problem effectively two-dimensional and hence it is always possible to choose a gauge in which the LBH has a hyper-spherical shape in the $(r,z)$ coordinates. Accordingly, the horizon is described by $r^2+z^2=\varrho _0^2$ with $\varrho _0$ denoting the coordinate radius of the horizon. Then, we end up with a domain of integration that is depicted in figure~\ref{fig:LBHint_domain_bare} and has five boundaries:
\begin{itemize}
	\item the asymptotic boundary $\mathcal I= \{ (r,z)\colon r\to \infty \, ,~ 0\leq z \leq L/2 \}$,
	\item the lower mirror boundary $\mathcal M_0 = \{ (r,z)\colon r\geq \varrho _0 \, ,~ z = 0 \}$,
	\item the upper mirror boundary $\mathcal M_1 = \{ (r,z)\colon r\geq  0 \, ,~ z = L/2 \}$,
	\item the exposed axis of symmetry $\mathcal A = \{ (r,z)\colon r = 0 \, ,~ \varrho _0 \leq z \leq L/2 \}$,
	\item the horizon $\mathcal H = \{ (r,z)\colon r\geq 0 \, ,~ z\geq 0 \, ,~ r^2+z^2=\varrho _0^2 \}$.
\end{itemize}
In addition, we refer to the $z=0$ plane as the equatorial plane and we denote the point $(r,z)=(0,\varrho _0)$ by the north pole of the horizon. 

In the following, we describe a sophisticated approach to find LBH solutions to Einstein's vacuum field equations with the above setup. In particular, we are interested in the critical regime where the poles of the black hole are about to touch each other on the compact dimension, cf.\ figure~\ref{fig:horizonsLBH3D}. The numerical techniques rely on a pseudo-spectral method that is outlined in appendix~\ref{sec:Pseudo-spectral_method}. Here, in section~\ref{sec:Metric_ansätze_and_boundary conditions}, we introduce two different metric ansätze that are suited for different regions of the domain of integration. The basic numerical strategy is outlined in section~\ref{sec:DeTurck method} and crucial adaptions are discussed in section~\ref{sec:Decomposition_of_the_domain_of_integration_LBH}. A discussion of physical and unphysical quantities is provided in section~\ref{sec:Parameters_and_physical_quantities_LBH}. Finally, we discuss the accuracy of the numerical results in section~\ref{sec:Accuracy_of_the_numerical_solutions_LBH}.

\section{Metric ansätze and boundary conditions}
\label{sec:Metric_ansätze_and_boundary conditions}

Due to its five boundaries it is not obvious which coordinates are most appropriate to cover the domain of integration. Reference~\cite{Kudoh:2003ki} provides a single coordinate system that contains all five boundaries located on constant coordinate lines. However, such a coordinate transformation is singular at some point, and thus has to be treated with special care. Moreover, due to the complexity of the transformation the resulting field equations will be lengthy. Instead, we follow the approach of reference~\cite{Headrick:2009pv} (see also reference~\cite{Sorkin:2003ka}): We consider two different coordinate charts, each one adapted to a different region of the domain of integration. More concretely, we introduce an asymptotic chart in subsection~\ref{subsec:Asymptotic_chart} and a near horizon chart in subsection~\ref{subsec:Near_horizon_chart}. 


\subsection{Asymptotic chart}
\label{subsec:Asymptotic_chart}

It is apparent that the $(r,z)$ coordinate system is already appropriate to describe the asymptotics. Within these coordinates, we utilize the general ansatz
\beq
	\D s^2_{\lowa\LBH} = - \lowa T \, \D t^2 + \lowa A \, \D r^2 + \lowa B \, \D z^2 + 2 \lowa F \, \D r \, \D z +  r^2 \lowa S \, \D \Omega ^2_{D-3} \, ,
	\label{eq:LBHmetric_asymptotic_chart}
\eeq
which incorporates the required symmetries. We omitted the explicit dependence of the five metric functions, $\lowa T$, $\lowa A$, $\lowa B$, $\lowa F$ and $\lowa S$, on $r$ and $z$. Note that if the function $\lowa F$ vanishes and all other functions are equal to one we recover the background metric~\eqref{eq:metricBG}, which describes the spacetime $\mathbb M^{D-1}\times \mathbb S^1$. We find the following boundary conditions on $\mathcal I$, $\mathcal A$, $\mathcal M_0$ and $\mathcal M_1$: 
\begin{itemize}
	\item 	The asymptotic boundary $\mathcal I$ ($r\to\infty$): \nopagebreak \\	
			In the asymptotic limit the spacetime shall approach the background~\eqref{eq:metricBG} that is
		  	\beq
		  	0 = \lowa T - 1 = \lowa A - 1 = \lowa B - 1 = \lowa S - 1 = \lowa F \, .
		  		\label{eq:LBHBCa_asymptotics}
		  	\eeq
	\item 	The exposed axis $\mathcal A$ ($r=0$): \nopagebreak \\
			This axis is the origin of the (hyper-)spherical symmetry and, accordingly, the metric degenerates there. However, in order to guarantee a regular spacetime the metric functions have to satisfy
		  	\beq
		  		0 = \lowa A - \lowa S = \frac{\partial \lowa T}{\partial r} = \frac{\partial \lowa A}{\partial r} = \frac{\partial \lowa B}{\partial r} = \frac{\partial \lowa S}{\partial r} = \lowa F \, .
		  		\label{eq:LBHBCa_axis}
		  	\eeq		  
	\item 	The mirror boundaries $\mathcal M_0$ ($z=0$) and $\mathcal M_1$ ($z=L/2$): \nopagebreak \\
		 	Mirror symmetry with respect to these two boundaries requires that 
		  	\beq
		  		0 = \frac{\partial \lowa T}{\partial z} = \frac{\partial \lowa A}{\partial z} = \frac{\partial \lowa B}{\partial z} = \frac{\partial \lowa S}{\partial z} = \lowa F \, .
		  		\label{eq:LBHBCa_mirror}
		  	\eeq
\end{itemize}
We note that the background metric~\eqref{eq:metricBG} already satisfies the conditions above. This fact will become important later. Obviously, it is rather cumbersome to find conditions on the horizon $\mathcal H$ in the $(r,z)$ coordinate. Hence, we proceed in describing a chart that is suitable for the horizon geometry.

\subsection{Near horizon chart}
\label{subsec:Near_horizon_chart}

The circular contour of the horizon in the $r$-$z$ plane calls for the introduction of polar coordinates 
\beq
	r =  \varrho \sin \varphi \, , \quad  z = \varrho \cos \varphi \, , 
	\label{eq:LBHpolar_coordinates}
\eeq
where the horizon is simply given by $\varrho = \varrho _0$. Then, we consider the metric ansatz
\beq
	\D s^2_{\lowh\LBH} = - \lowh T \, \D t^2 + \lowh A \, \D \varrho ^2 + \varrho ^2 \lowh B \, \D \varphi ^2 + 2 \, \varrho \lowh F \, \D \varrho \, \D \varphi + \varrho ^2 \sin ^2 \varphi \, \lowh S \, \D \Omega ^2_{D-3} \, ,
	\label{eq:LBHmetric_horizon_chart}
\eeq
where $\lowh T$, $\lowh A$, $\lowh B$, $\lowh F$ and $\lowh S$ are now functions of $\varrho$ and $\varphi$. The background metric in polar coordinates is obtained if $\lowh F$ vanishes and all other metric functions are equal to one. Furthermore, by comparing the $(r,z)$ and $(\varrho ,\varphi )$ coordinate system and the corresponding line elements~\eqref{eq:LBHmetric_asymptotic_chart} and~\eqref{eq:LBHmetric_horizon_chart} we find that $\lowh T = \lowa T$ and $\lowh S = \lowa S$, and that there is a linear connection between the functions $\lowh A$, $\lowh B$, $\lowh F$ and $\lowa A$, $\lowa B$, $\lowa F$:
\refstepcounter{equation}\label{eq:LBHrelatioABF}
\begin{align}
	\lowh A &= \sin ^2\varphi \, \lowa A + \cos ^2\varphi \, \lowa B + 2 \, \sin \varphi \, \cos \varphi \, \lowa F \, , \tag{\theequation a} \label{eq:LBHrelatioABFa} \\
	\lowh B &= \cos ^2\varphi \, \lowa A + \sin ^2\varphi \, \lowa B - 2\, \sin \varphi \, \cos \varphi \, \lowa F \, , \tag{\theequation b} \label{eq:LBHrelatioABFb} \\
	\lowh F &= \sin \varphi \, \cos \varphi \left( \lowa A - \lowa B \right) + \left( \cos ^2 \varphi - \sin ^2 \varphi \right) \lowa F \, . \tag{\theequation c} \label{eq:LBHrelatioABFc} 
\end{align}
As mentioned before, we assume the black hole horizon to be located at $\varrho =\varrho _0$ and ensure this by rewriting $\lowh T$ as 
\beq
	\lowh T = \kappa ^2 (\varrho -\varrho _0)^2 \lowh{\tilde T} \, ,
	\label{eq:LBHfuncT}
\eeq
with the new function $\lowh{\tilde T}$ being regular at the horizon. As a result, the following boundary conditions on $\mathcal H$, $\mathcal A$, $\mathcal M_0$ and $\mathcal M_1$ arise:
\begin{itemize}
	\item 	The horizon boundary $\mathcal H$ ($\varrho =\varrho _0$):  \nopagebreak \\
			We require regularity of the spacetime at the horizon and, moreover, that $\kappa$ is the surface gravity of the LBH. This yields the conditions
		 	\beq
		  		0 = \lowh{\tilde T} -\lowh A = \frac{\partial \lowh{\tilde T}}{\partial \varrho} = \frac{\partial \lowh A}{\partial \varrho} =  \frac{\partial }{\partial \varrho} (\varrho ^2\lowh B) = \frac{\partial }{\partial \varrho} (\varrho ^2\lowh S)= \lowh F \, .
		  		\label{eq:LBHBCh_horizon}
		 	 \eeq
	\item 	The exposed axis $\mathcal A$ ($\varphi =0$): \nopagebreak \\ 
		  	Again, regularity of the metric requires that
		  	\beq
		  		0 = \lowh B - \lowh S = \frac{\partial \lowh{\tilde T}}{\partial \varphi} = \frac{\partial \lowh A}{\partial \varphi} = \frac{\partial \lowh{B}}{\partial \varphi} = \frac{\partial \lowh S}{\partial \varphi} = \lowh F \, .
		  	\label{eq:LBHBCh_axis}
		  	\eeq
	\item 	The lower mirror boundary $\mathcal M_0$ ($\varphi = \pi /2$):  \nopagebreak \\
			We retain mirror symmetry by finding
			\beq
		  		0 = \frac{\partial \lowh T}{\partial \varphi} = \frac{\partial \lowh A}{\partial \varphi} = \frac{\partial \lowh B}{\partial \varphi} = \frac{\partial \lowh S}{\partial \varphi} = \lowh F \, .				
				\label{eq:LBHBCh_mirror0}
			\eeq
	\item 	The upper mirror boundary $\mathcal M_1$ ($\varrho \cos \varphi = L/2$): \nopagebreak \\
			Using the relations~\eqref{eq:LBHrelatioABF} it is straightforward to convert the conditions~\eqref{eq:LBHBCa_mirror} at the upper mirror boundary into the corresponding conditions in the near horizon chart. We do not give their explicit expressions here as they are rather lengthy.
			
\end{itemize}

\section{DeTurck method}
\label{sec:DeTurck method}

This time, instead of solving Einstein's vacuum field equations~\eqref{eq:EFEvac} directly, we employ the well-established DeTurck method~\cite{Headrick:2009pv,Figueras:2011va}. Since its first formulation in 2009, the DeTurck method has become the main strategy to find numerical solutions to Einstein's equations in static or stationary situations. Besides other advantages, the DeTurck method most importantly leads to a system of partial differential equations that is strictly elliptic. Recall that this is not the case for Einstein's equations in their original form~\eqref{eq:EFEvac}. We refer to references~\cite{Wiseman:2011by,Dias:2015nua} for detailed reviews of the DeTurck method. 

The method relies on the so-called Einstein-DeTurck equations\footnote{They are also referred to as generalized harmonic equations.}
\beq
	R_{\mu\nu} - \nabla _{(\mu}\xi_{\nu )} = 0 \, ,
	\label{eq:LBHEinstein-DeTurck_equations}
\eeq
where the DeTurck vector field is given by
\beq
	\xi ^\mu := g^{\alpha\beta} ( \Gamma ^\mu_{\alpha\beta} - \bar\Gamma ^\mu_{\alpha\beta} ) \, .
	\label{eq:LBHDeTurck_vector}
\eeq
The Christoffel connection $\bar\Gamma^\mu_{\alpha\beta}$ is associated with a prescribed reference metric $\bar g_{\mu\nu}$, whereas the other geometrical objects are constructed with respect to the unknown but desired target metric $g_{\mu\nu}$.

If we have a solution $g_{\mu\nu}$ to the Einstein-DeTurck equations at hand \textit{and} the corresponding DeTurck vector vanishes, the metric $g_{\mu\nu}$ obviously solves Einstein's vacuum field equations as well. The necessary conditions in order to obtain a vanishing DeTurck vector field are that the reference metric $\bar g_{\mu\nu}$ exhibits the same causal structure and boundary conditions as the target spacetime. In particular, the reference metric itself does not have to be a solution to Einstein's equations. However, given that the reference metric $\bar g_{\mu\nu}$ indeed satisfies the above requirements and that $g_{\mu\nu}$ is a solution to the Einstein-DeTurck equations, one can not generally guarantee that $g_{\mu\nu}$ solves Einstein's field equations. There could arise undesired solutions called Ricci solitons. Nevertheless, in the static case considered here, reference~\cite{Figueras:2011va} rules out the occurrence of Ricci solitons. Regardless of this, it is a necessary consistency test to check whether the DeTurck vector field vanishes for any numerically obtained solution to the Einstein-DeTurck equations. 

An appropriately chosen reference metric is of particular importance for the DeTurck method to work. In the next subsection~\ref{subsec:Construction_of_the_reference_metric} we discuss our choice of reference metric in the LBH context. Afterwards, we will describe the overall numerical scheme in subsection~\ref{subsec:Overall_scheme}.

\subsection{Construction of the reference metric}
\label{subsec:Construction_of_the_reference_metric}

We construct an appropriate reference metric, which is consistent with the boundary conditions described in section~\ref{sec:Metric_ansätze_and_boundary conditions}, by following the strategy of reference~\cite{Headrick:2009pv}. For this purpose, we come back to the observation that the background metric~\eqref{eq:metricBG} already satisfies the boundary conditions on four of the five boundaries. At this point it is convenient to rewrite the background metric in terms of the polar coordinates~\eqref{eq:LBHpolar_coordinates}:
\beq
	\D s^2_{\mathrm{BG}} = -\D t^2 + \D \varrho ^2 + \varrho ^2 \left( \D \varphi ^2 + \sin ^2 \varphi \, \D \Omega ^2_{D-3} \right) \, .
	\label{eq:LBHBGmetric_polar}
\eeq 
Now, the idea is to use this background metric as a reference only in a region of the domain of integration where $\varrho \geq \varrho _1$ with $\varrho _0 < \varrho _1< L/2$. Then, in the complement region where $\varrho _0 \leq \varrho < \varrho _1$, we only have to find a reference metric that satisfies the boundary conditions on the horizon $\mathcal H$, the exposed axis $\mathcal A$ and the lower mirror boundary $\mathcal M_0$, while it has to match the background metric at $\varrho = \varrho _1$. In particular, a metric that is independent of $\varphi$ and takes the form
\beq
	\D \bar s^2_{\LBH_\text{ref}} = - \bar H(\varrho )\, \D t^2 + \D \varrho ^2 + \bar G(\varrho ) \, \D \Omega ^2_{D-2}  \, .
	\label{eq:LBHrefmetric}	
\eeq
This metric already satisfies the conditions on $\mathcal A$ and $\mathcal M_0$. Here, we utilized $\D \Omega ^2_{D-2} = \D \varphi ^2 + \sin ^2 \varphi \, \D \Omega ^2_{D-3}$. In order to recover the background metric~\eqref{eq:LBHBGmetric_polar} for $\varrho \geq \varrho _1$, the functions $\bar H$ and $\bar G$ are written as
\beq
	\bar H(\varrho ) = \begin{cases}
						\bar H_\text{hor} (\varrho ) & \text{if} \quad \varrho <\varrho _1 \, , \\
						1	 						 & \text{if} \quad  \varrho \geq \varrho _1 \, ,
			      \end{cases}
			      \quad \text{and} \quad
	\bar G(\varrho ) = \begin{cases}
						\bar G_\text{hor} (\varrho ) & \text{if} \quad \varrho <\varrho _1 \, , \\
						\varrho ^2  				 & \text{if} \quad  \varrho \geq \varrho _1 \, .
			      \end{cases}
	\label{eq:LBHrefmetric_functions}
\eeq 
The obvious way to create a reference metric that contains a horizon is to start with a $D$-dimensional Schwarzschild-Tangherlini (ST) solution, since it does not have a $\varphi$-dependence as well. Of course, an ST metric does not match the background metric~\eqref{eq:LBHBGmetric_polar} and we thus have to make some adaptions. On the horizon, in order to satisfy the boundary conditions, the leading behavior of $\bar H_\text{hor}$ and $\bar G_\text{hor}$ reads
\refstepcounter{equation}\label{eq:LBHrefmetric_functions_horizon_expansion}
\begin{align}
	\bar H_\text{hor} &= \kappa ^2 (\varrho - \varrho _0 )^2 + \mathcal O \left[ (\varrho - \varrho _0 )^4 \right] \, , \tag{\theequation a} \label{eq:LBHrefmetric_functions_horizon_expansion_a}  \\
	\bar G_\text{hor} &= \frac{(D-3)^2}{4\,\kappa ^2} + \frac{D-3}{2}(\varrho - \varrho _0)^2 + \mathcal O \left[ (\varrho - \varrho _0 )^3 \right] \tag{\theequation b} \label{eq:LBHrefmetric_functions_horizon_expansion_b} \, .
\end{align} 
Note that this is an approximation of the ST metric near the horizon $\varrho = \varrho _0$ with surface gravity $\kappa$. In contrast, at $\varrho = \varrho _1$, it is required that
\refstepcounter{equation}\label{eq:LBHrefmetric_functions_rho1_expansion}
\begin{align}
	\bar H_\text{hor} &= 1 + \mathcal O \left[ (\varrho - \varrho _1 )^{k+1} \right] \, , \tag{\theequation a} \label{eq:LBHrefmetric_functions_rho1_expansion_a}  \\
	\bar G_\text{hor} &= \varrho ^2 + \mathcal O \left[ (\varrho - \varrho _1 )^{k+1} \right] \tag{\theequation b} \label{eq:LBHrefmetric_functions_rho1_expansion_b} \, ,
\end{align} 
where $k\geq 2$, i.e.\ we want the first $k$ derivatives of $\bar H_\text{hor}$ and $\bar G_\text{hor}$ to match the background metric. At this point, we have to emphasize that we construct the numerical solution by means of a pseudo-spectral multi-domain method. In each subdomain we essentially use Chebyshev polynomials to approximate the solution therein. The global solution is obtained by demanding continuity of the functions and their first normal derivatives on the inner boundaries between two domains. Based on this, if we place an inner boundary exactly at $\varrho = \varrho _1$, we can fix the gauge of the full solution by choosing a reference metric that is only $k$ times differentiable at $\varrho = \varrho _1$, as expressed by equations~\eqref{eq:LBHrefmetric_functions_rho1_expansion}.

In practice, we implemented two approaches: one that considers a smooth reference metric ($k\to\infty$) and one that considers a reference metric that is only twice continuously differentiable ($k=2$). In the former case we choose 
\refstepcounter{equation}\label{eq:LBHrefmetric_functions_horizon_exp}
\begin{align}
	\bar H_\text{hor} &= 1 - E(\varrho ) \, ,  \tag{\theequation a} \label{eq:LBHrefmetric_functions_horizon_exp_a} \\
	\bar G_\text{hor} &= \varrho ^2 - E(\varrho ) \left[ \varrho ^2 - \frac{(D-3)^2}{4\,\kappa ^2} - (\varrho - \varrho _0)^2 \left( \frac{D^2}{4} -D +\frac{3}{4} - \kappa ^2 \varrho _0^2 \right)  \right] \, ,  \tag{\theequation b} \label{eq:LBHrefmetric_functions_horizon_exp_b}
\end{align}
where the auxiliary function $E(\varrho)$ reads
\beq
	E(\varrho ) = \exp \left[ -\kappa ^2 \frac{(\varrho -\varrho _0 )^2}{1-(\varrho -\varrho _0)^2 / (\varrho _1 - \varrho _0 )^2} \right] \, .
	\label{eq:LBHauxexp_function}
\eeq
Obviously, we have $E(\varrho _0) = 1$, while for $\varrho\to\varrho _1$ the function $E( \varrho )$ shows an exponential decay to zero. This approach is similar to the one utilized in reference~\cite{Headrick:2009pv}. 

In case of the $k=2$ matching, we consider the following simplified ansatz
\refstepcounter{equation}\label{eq:LBHrefmetric_functions_horizon_poly}
\begin{align}
 	\bar H_\text{hor} &=  \kappa ^2 (\varrho -\varrho _0)^2 + \bar h_1 (\varrho -\varrho _0)^4 + \bar h_2 (\varrho -\varrho _0 )^6 + \bar h_3 (\varrho -\varrho _0 )^8 \, , \tag{\theequation a} \label{eq:LBHrefmetric_functions_horizon_poly_a} \\
	\bar G_\text{hor} &=  \frac{(D-3)^2}{4\kappa ^2} + \frac{D-3}{2}(\varrho - \varrho _0)^2  + \bar g_1 (\varrho -\varrho _0)^4 + \bar g_2 (\varrho -\varrho _0 )^6 + \bar g_3 (\varrho -\varrho _0 )^8 \, . \tag{\theequation b} \label{eq:LBHrefmetric_functions_horizon_poly_b}
\end{align}
We calculate the coefficients $\bar h_1$, $\bar h_2$, $\bar h_3$, $\bar g_1$, $\bar g_2$ and $\bar g_3$ by matching $\bar H_\text{hor}$ and $\bar G_\text{hor}$ with the background metric functions up to the second derivative at $\varrho =\varrho _1$. 

Let us investigate the crucial differences between these two approaches. Most importantly, the auxiliary function $E(\varrho )$ is smooth but not analytic at $\varrho = \varrho _1$. As pointed out in the section~\ref{sec:Numerical_method}, this considerably slows down the convergence of the spectral representation compared to an analytic function. In the $k=2$ approach~\eqref{eq:LBHrefmetric_functions_horizon_poly} we circumvent this by choosing reference metric functions that are perfectly analytic within the subdomain where $\varrho <\varrho _1$ as well as in the composite domain. In practice, when extracting physical quantities at sufficiently large resolutions, we saw no difference between the two approaches, apart from numerical fluctuations. Nevertheless, as expected, the resolution to reach a certain accuracy is substantially smaller for the $k=2$ approach. For this reason, the $k=2$ approach is preferable. 

However, the $k=2$ approach rises the question about the smoothness of the desired target metric. Once again, we emphasize that, basically, we numerically construct a spectral spline approximation of the desired smooth solution (if existing). Thus, the reference metric does not necessarily need to be smooth. Moreover, recall that as long as the reference spacetime exhibits the same causal structure and boundary behavior as the target spacetime, its explicit form only influences the gauge of the target metric. Consequently, we only need to ensure that the reference metric gives rise to a reasonable cover of the underlying manifold and, in addition, that it is of sufficient regularity in order to extract the physical quantities. 

\subsection{Overall scheme}
\label{subsec:Overall_scheme}

\begin{figure}[ht]
	\centering
	\includegraphics[scale=1.0]{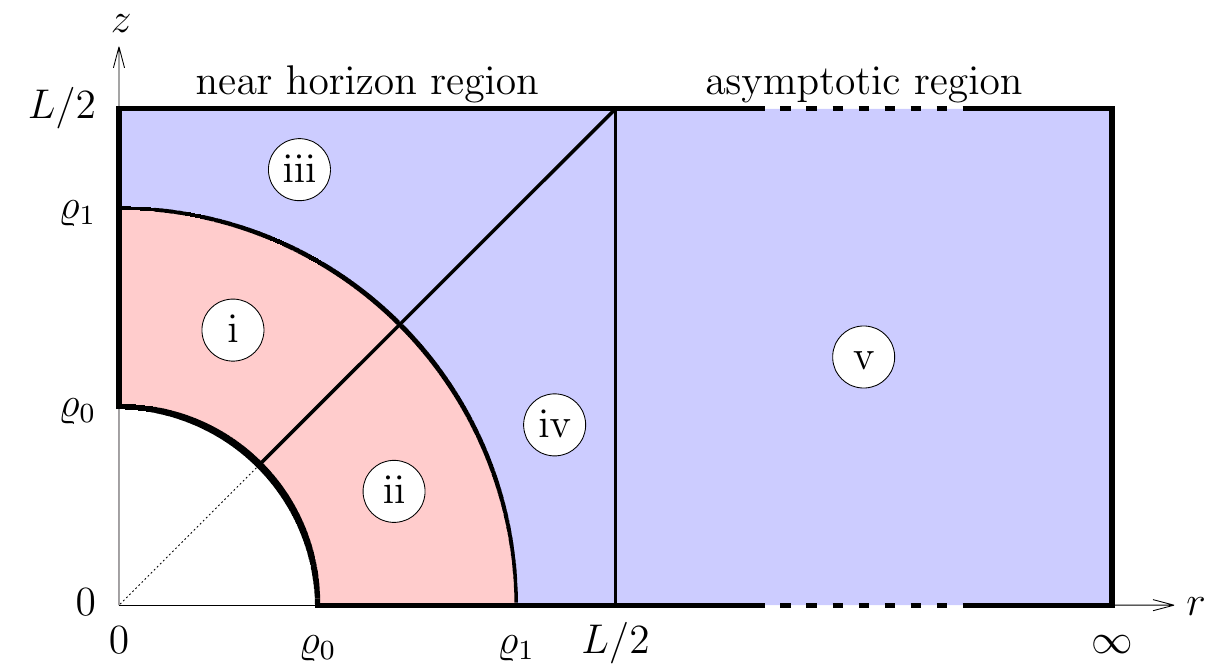}
	\caption{Basic decomposition of the LBH domain of integration. We denote the region where $r\geq L/2$ as the asymptotic region and where $r\leq L/2$ as the near horizon region. The contours $r=z$ (or $\varphi =\pi /4$) and $r^2+z^2=\varrho ^2=\varrho _1^2$ divide the near horizon region into four subdomains with four edges each. For $\varrho \geq \varrho _1$ (blue shaded region corresponding to subdomains (iii), (iv) and (v)) we use the background metric as reference for the DeTurck method.  In the $\varrho \leq \varrho _1$ region (red shaded region corresponding to subdomains (i) and (ii)) we build up a reference metric that approximates an ST solution at the horizon $\varrho =\varrho _0$ and matches the background metric at $\varrho =\varrho _1$.}
	\label{fig:LBHint_domain_split}
\end{figure}

Having a reference metric at hand, we build up two versions of the field equations out of the Einstein-DeTurck equations~\eqref{eq:LBHEinstein-DeTurck_equations}: one version that incorporates the asymptotic chart~\eqref{eq:LBHmetric_asymptotic_chart} and one version that incorporates the near horizon chart~\eqref{eq:LBHmetric_horizon_chart}. For this purpose, we divide the domain of integration into an asymptotic region, where $r\geq L/2$, and a near horizon region, where $r\leq L/2$. Moreover, due to the structure of the reference metric we decompose the near horizon region into a subdomain where $\varrho \leq \varrho _1$ and a subdomain where $\varrho \geq \varrho _1$. Finally, to obtain subdomains with only four boundaries we introduce an additional inner domain boundary at the coordinate line $\varphi = \pi /4$ (equivalent to $r=z$). The advantage of subdomains with only four boundaries is that there are rather simple coordinate transformations to cover the corresponding domain smoothly. Figure~\ref{fig:LBHint_domain_split} depicts the described basic arrangement of domains. 

Besides the overall length scale $L$ and the gauge fixing parameters $\varrho _0$ and $\varrho _1$ we have to prescribe the value of the surface gravity $\kappa$.
Then, by varying $\kappa$ we construct physically inequivalent LBH solutions. However, we need a good initial guess to find a first LBH solution from which we can go to different solutions. We find that the corresponding reference metric may serve as a sufficient initial guess if the relevant parameters are chosen appropriately. 

Another technical detail arises from the boundary conditions at the horizon $\mathcal H$, cf.\ equation~\eqref{eq:LBHBCh_horizon}, and the exposed axes $\mathcal A$, cf.\ equation~\eqref{eq:LBHBCh_axis}. In each case, we count six conditions for only five functions suggesting that one condition has to be dropped in the numerical implementation. According to reference~\cite{Dias:2015nua}, we are free to drop any of these. Then, the disregarded condition manifests itself \textit{a posteriori} as a property of the numerical solution, at least up to the numerical accuracy. In practice we omit the condition $\partial \lowh{\tilde T}/\partial \varrho=0$ at the horizon $\mathcal H$ and the condition $\partial \lowh S/\partial\varphi=0$ at the exposed axis $\mathcal A$.

\section{Decomposition of the domain of integration}
\label{sec:Decomposition_of_the_domain_of_integration_LBH}

There are still some numerical details to be discussed, in particular, adaptions of the method that guarantee high accuracy at reasonably small computational resources even in the critical regime of nearly touching poles of the LBHs. 
Below, we discuss appropriate adaptions in the asymptotic region, see subsection~\ref{subsec:Asymptotic_region}, and in the near horizon region, see subsection~\ref{subsec:Near_horizon_region}. 

\subsection{Asymptotic region}
\label{subsec:Asymptotic_region}

To cover the asymptotic region up to infinity we utilize the coordinate transformation
\beq
	r=\frac{L}{1-s} \, ,
	\label{eq:LBHcoordinate_s}
\eeq
where infinity is compactified to $s=1$ and the coordinate value $r=L/2$ corresponds to $s=-1$. Recall that in the construction of NBSs we came up with a sophisticated ansatz that explicitly took care of the specific behavior of the metric functions at infinity. As a benefit we were able to obtain the values of the asymptotic charges without performing derivatives on the numerical solution and hence with high accuracy. This time, however, we refrain from doing this effort and simply take the metric functions in the asymptotic chart~\eqref{eq:LBHmetric_asymptotic_chart} as they are. In practice, according to equations~\eqref{eq:asymptotic_corrections} we have to take $D-4$ derivatives of the metric functions $\lowa T$ and $\lowa B$. Thus, for the cases $D=5$ and $D=6$ considered here, the accuracy loss due to the numerical derivatives stays acceptable. 

\begin{figure}[ht]
	\centering
	\includegraphics[scale=1]{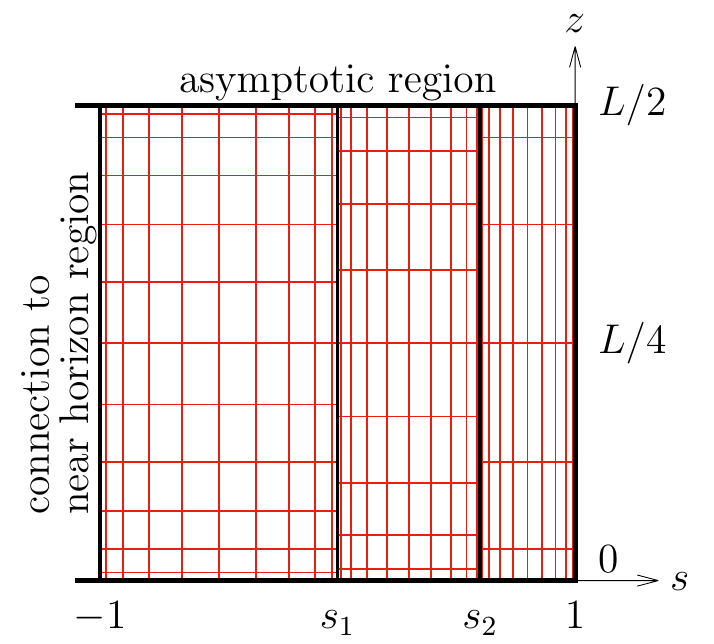}
	\caption{Numerical grid in the asymptotic region. The coordinate $s$ compactifies infinity to the coordinate value $s=1$. We decompose the asymptotic region into three linearly connected subdomains separated by the coordinate lines $s= s_1$ and $s=s_2$. This region is connected to the near horizon region at $s=-1$ corresponding to $r=L/2$.}
	\label{fig:LBHdomain_decomposition_asymptotic}
\end{figure}

Nonetheless, we adopt the grid structure that we already incorporated in the NBS context: We divide the asymptotic region into several linearly connected subdomains, see figure~\ref{fig:LBHdomain_decomposition_asymptotic}. Therefore, by choosing narrow windows near $s=1$, we take into account the non-analytic behavior of the metric functions. Again, this behavior is caused by logarithmic functions and the exponential term $\expterm$ that suppresses the $z$ dependence at infinity. As before, this setup allows us to adapt the resolution in each of the subdomains leading to a considerable reduction of the total number of grid points. 

\subsection{Near horizon region}
\label{subsec:Near_horizon_region}

In figure~\ref{fig:LBHint_domain_split} the basic domain structure in the near horizon region is already illustrated. Obviously, polar coordinates~\eqref{eq:LBHpolar_coordinates} are well suited to cover the domains (i) and (ii). However, to cover the domains (iii) and (iv), we modify the radial coordinate according to 
\beq
	r = \tilde\varrho (v,\varphi ) \sin \varphi \, , \quad  z = \tilde \varrho ( v,\varphi ) \cos \varphi \, . 
	\label{eq:LBHmodified_polar_coordinates}
\eeq
The radial function $\tilde \varrho$ takes the form
\beq
	\tilde \varrho (v,\varphi )= \varrho _1 \frac{L/2-v}{L/2-\varrho _1} + L/2 \frac{v-\varrho _1}{L/2-\varrho _1}
		\begin{cases}
			(\cos \varphi )^{-1} & \text{for domain (iii),}  \\
			(\sin \varphi )^{-1} & \text{for domain (iv),} 			
		\end{cases}
	\label{eq:LBHmodified_radial_coordinate}
\eeq
where the modified radial coordinate $v$ lies within $v\in[\varrho _1,L/2]$. If $v=\varrho _1$, we are at the contour $\varrho = \varrho _1$. In domain (iii) the coordinate value $v=L/2$ corresponds to $z=L/2$, while in domain (iv) the value $v=L/2$ corresponds to $r=L/2$. 

\begin{figure}[ht]
	\centering	
	\includegraphics[scale=1]{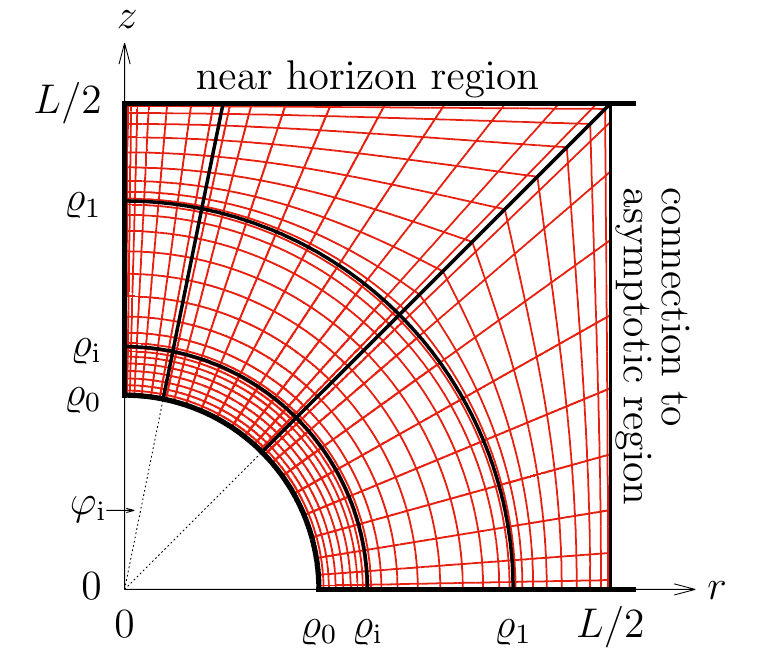}	
	\caption{Numerical grid in the near horizon region. We decompose the near horizon region into nine subdomains with inner boundaries at $\varrho =\varrho _1$, $\varrho =\varrho _\text{i}$, $\varphi = \pi /4$ and $\varphi =\varphi _\text{i}$. We adapt polar coordinates, but for $\varrho >\varrho _1$ we utilize an appropriately modified radial coordinates. Moreover, for $\varphi \leq\varphi _\text{i}$ we incorporate an analytic mesh refinement with respect to the angular coordinate $\varphi$. This region is connected to the asymptotic region at $r=L/2$.}
	\label{fig:LBHdomain_decomposition_near_horizon}
\end{figure}

With this setup we are already in a comfortable situation to construct LBH solutions. Nevertheless, to approach the critical regime of nearly touching poles we have to take special care of the functions' specific behavior, especially near the horizon $\mathcal H$ and the exposed axis $\mathcal A$. We increase the resolution in the vicinity of the horizon by dividing the subdomains (i) and (ii) further along a contour $\varrho = \varrho _\text{i}$ where $\varrho _0 < \varrho _\text{i} < \varrho _1$. It turns out that in the critical regime of nearly touching poles the highest gradients appear close to the exposed axis. Therefore, we utilize the same trick as before: We choose a $\varphi _\text{i}$ obeying $0<\varphi _\text{i}< \pi /4$ and split all domains along the contour $\varphi = \varphi _\text{i}$. The bottom line is that instead of having four subdomains in the near horizon region we end up with nine subdomains, as depicted in figure~\ref{fig:LBHdomain_decomposition_near_horizon}.

We highlight two further crucial adaptions that deal with the steep gradients near $\varphi =0$. First, we observe that the functions $\lowh B$ and $\lowh S$ run towards exceedingly high values on the axis when the critical transit solution is approached. We avoid these high values in the numerical computations by performing all computations with respect to the respective inverse function, namely
\beq
	\lowh{\tilde B} := \frac{1}{\lowh{ B}} \quad \text{and} \quad \lowh{\tilde S} := \frac{1}{\lowh{ S}} \, .
	\label{eq:LBHredefinition_Bh_and_Sh}
\eeq
Second, we employ an analytic mesh refinement with respect to the coordinate $\varphi$ in the interval $\varphi \in [0,\varphi _\text{i} ]$, i.e.\
\beq
	\varphi = \varphi _\text{i} \frac{\sinh \left( \lambda \, \bar\varphi / \varphi _\text{i} \right)}{\sinh \lambda} \, ,
	\label{eq:LBHanalytic_mesh_refinement}
\eeq
with the new coordinate $\bar \varphi \in [0,\varphi _\text{i}]$. The parameter $\lambda$ has to be chosen appropriately but is normally $\mathcal O(1)$. Note that this trick was already employed in the NBS context, cf.\ equation~\eqref{eq:NBSanalytic_mesh_refinement}, and is discussed in more detail in appendix~\ref{subsec:Further_common_techniques}.

\section{Parameters and physical quantities}
\label{sec:Parameters_and_physical_quantities_LBH}

We defined a couple of parameters that enter the numerical scheme, thus, in subsection~\ref{subsec:Parameter_values_LBH} we specify appropriate values of these parameters. Furthermore, in subsection~\ref{subsec:Extraction_of_physical_quantities_LBH} we discuss the relevant physical quantities. 

\subsection{Parameter values}
\label{subsec:Parameter_values_LBH}

The asymptotic length of the compact dimension $L$ and the surface gravity $\kappa$ are important physical parameters of the spacetime, while the radii $\varrho _0$ and $\varrho _1$ influence the reference metric and therefore the gauge. In addition, we control the numerical grid with the parameters $L$, $\varrho _0$, $\varrho _1$, $\varrho _\text{i}$, $\varphi _\text{i}$, $s _1$, $s _2$ and $\lambda$.\footnote{We emphasize that we set an explicit value for $L$ in the numerical implementation. Instead, we could also completely get rid of this length scale in all computations by scaling each quantity with appropriate powers of $L$.} Table~\ref{tab:LBHParameters} lists the values of these parameters that were used to approach the critical regime of nearly touching LBH poles. The value of $\lambda$ has to be adjusted accordingly when the critical transition is approached. We increased it up to $\lambda \approx 10$ for the solutions closest to the transition, while smaller values are adequate otherwise. 
\begin{table}
	\centering
	\caption{Parameter values that shape the numerical grid. In particular, these values were used to find LBH solutions in the critical regime of nearly touching poles. See figures~\ref{fig:LBHdomain_decomposition_asymptotic}, \ref{fig:LBHdomain_decomposition_near_horizon} and the text for the meaning of the parameters. We used these values both for $D=5$ and $D=6$.}
	\label{tab:LBHParameters}
	\begin{tabularx}{\textwidth}{*{7}{Y}}
 		\toprule
		 $L$	& $\varrho _0$ 	& $\varrho _1$	& $\varrho _\text{i}$	& $\varphi _\text{i}$ 	& $s _1$ 	& $s _2$		\\
		\midrule
		 8 		& 0.5			& 1.5	 		& 1		 				& 0.1  					& 0		   	& 0.8			\\
	 	\bottomrule
	\end{tabularx}	
\end{table} 

However, the values listed in table~\ref{tab:LBHParameters} are not necessarily appropriate for the construction of a first solution. For this purpose, we have to find a reasonably good approximation of an actual LBH solution in order to provide a good initial guess for the Newton-Raphson method. In fact, we simply use the reference metric, see equations~\eqref{eq:LBHrefmetric_functions} and~\eqref{eq:LBHrefmetric_functions_horizon_poly}, as an initial guess for a relatively small LBH solution, i.e.\ with $\kappa \approx 2$ (in units where $L=8$). Then, we are rather flexible to change the reference metric and therefore the initial guess by varying $\varrho _0$ and $\varrho _1$. Once a first solution is obtained, we slightly modify $\kappa$ to find another physically inequivalent solution, while the former serves as the new initial guess. This procedure works well until we reach a turning point in $\kappa$. We overcome such an extreme point with the trick presented in reference~\cite{Dias:2015nua} section VII.B.

\subsection{Extraction of physical quantities}
\label{subsec:Extraction_of_physical_quantities_LBH}

In this section, we discuss the physical quantities of interest accessible within the framework described above. On the one side, we consider the thermodynamic quantities that were already introduced in subsection~\ref{subsec:Physical_quantities}. On the other side, we define relevant geometric quantities.

\subsubsection{Thermodynamic quantities}

In contrast to our approach in the NBS context we can not directly read off the asymptotic charges, mass $M$ and relative tension $n$, from the metric functions. To obtain their values, we first have to get the asymptotic coefficients $c_t$ and $c_z$, cf.\ equations~\eqref{eq:Mass} and~\eqref{eq:RelativeTension}. Comparing the asymptotic corrections of the metric~\eqref{eq:asymptotic_corrections} and the ansatz in the asymptotic chart~\eqref{eq:LBHmetric_asymptotic_chart} we see that $c_t$ and $c_z$ are encoded in the metric functions $\lowa T$ and $\lowa B$. We consider these functions with respect to the compactified coordinate $s$. Thus we find at the asymptotic boundary $s=1$:
\begin{align}
	c_t &= (-1)^{D-3} \frac{L^{D-4}}{(D-4)!} \frac{\partial ^{D-4} \lowa T}{\partial s^{D-4}} \, , \label{eq:LBHct} \\
	c_z &= (-1)^{D-4} \frac{L^{D-4}}{(D-4)!} \frac{\partial ^{D-4} \lowa B}{\partial s^{D-4}} \, . \label{eq:LBHcz}
\end{align}
Then, we obtain
\begin{align}
	M/M_\GL & = \frac{K_\GL ^{D-4}}{D-3} \left[ (D-3) c_t - c_z  \right] \, , \label{eq:LBHMass} \\
	n/n_\GL & = (D-3) \frac{c_t - (D-3) c_z}{(D-3) c_t - c_z}  \label{eq:LBHRelativeTension} \, .
\end{align}	
The quantities $M_\GL$ and $n_\GL$ are the corresponding values of a marginally stable uniform black string (UBS) at the Gregory-Laflamme (GL) point, and $K_\GL$ is given in equation~\eqref{eq:NBSKGL5D} for $D=5$ and in equation~\eqref{eq:NBSKGL6D} for $D=6$. This normalization allows us to compare the values of $M$ and $n$ with the NBS results straightforwardly. 

The temperature does not have to be extracted from the numerical data, since it is directly related to the surface gravity $\kappa$, which we manually impose for each solution. We have
\beq
	T/T_\GL = \frac{2\, L}{(D-4)\, K_\GL} \kappa \, ,
	\label{eq:LBHTemperature}
\eeq
where $T_\GL$ is again the corresponding value of a UBS at the GL point. The entropy is proportional to the surface area of the horizon leading to the following integral 
\beq
	S/S_\GL =  2 \, \frac{\varrho _0^{D-2}}{L^{D-2}} K_\GL^{D-3} \int_{0}^{\pi /2} \sqrt{\lowh B \lowh S^{D-3}} (\sin \varphi )^{D-3} \, \D \varphi \,  
	\label{eq:LBHEntropy} 
\eeq
evaluated at $\varrho = \varrho _0$ and normalized by $S_\GL$. 

\subsubsection{Geometric quantities}

In analogy to the NBSs we consider the following quantities on the horizon $\varrho =\varrho _0$: the maximal horizon areal radius
\beq
	R_{\text{max}} = \varrho _0 \sqrt{\lowh S} \quad \text{at} \quad  \varphi =  \pi /2  \, , 
	\label{eq:LBHReq}
\eeq
which is measured at the equator, and the proper length of the horizon from north to south pole 
\beq
	L_\mathcal{H} = 2\varrho _0 \int _0^{\pi /2} \sqrt{\lowh B} \, \D \varphi \, . 
	\label{eq:LBHLpolar}
\eeq
Of particular interest is the proper length of the exposed axis of symmetry $\mathcal A$, i.e.\ the proper distance between north and south pole when moving along the $\varphi = 0$ (or $r=0$) axis, 
\beq
	L_{\mathcal A} = 2 \int _{\varrho _0}^{L/2} \sqrt{\lowh A} \, \D \varrho \, .
	\label{eq:LBHLaxis}
\eeq
With this quantity we are able to characterize the limit of infinitesimal LBHs by taking $L_\mathcal{A}\to L$ and the limit of touching poles by taking $L_\mathcal{A} \to 0$. 

Again, for the purpose of illustration we embed the LBH horizons into the $(D-1)$-dimensional flat space. Comparing the flat metric~\eqref{eq:metricFlatSpace} to equation~\eqref{eq:LBHmetric_horizon_chart} we deduce 
\refstepcounter{equation}\label{eq:LBHEmbedding}
\begin{align}
	R(\varphi ) &= \varrho _0 \sin \varphi \sqrt{\lowh S}  \, , \tag{\theequation a} \label{eq:LBHEmbedding_R} \\
	Z(\varphi ) &= \int _\varphi ^{\pi /2} \sqrt{\varrho _0^2 \lowh B - (\partial R / \partial \tilde\varphi )^2} \, \D \tilde \varphi  \, , \tag{\theequation b} \label{eq:LBHEmbedding_Z} 
\end{align}
where both $R$ and $Z$ are evaluated at the horizon $\varrho =\varrho _0$. Note that we fixed an arbitrary constant of integration such that $Y=0$ at the equator, i.e.\ at $\varphi = \pi / 2$ (or $z=0$).

\section{Accuracy of the numerical solutions}
\label{sec:Accuracy_of_the_numerical_solutions_LBH}

\begin{figure}[ht]
	\centering
	\includegraphics[scale=1]{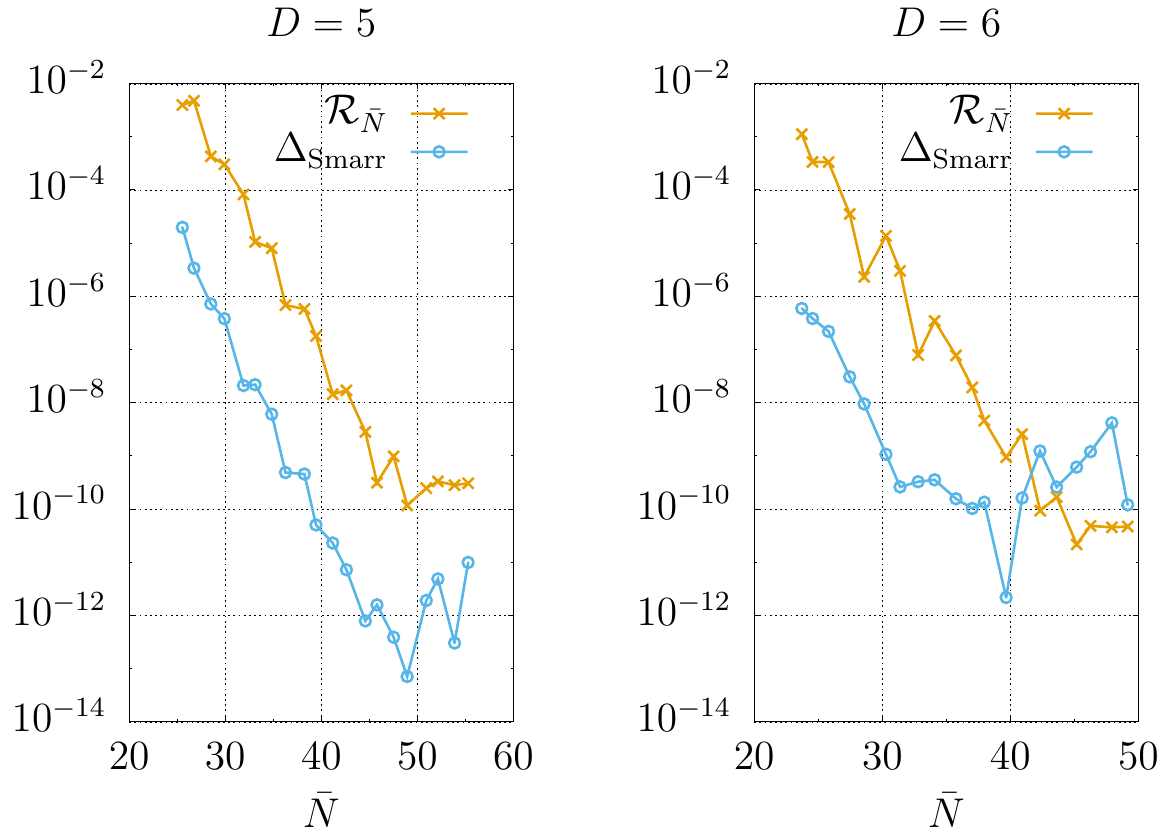}
	\caption{Convergence of the residue $\mathcal R_{\bar N}$ and the deviation from Smarr's relation $\Delta _\text{Smarr}$ as a function of the mean resolution $\bar N$. The respective numerical solutions correspond to LBHs with proper separation of the poles of $L_\mathcal{A} /L \approx 0.0029$ for $D=5$ and $L_\mathcal{A} /L \approx 0.00083$ for $D=6$. }
	\label{fig:LBHconvergence}
\end{figure}

Finally, we discuss the accuracy of the LBH solutions in a similar manner as for NBSs, cf.\ section~\ref{sec:Accuracy_of_the_numerical_solutions_NBS}. We analyze how the residue $\mathcal R_{\bar N}$ and the deviation $\Delta _\text{Smarr}$ from Smarr's relation~\eqref{eq:SmarrRelation} converges as the resolution $\bar N$ is increasing, where $\bar N$ denotes the mean resolution averaged over all subdomains and directions. In particular, figure~\ref{fig:LBHconvergence} displays the convergence plots for a numerical LBH solution close to the critical transition in $D=5$ and $D=6$, respectively. We observe that the residue rapidly falls down and saturates at values of the order of $10^{-10}$ due to numerical limitations that are caused by finite machine precision and rounding errors.\footnote{Again, we note that the numerical calculations are performed in long double precision (80-bit extended precision).} It is apparent that the saturation value of $\Delta _\text{Smarr}$ in $D=6$ is considerably higher than in $D=5$ with a difference of about two orders of magnitude. However, this is not a big surprise since in $D=6$ we have to perform two numerical derivatives to get access to the asymptotic coefficient $c_t$ that enters Smarr's relation, cf.\ equation~\eqref{eq:LBHct}. On the contrary, in $D=5$, only the first derivative is needed.

As stated above, a necessary consistency check for a numerical solution obtained by the DeTurck method involves the DeTurck vector field~\eqref{eq:LBHDeTurck_vector} itself. If the DeTurck vector field vanishes then a solution to the Einstein-DeTurck equations~\eqref{eq:LBHEinstein-DeTurck_equations} is a solution to Einstein's equations as well. Indeed, for our numerical solutions the non-trivial components of the DeTurck vector are always smaller than $10^{-10}$ in magnitude, which is negligible compared to typical values of the metric functions.

\chapter{Completion of the phase diagram of static Kaluza-Klein black holes}
\label{chap:Complete_phase_diagram_of_static_Kaluza-Klein_black_holes}

In the previous chapters we have thoroughly discussed our pseudo-spectral numerical scheme to find solutions of Einstein's vacuum field equations that describe static Kaluza-Klein (KK) black holes. The solutions of interest are localized black holes (LBHs) and non-uniform black strings (NBSs), whereas the latter emanate from the Gregory-Laflamme (GL) instability of the analytically known uniform black strings (UBSs). In particular we emphasized the crucial adaptions of the method to find numerical solutions even in the critical regime, where the LBH and NBS branches are about to meet. Indeed, these sophisticated approaches allow us to construct solutions that are unprecedentedly close to the critical transit solution between the branches. Therefore, with the corresponding numerical data at hand, we are now able to explore this critical regime, which was not accessible by previous implementations. However, prior work already gave strong evidence in favor of the picture that both branches meet, cf.\ for example figure~\ref{fig:PhaseDiagram5Dold}. Nevertheless, it was still unclear how this transit solution is approached and what role the double cone metric plays, see subsections~\ref{subsec:Phase_diagram_of_static_Kaluza-Klein_black_holes} and~\ref{subsec:Double-cone_metric}. Here, we will give answers to both of those questions. 

First, we show the qualitative behavior of thermodynamic quantities and the corresponding phase diagram in section~\ref{sec:Thermodynamics}. A similar discussion of relevant geometric quantities is provided in section~\ref{sec:Geometry}. Moreover, we investigate in detail the deformations of the horizon when moving along the two branches. Finally, in section~\ref{sec:Critical_behavior} we quantitatively analyze the critical regime close to the transition. In particular, we examine the validity of the conjectured critical scaling~\eqref{eq:dp_perturbed_cone_real} of physical quantities, which was derived from the double-cone metric. 

We present the findings obtained in $D=5$ and $D=6$ and we note that the results are qualitatively similar. Therefore, we discuss the cases separately only if there is a notable difference.

\section{Thermodynamics}
\label{sec:Thermodynamics}

\begin{figure}[ht]
	\centering
	\includegraphics[scale=1]{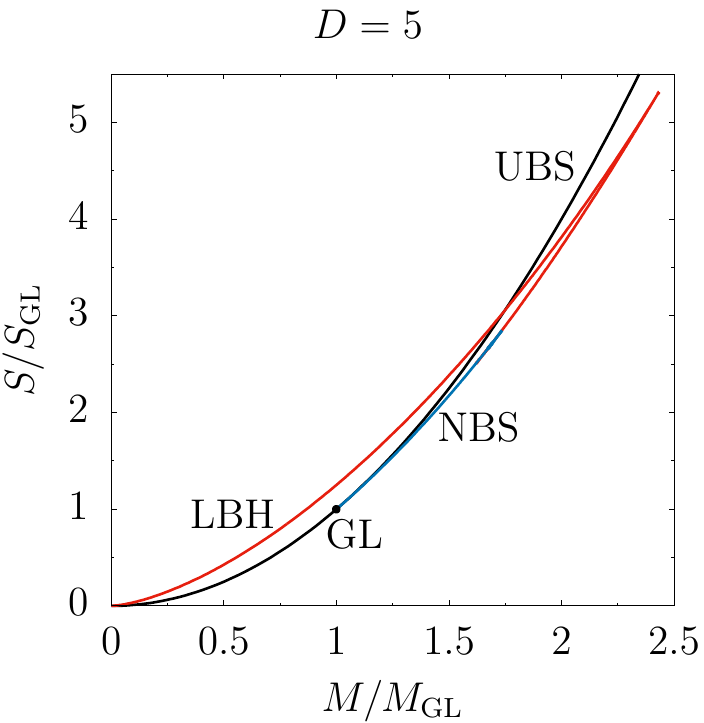}
	\includegraphics[scale=1]{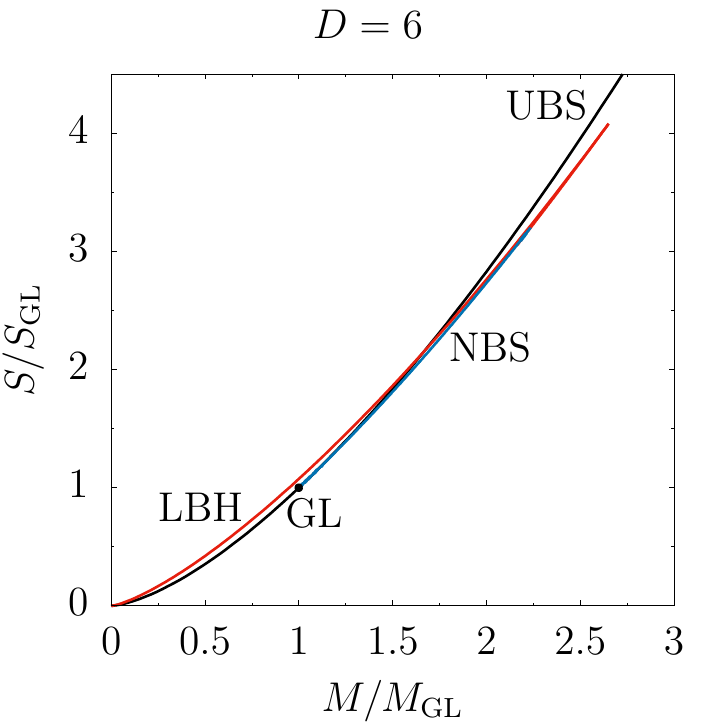}
	\caption{Phase diagram in the microcanonical ensemble. We normalize the values of entropy $S$ and mass $M$ with respect to their corresponding values of a UBS (black line) at the GL point. We find a critical region where the LBH (red line) and NBS (blue line) branches approach each other. The left diagram corresponds to $D=5$ and the right one to $D=6$.}
	\label{fig:PhaseDiagram}
\end{figure}

\begin{figure}[ht]
	\centering
	\includegraphics[scale=1]{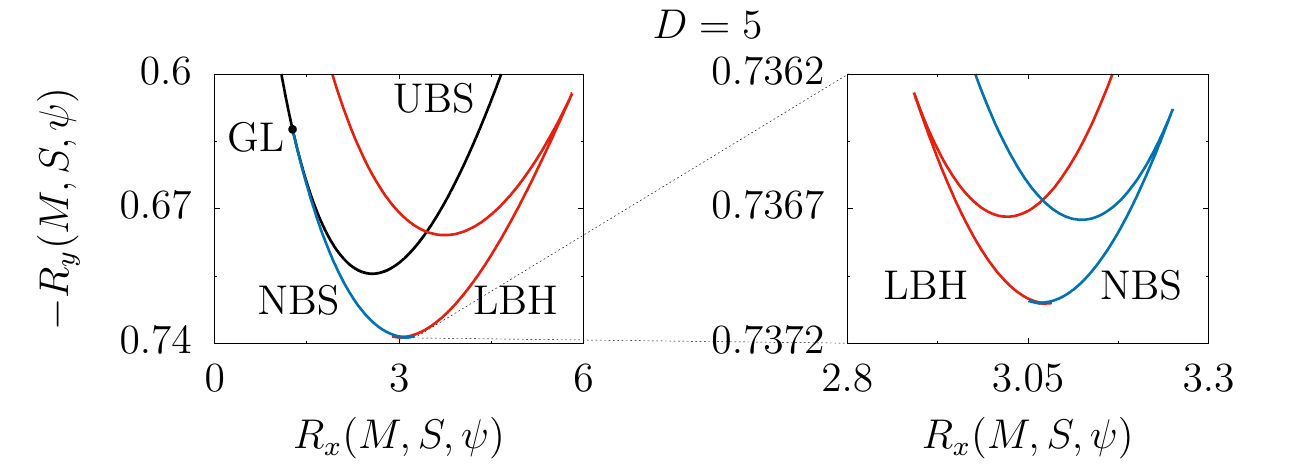} \ 
	\includegraphics[scale=1]{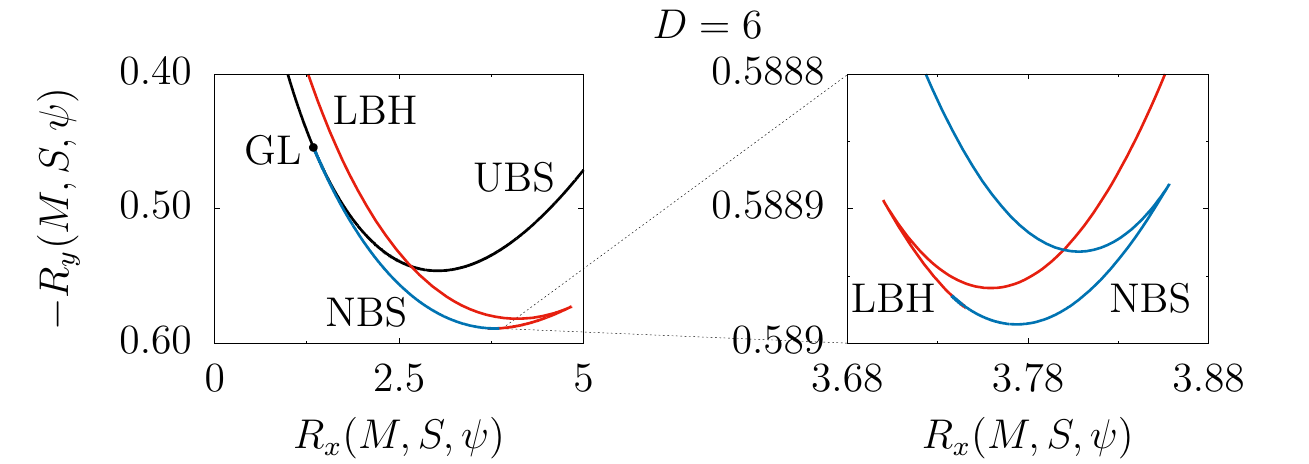} 
	\caption{Rotated phase diagram in the microcanonical ensemble. The rotation is performed with respect to the origin, i.e.\ we have for the abscissa $R_x(M,S,\psi ) = M/M_\GL \cos \psi - S/S_\GL \sin \psi$ and also for the ordinate $R_y(M,S,\psi ) = M/M_\GL \sin \psi + S/S_\GL \cos \psi$. We rotate by an angle $\psi \approx -71.4^\circ$ for $D=5$ and $\psi = -63.7^\circ$ for $D=6$. The upper diagrams correspond to $D=5$ and the lower ones to $D=6$. The left diagrams show quite a big portion of the phase diagram, whereas in the right diagrams the critical region is magnified. }
	\label{fig:PhaseDiagramZoom}
\end{figure}

\begin{figure}[ht]
	\centering
	\includegraphics[scale=1]{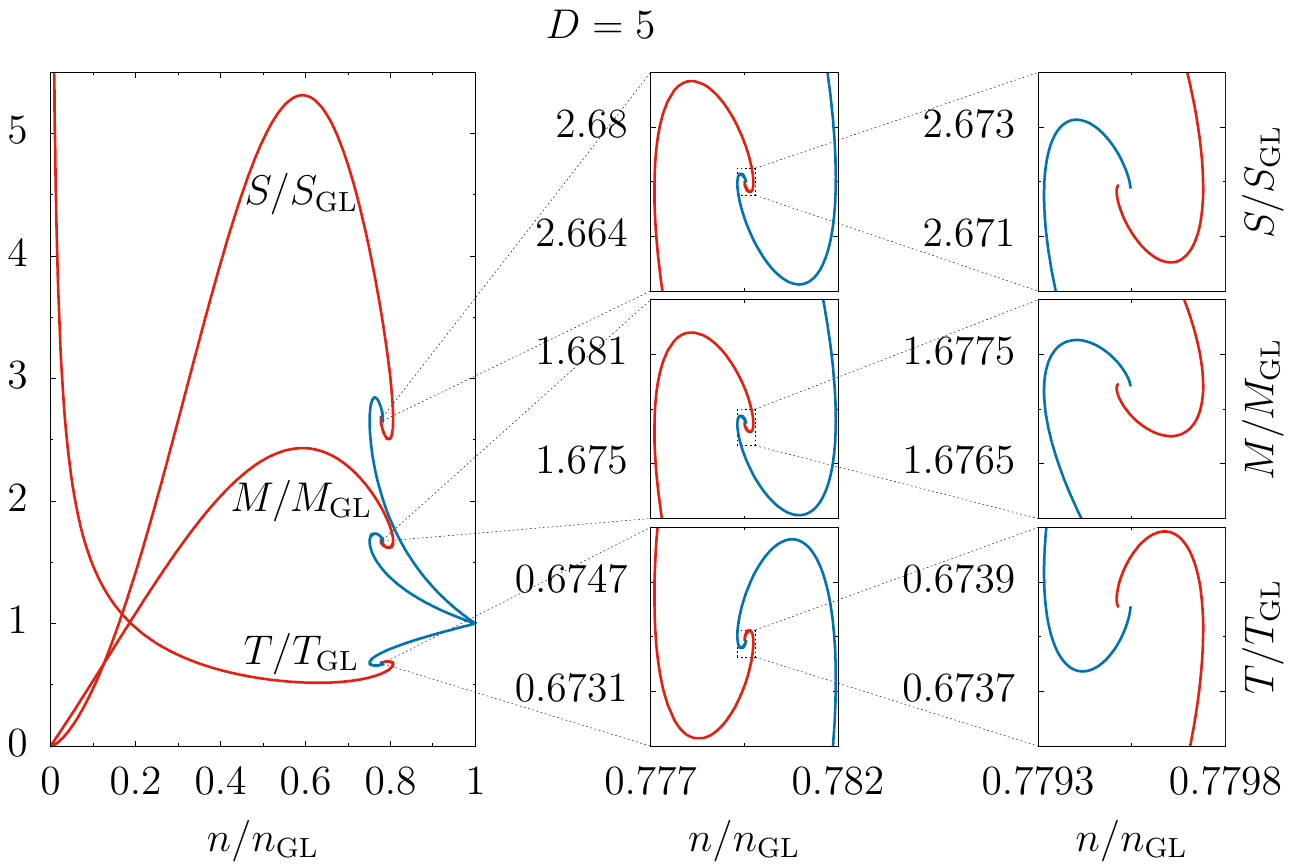} \  \\ \ \\
	\includegraphics[scale=1]{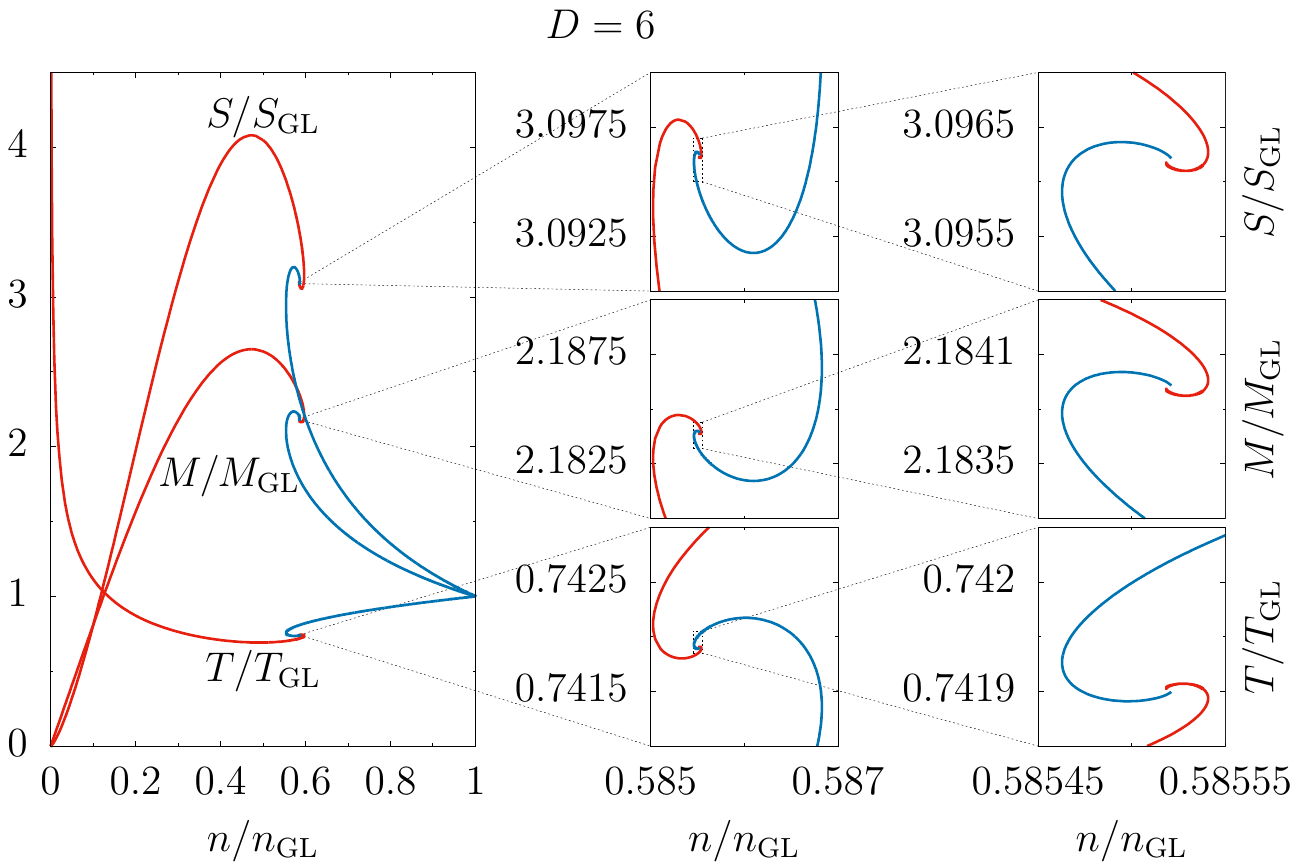} 
	\caption{Entropy $S$, mass $M$ and temperature $T$ as functions of the relative tension $n$. We normalize all quantities with respect to their corresponding values of a UBS at the GL point. The region where the LBH (red line) and NBS (blue line) branches approach each other is magnified twice in the center and right column. The upper diagrams corresponds to $D=5$ and the lower ones to $D=6$.  }
	\label{fig:Thermodynamics}
\end{figure}

We already gave the phase diagram of KK black holes in the microcanonical ensemble in five dimensions according to previous results, see figure~\ref{fig:PhaseDiagram5Dold}. As depicted in figure~\ref{fig:PhaseDiagram}, our results close the gap between the LBH and the NBS branch. At first glance, it seems as though there is nothing special going on and both branches would meet straightforwardly. However, if we look closer into the critical region where the two branches approach each other we see that, actually, they turn back and forth. To see this in more detail we rotate the phase diagram around its origin in figure~\ref{fig:PhaseDiagramZoom} and magnify the critical region. We see that entropy and mass start to oscillate when approaching the common end point of both phases, which leads to turning points in the phase diagram. Apart from the strongly pronounced turning point in the LBH branch we are able to resolve three further turning points in both branches. We emphasize that this confirms the results of reference~\cite{Kleihaus:2006ee} already showing the first of these turning points in the NBS branch. Again, due to the first law $\delta M = T\, \delta S$ the extremal points in the mass-entropy-diagram are spiky. 
\newpage
A clearer way to analyze the behavior of physical quantities is to consider them as functions of the relative tension $n$. Moreover, this has the additional advantage that small LBHs have $n\gtrsim 0$, while all UBSs have $n=1/(D-3)$, cf.\ equation~\eqref{eq:tensionUBS}. Consequently, slightly deformed NBSs have $n\lesssim 1/(D-3)$. We plot the entropy $S$, the mass $M$ and the temperature $T$ as functions of the relative tension $n$ in figure~\ref{fig:Thermodynamics} with appropriate magnifications of the critical region. Obviously, in the critical region, both branches begin to describe a spiral curve due to the oscillating behavior already observed above. Moreover, the spirals of both branches adapt perfectly to each other and we clearly see that our data confirms nearly two complete turns of each spiral curve. This is in remarkable accordance with the expectation of a common end point. In addition, it is apparent that the spirals shrink very rapidly with each turn. We will investigate this observation in more detail in section~\ref{sec:Critical_behavior}. 

\section{Geometry}
\label{sec:Geometry}

\begin{figure}[ht]
	\centering
	\includegraphics[scale=1]{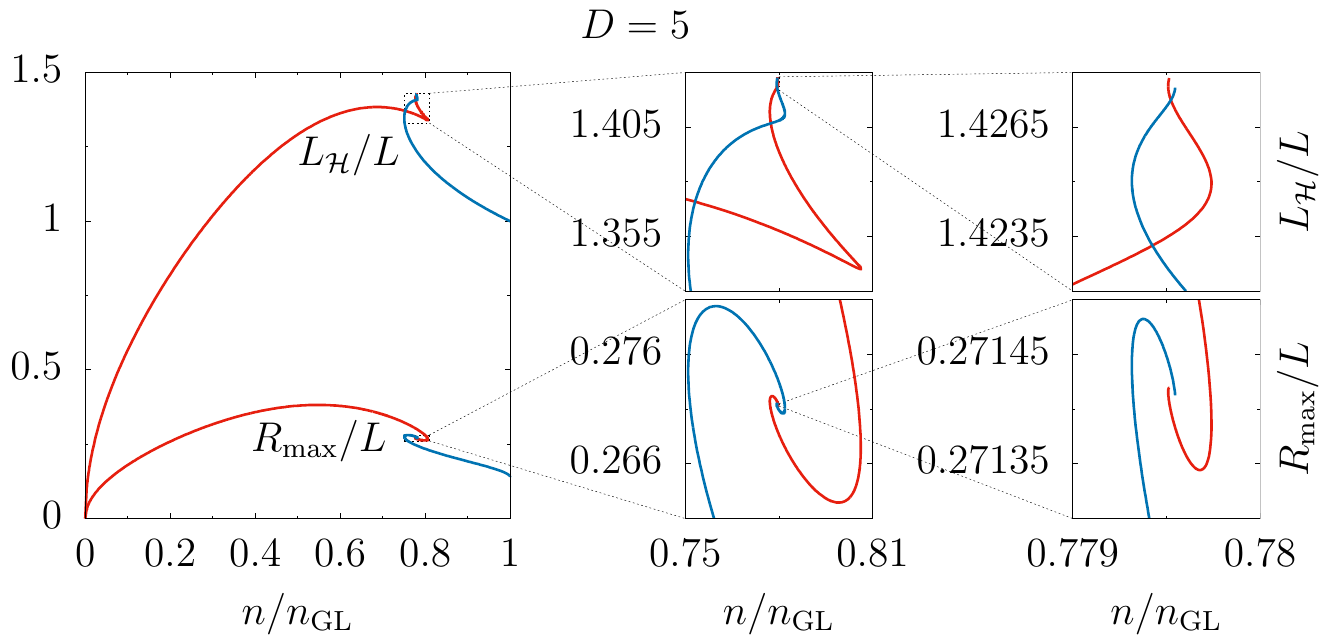} \  \\ 
	\includegraphics[scale=1]{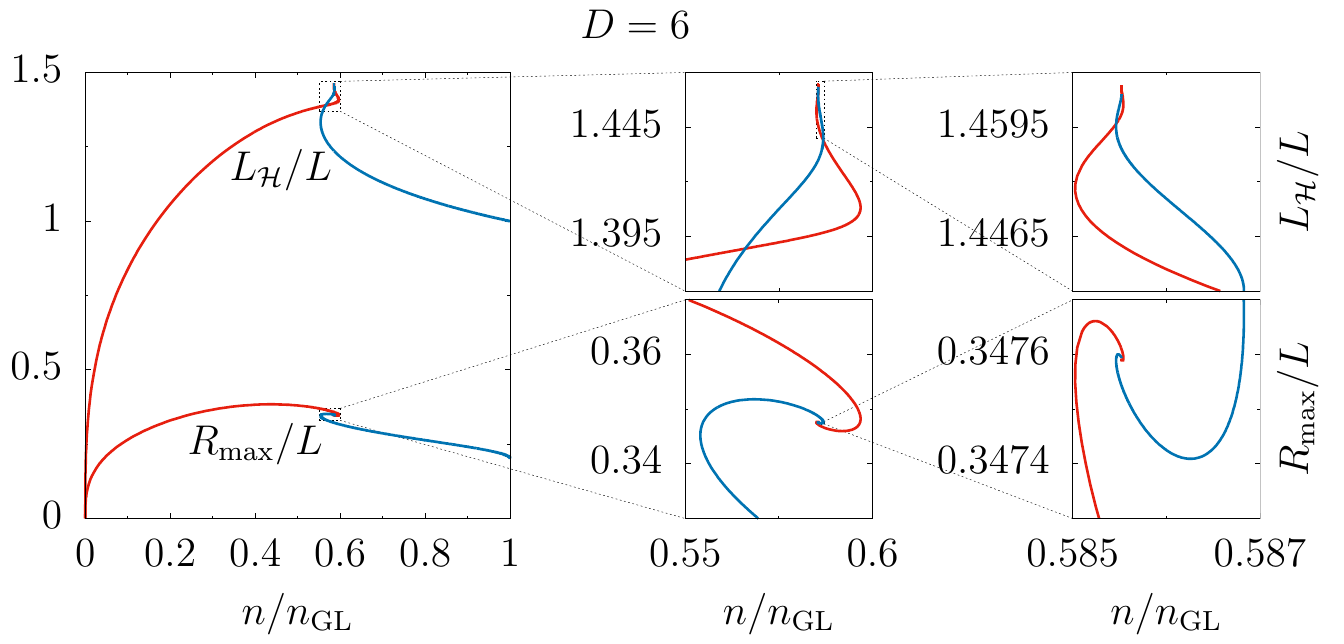} 
	\caption{Proper horizon length $L_\mathcal H$ and maximal horizon areal radius $\Rmax$ as functions of the relative tension $n$. The region where the LBH (red line) and NBS (blue line) branches approach each other is magnified twice in the center and right column. The upper diagrams correspond to $D=5$ and the lower ones to $D=6$. }
	\label{fig:Geometry}
\end{figure}

In this section we show how the geometry of LBHs and NBSs changes when moving along their branches and approaching the critical transition. Figure~\ref{fig:Geometry} illustrates the behavior of the proper horizon length $L_\mathcal{H}$ and the maximal horizon areal radius $\Rmax$ as functions of the relative tension $n$, see sections~\ref{subsec:Extraction_of_physical_quantities_NBS} and~\ref{subsec:Extraction_of_physical_quantities_LBH} for definitions. Again, we see spiral curves appearing for $\Rmax$ with the spirals of both branches converging towards each other. In contrast, the $L_\mathcal H$ curve does not exhibit a spiraling behavior but rather runs towards a global maximum when the transition is approached. 

\begin{figure}[ht]
	\centering
	\includegraphics[scale=1]{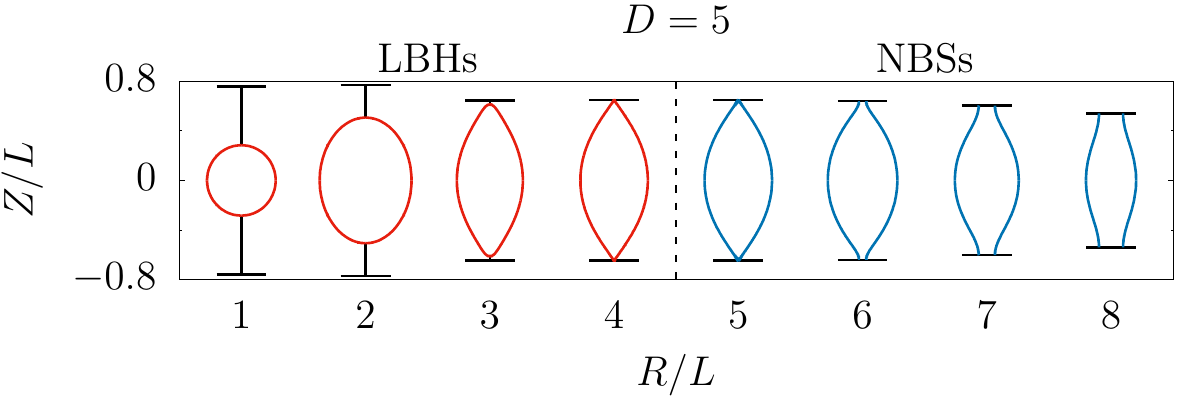} 
	\includegraphics[scale=1]{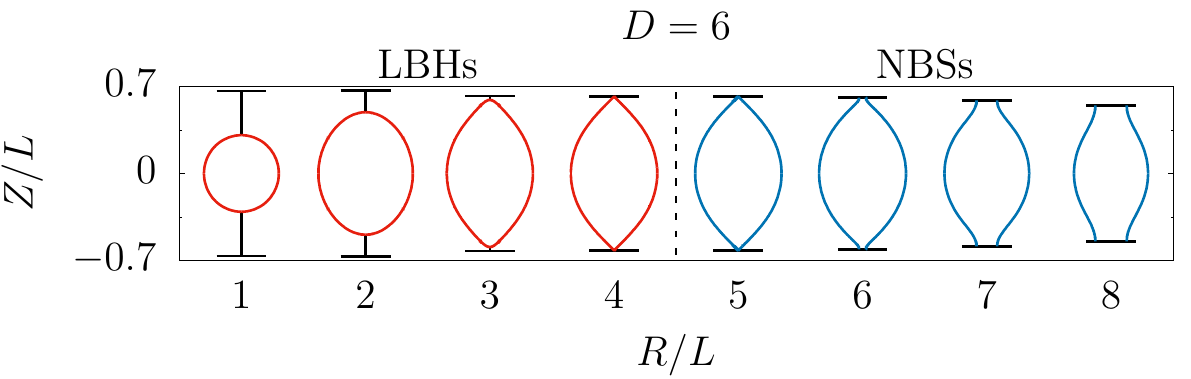} 
	\caption{Spatial embeddings of the horizons of different LBH and NBS solutions. The embeddings are accordingly shifted along the $R$-axis. The upper diagram corresponds to $D=5$ and the lower one to $D=6$.}
	\label{fig:Horizons}
\end{figure}

The spatial embedding of different LBH and NBS horizons is depicted in figure~\ref{fig:Horizons}, cf.\ equations~\eqref{eq:NBSHorizonEmbedding} and~\eqref{eq:LBHEmbedding}. When following the LBH branch we see that the horizons spread more and more along the compact periodic dimension until it is almost completely wrapped. From the NBS point of view the horizon becomes more and more deformed and develops a bulge as well as a waist region. In particular, the waist is more and more shrinking until it is about to pinch off. In fact, in figure~\ref{fig:Horizons} we can hardly distinguish a difference between the embeddings 4 and 5. Both correspond to solutions very close to the transition. However, since both horizons have different topology, there is a difference at least in the region near the periodic boundary, where the poles of the LBH and the waist of the NBS are located. 

\begin{figure}[ht]
	\centering
	\includegraphics[scale=1]{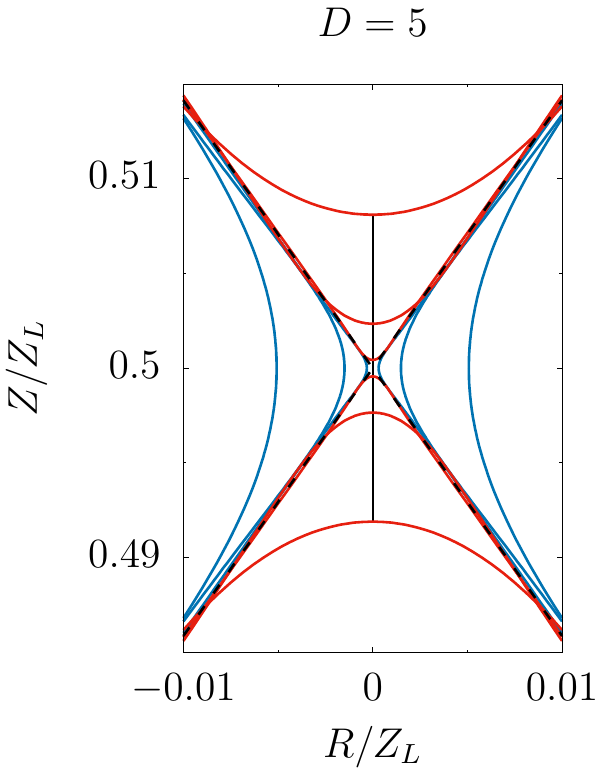} 
	\includegraphics[scale=1]{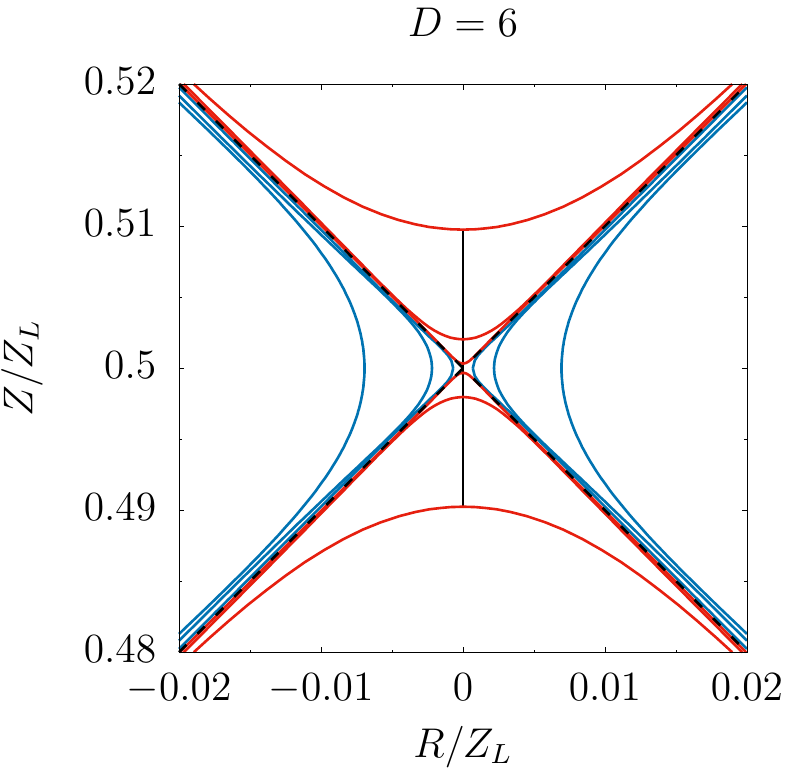} 
	\caption{Magnification of the region where the poles of LBHs are about to merge or NBSs are about to pinch-off. The dashed lines correspond to the double-cone geometry. Here, LBH horizons (red lines) approach the double-cone shape from above/below while NBS horizons (blue lines) approach it from left/right. The embedding coordinates $R$ and $Z$ are normalized by $Z_L$, the length of the compact dimension measured in $Z$. We mirror the plots with respect to the periodic boundary $Z/Z_L = 1/2$. The left plot corresponds to $D=5$ and the right one to $D=6$.}
	\label{fig:Cones}
\end{figure}

We magnify the region around the periodic boundary in figure~\ref{fig:Cones} and illustrate several LBH and NBS horizons that are close to the transition. It is apparent, that the horizons of both types of solutions locally converge towards straight lines, when approaching the transition. Moreover, we see that these straight lines correspond exactly to the double-cone geometry discussed in subsection~\ref{subsec:Double-cone_metric}. In particular, we note that the $D$-dependent opening angle of the double-cone geometry~\eqref{eq:doubleconeEmbedding} is nicely approached from both types of solutions. Consequently, this provides strong qualitative evidence in favor of Kol's conjecture, which states that the double-cone metric~\eqref{eq:metricDoubleCone} is a local model of the transit solution between LBHs and NBSs~\cite{Kol:2002xz}. 

\section{Critical behavior}
\label{sec:Critical_behavior}

Finally, we are able to test another fundamental prediction that follows from Kol's conjecture, namely the critical scaling~\eqref{eq:dp_perturbed_cone_real} of physical quantities when the transition is approached~\cite{Kol:2005vy}. More concretely, it was argued that different physical quantities are expressed by the same \textit{critical exponents}. Generically, these exponents are universal and do only depend on the number of spacetime dimensions $D$. This \textit{critical behavior} is originally known to appear in quantum and statistical field theories close to certain phase transitions. However, some gravitational systems exhibit critical behavior as well. Most famously, already in 1992 Choptuik showed that the spherical collapse of a scalar field is controlled by a critical exponent at the threshold of black hole formation~\cite{Choptuik:1992jv}. Surprisingly, there is a formal relation between Choptuik's system in $D-1$ dimensions and the LBH/NBS system in $D$ dimensions~\cite{Kol:2005vy}.\footnote{In reference~\cite{Sorkin:2005vz} the hyper-spherical collapse of a scalar field in higher dimensions up to $D=11$ is investigated showing qualitatively similar results as in the four-dimensional case.} Nevertheless, since the regarding systems are subject to different boundary conditions, one can not necessarily infer from one system to the other. 

In order to test the LBH/NBS system with respect to a critical behavior, we first need to identify an appropriate length scale that parametrizes the LBH and NBS branch and approaches zero for the transition. Fortunately, we already defined such a quantity, namely the proper distance between the poles $L_\mathcal{A}$~\eqref{eq:LBHLaxis} for LBHs and the minimal horizon areal radius $\Rmin$~\eqref{eq:NBSHorizonArealRadiusmin} for NBSs. For convenience, we write
\refstepcounter{equation}\label{eq:Q}
\begin{align}
	Q_\text{LBH} &= L_\mathcal{A}/L 	 \, , \tag{\theequation a} \label{eq:QLBH} \\
	Q_\text{NBS} &= \Rmin /R_\GL  \, , \tag{\theequation b} \label{eq:QNBS} 
\end{align}
where $R_\GL$ is the corresponding horizon areal radius of a UBS at the GL point. As a result of the chosen normalization $Q$ approaches one at the starting point of the corresponding branch, i.e. $Q_\LBH \lesssim 1$ for small LBHs and $Q_\NBS \lesssim 1$ for slightly deformed NBSs. As requested, the critical transition is reached in the limit $Q\to 0$.

According to equation~\eqref{eq:dp_perturbed_cone_real} physical quantities scale as 
\beq
	f(Q) = f_\text{c} + a\, Q^b \cos \left(c \log Q + d \right) \, ,
	\label{eq:fit_ansatz_mass}
\eeq
for small $Q$. Here, $f$ stands for any quantity such as mass, relative tension, temperature or entropy. Consequently, the parameter $f_\text{c}$ denotes the critical value of this quantity at the transition $Q = 0$. We refer to the parameters $b$ and $c$ as the real critical exponent and log-periodicity, respectively. Their values are predicted from the analysis of perturbations of the double-cone metric, cf.\ equation~\eqref{eq:perturb_cone} and~\eqref{eq:complex_exponents}. In our case the prediction is $b=3/2$ and $c=\sqrt{15} /2 \approx 1.9365$ for $D=5$ and $b=c=2$ for $D=6$.  On the contrary, the parameters $a$ and $d$ are different for each physical quantity and do not have an explicit physical meaning. 

\begin{figure}
	\centering
	\includegraphics[scale=1]{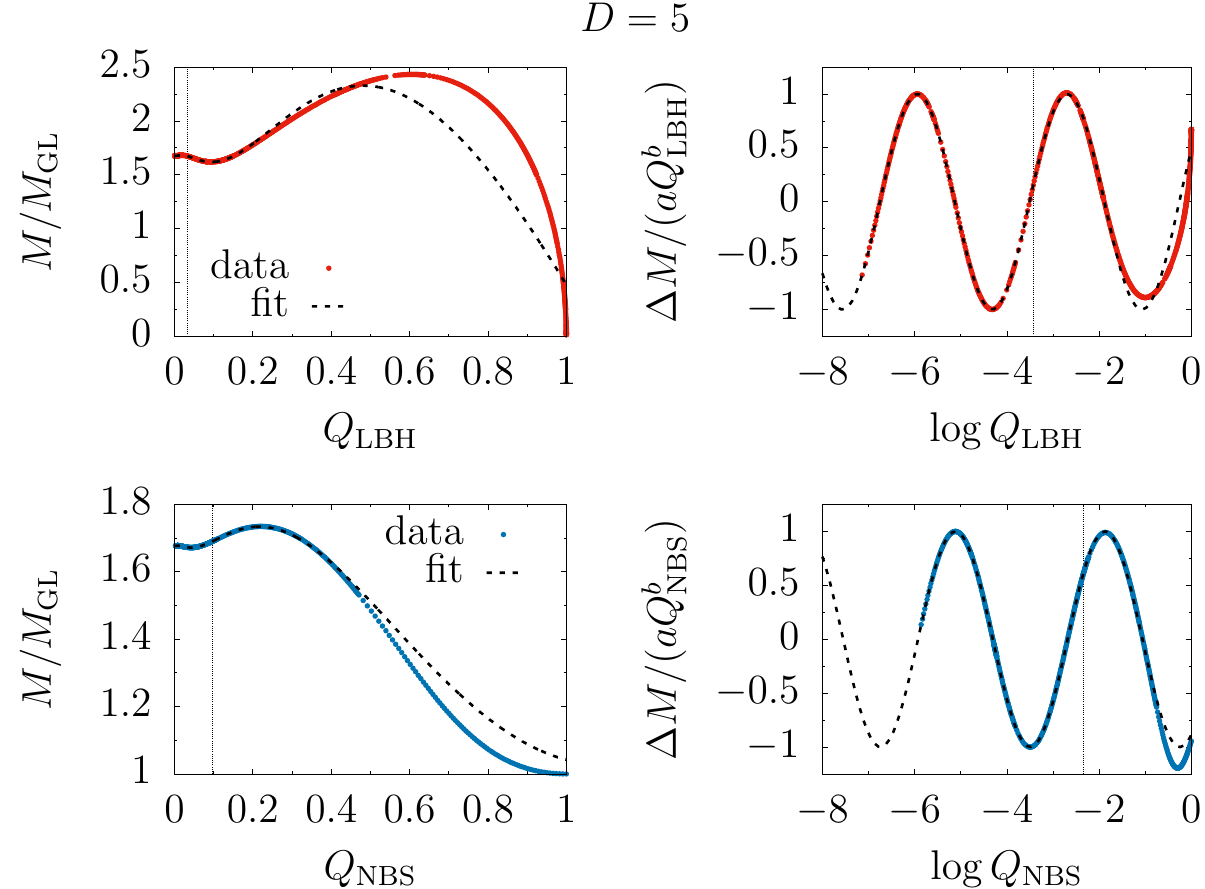} \ \\
	\includegraphics[scale=1]{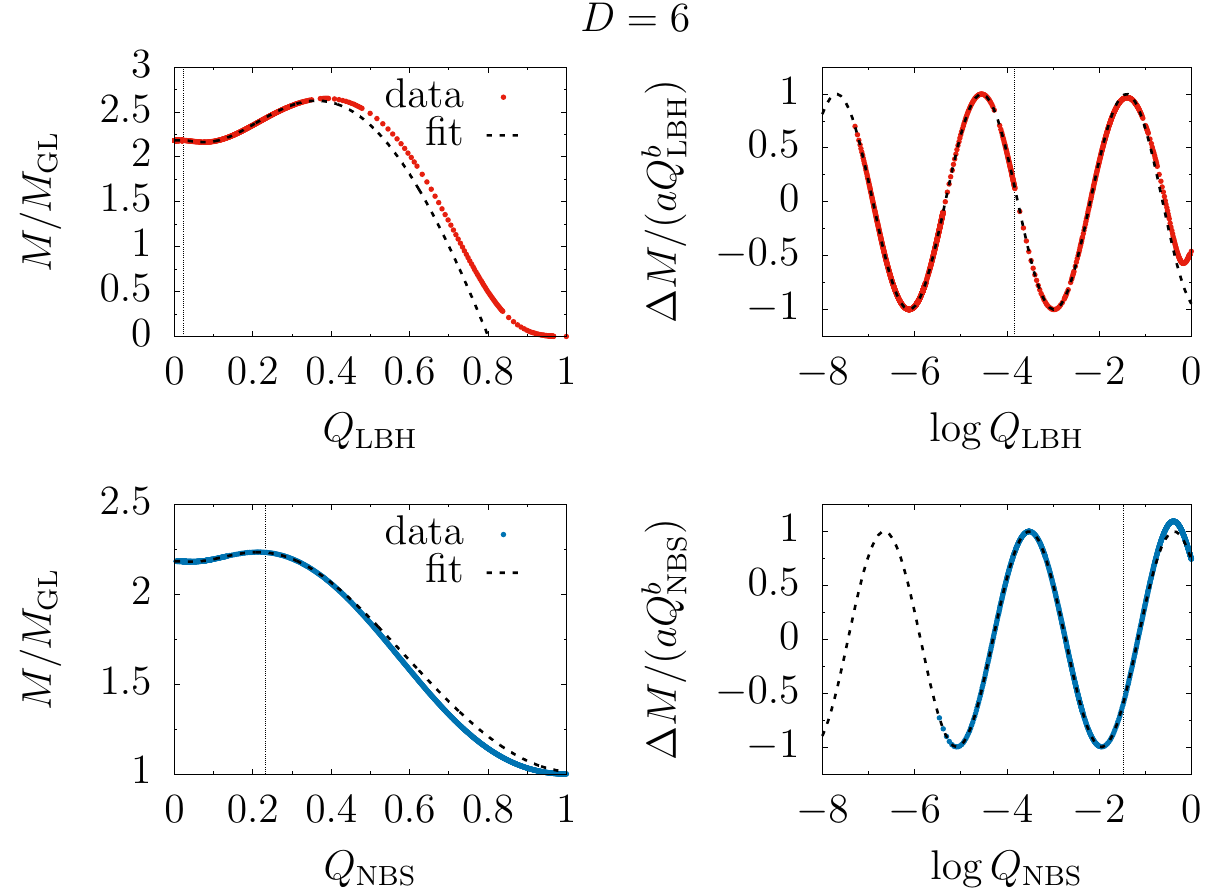}
	\caption{Data points (red dots for LBHs and blue dots for NBSs) and fit (dashed line) of the mass $M$ as a function of $Q_\text{LBH}$ or $Q_\text{NBS}$, respectively. The left column shows the explicit functional dependence. We resolve the tiny oscillations of the functions in the right column, where $\Delta M = M/M_\GL -f_\text{c}$ is plotted with a logarithmic rescaling of the abscissa. The first two rows correspond to $D=5$, while the last two rows correspond to $D=6$. In each plot the thin dotted vertical line indicates that all data points left to this line were used to produce the fit. }
	\label{fig:Scaling_Mass}
\end{figure}

We analyze our data in regard of the critical behavior by fitting the data points with the ansatz~\eqref{eq:fit_ansatz_mass} and treating $f_\text{c}$, $a$, $b$, $c$ and $d$ as free parameters. We utilized Mathematica's fit routine to carry out the fit. In figure~\ref{fig:Scaling_Mass} we show data points and fit on the example of the mass $M/M_\GL$. The exponentially suppressed oscillating behavior of the functions is shown more explicitly in the right column of figure~\ref{fig:Scaling_Mass} by an appropriate rescaling. Close to the transition, i.e.\ for small $Q$, we observe a remarkable agreement of data and fit. In fact, considering the other thermodynamic quantities we find a similar picture. 

We summarize the obtained fit parameters for the mass $M$, the relative tension $n$, the temperature $T$ and the entropy $S$ in the tables~\ref{tab:param_thermo5D} and~\ref{tab:param_thermo6D}. Remarkably, the predicted values of the critical exponent and the log-periodicity, $b$ and $c$, are excellently reproduced and deviate from the predictions by only less than $0.5\%$.\footnote{One may wonder why this value is much greater than the accuracy of the numerical solutions, cf.\ sections~\ref{sec:Accuracy_of_the_numerical_solutions_NBS} and~\ref{sec:Accuracy_of_the_numerical_solutions_LBH}. In fact, the fit ansatz~\eqref{eq:fit_ansatz_mass} is not an exact model of the functions. It merely describes the leading behavior for small $Q$. } Moreover, for a given dimension they are indeed the same for different quantities and for both types of solutions. Furthermore, we observe that the respective values $f_\text{c}$ coincide up to seven digits after the decimal point for both branches. This is by far the best approximation of the values of these quantities at the transition. 
\begin{table}[ht]
	\centering
	\caption{Parameter values of the fit $f(Q) = f_\text{c} + a\, Q^b \cos \left(c \log Q + d \right)$ for the thermodynamic quantities in $D=5$. We give the values obtained by fitting both the LBH and NBS branch near the transition.  }
	\label{tab:param_thermo5D}
	\begin{tabularx}{\textwidth}{c *{6}{Y}}
		\toprule
		&	$f$		 	& $f_\text{c}$ 	& $a$		& $b$		& $c$		& $d$ 		\\
		\midrule
		\multirow{ 4}{*}{LBH}
		&	$M/M_\GL$ 	& 1.6771933 	& 2.4700	& 1.4997 	& 1.9362 	& 2.0766 	\\
	 	&	$n/n_\GL$	& 0.7795283 	& 0.5762	& 1.4986 	& 1.9359 	& 4.2842 	\\
	 	&	$T/T_\GL$ 	& 0.6738645 	& 0.7869	& 1.4990 	& 1.9367 	& 5.3444 	\\
		&	$S/S_\GL$ 	& 2.6718298 	& 7.3502	& 1.5001 	& 1.9359 	& 2.0752 	\\
		\midrule 
		\multirow{ 4}{*}{NBS}
		&	$M/M_\GL$ 	& 1.6771932 	& 0.7161 	& 1.4995 	& 1.9364 	& 3.6215	\\
	 	&	$n/n_\GL$ 	& 0.7795282	 	& 0.1691 	& 1.5010 	& 1.9375 	& 5.8367	\\
		&	$T/T_\GL$	& 0.6738646 	& 0.2295	& 1.4998	& 1.9358 	& 0.6010 	\\
		&	$S/S_\GL$ 	& 2.6718297 	& 2.1232 	& 1.4994 	& 1.9369 	& 3.6237	\\
		\bottomrule
	\end{tabularx}	
\end{table}
\begin{table}[ht]
	\centering
	\caption{Parameter values of the fit $f(Q) = f_\text{c} + a\, Q^b \cos \left(c \log Q + d \right)$ for the thermodynamic quantities in $D=6$. We give the values obtained by fitting both the LBH and NBS branch near the transition. }
	\label{tab:param_thermo6D}
	\begin{tabularx}{\textwidth}{c *{6}{Y}}											
 		\toprule
		&	$f$			& $f_\text{c}$ 	& $a$		& $b$		& $c$		& $d$ 	 	\\
		\midrule
		\multirow{ 4}{*}{LBH}
		&	$M/M_\GL$ 	& 2.1839096 	& 4.75319 	& 1.99999 	& 1.99993 	& 5.95517 	\\
 		&	$n/n_\GL$ 	& 0.5855194 	& 0.93638 	& 1.99991 	& 1.99994 	& 1.70328 	\\
 		&	$T/T_\GL$ 	& 0.7419027 	& 0.65522 	& 1.99991 	& 1.99996 	& 2.92683 	\\
		&	$S/S_\GL$ 	& 3.0961719 	& 9.61169 	& 2.00001 	& 1.99992 	& 5.95511 	\\
		\midrule
		\multirow{ 4}{*}{NBS}
	 	&	$M/M_\GL$ 	& 2.1839096 	& 1.59247 	& 1.99923 	& 1.99932 	& 0.74457 	\\
 		&	$n/n_\GL$ 	& 0.5855195		& 0.30918 	& 1.99487 	& 1.99655 	& 2.76608 	\\
 		&	$T/T_\GL$ 	& 0.7419027	 	& 0.21640 	& 1.99512 	& 2.00111 	& 4.00513 	\\
 		&	$S/S_\GL$ 	& 3.0961720 	& 3.23682 	& 2.00071 	& 1.99891 	& 0.74332	\\
		\bottomrule
	\end{tabularx}
\end{table}

We stress that the values listed in the tables~\ref{tab:param_thermo5D} and~\ref{tab:param_thermo6D} were obtained by only taking data points of about the last cycle into account. Including more data points with greater $Q$ leads to slightly bigger deviations of the fitting parameters from their predicted values, since the fit ansatz~\eqref{eq:fit_ansatz_mass} becomes less appropriate for greater $Q$. We note that the standard error arising within the fit routine is mostly of the order of the last digit (or even smaller) that is printed in the tables~\ref{tab:param_thermo5D} and~\ref{tab:param_thermo6D}. 

With the values of the tables~\ref{tab:param_thermo5D} and~\ref{tab:param_thermo6D} at hand we are able to perform some consistency checks. First, we verify that the critical values of the thermodynamic quantities $f_\text{c}$ at the transition indeed satisfy Smarr's relation~\eqref{eq:SmarrRelation}, with deviations only of the order of $10^{-7}$. Moreover, the first law of black hole thermodynamics $\delta M = T\, \delta S$ implies that the extreme points of mass and entropy coincide. Here, this means that the phase shifts $d$ of mass and entropy are the same, which is satisfied with an error of less than $1\%$. When we plug in the ansatz~\eqref{eq:fit_ansatz_mass} into Smarr's relation and the first law we derive three further conditions on the parameters of the fit functions of the thermodynamic quantities. The values of the tables~\ref{tab:param_thermo5D} and~\ref{tab:param_thermo6D} give rise to a violation of these conditions of the order of $1\%$.  

Furthermore, the above analysis allows us to definitely answer a question raised in reference~\cite{Headrick:2009pv}. Therein, the authors provided evidence that there is a small window of LBH solutions with positive specific heat.\footnote{The specific heat of a black hole is proportional to $\partial M / \partial T$. Normally, black holes have negative specific heat, i.e.\ they are hotter the less massive they are.} Our results are in agreement with this observation and, moreover, we find evidence for infinitely many tiny regions with positive specific heat. We follow this from the significant difference of the phase shifts $d$ of mass and temperature (modulo $\pi$) compared to the small discrepancy in the phase shifts of mass and entropy, cf.\ tables~\ref{tab:param_thermo5D} and~\ref{tab:param_thermo6D}. Naturally, the same argument holds for the NBS solutions.

\begin{figure}
	\centering
	\includegraphics[scale=1]{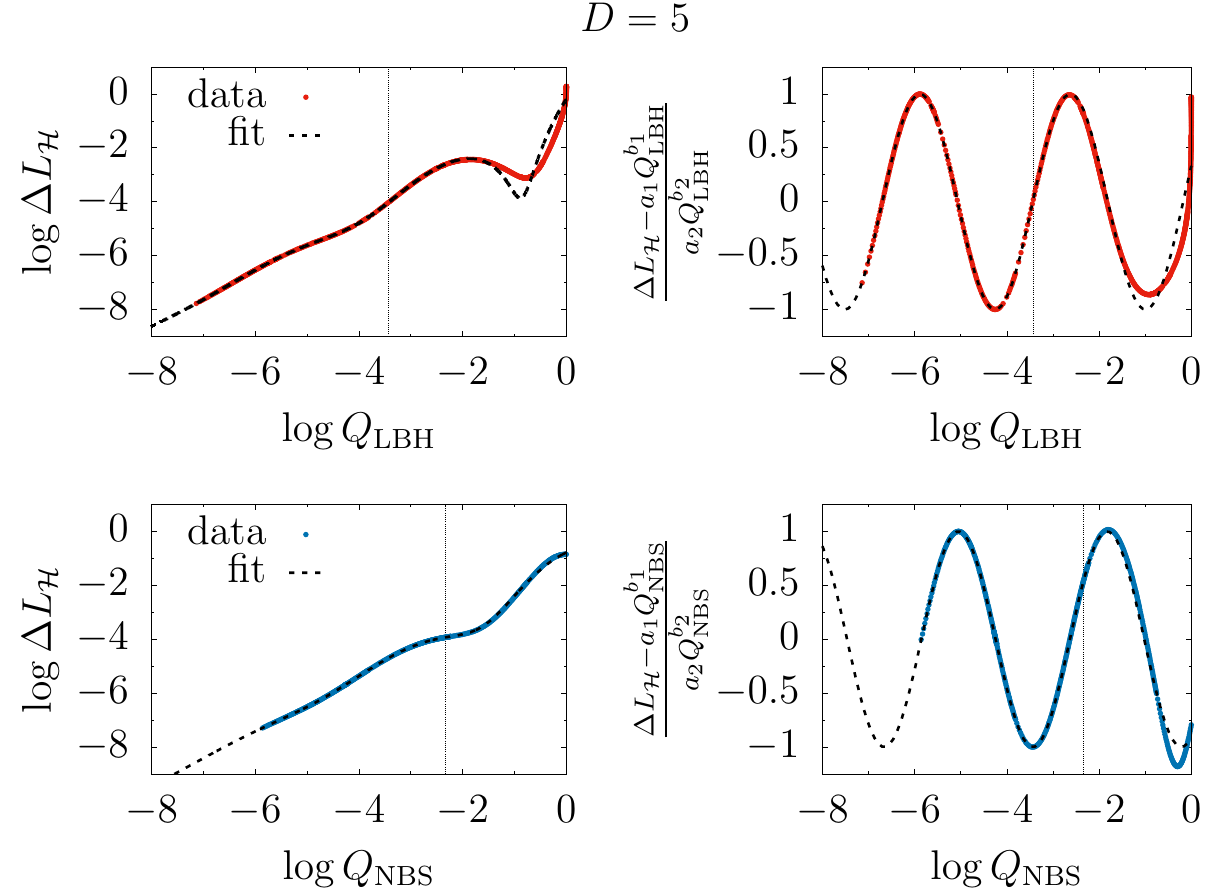} \ \\
	\includegraphics[scale=1]{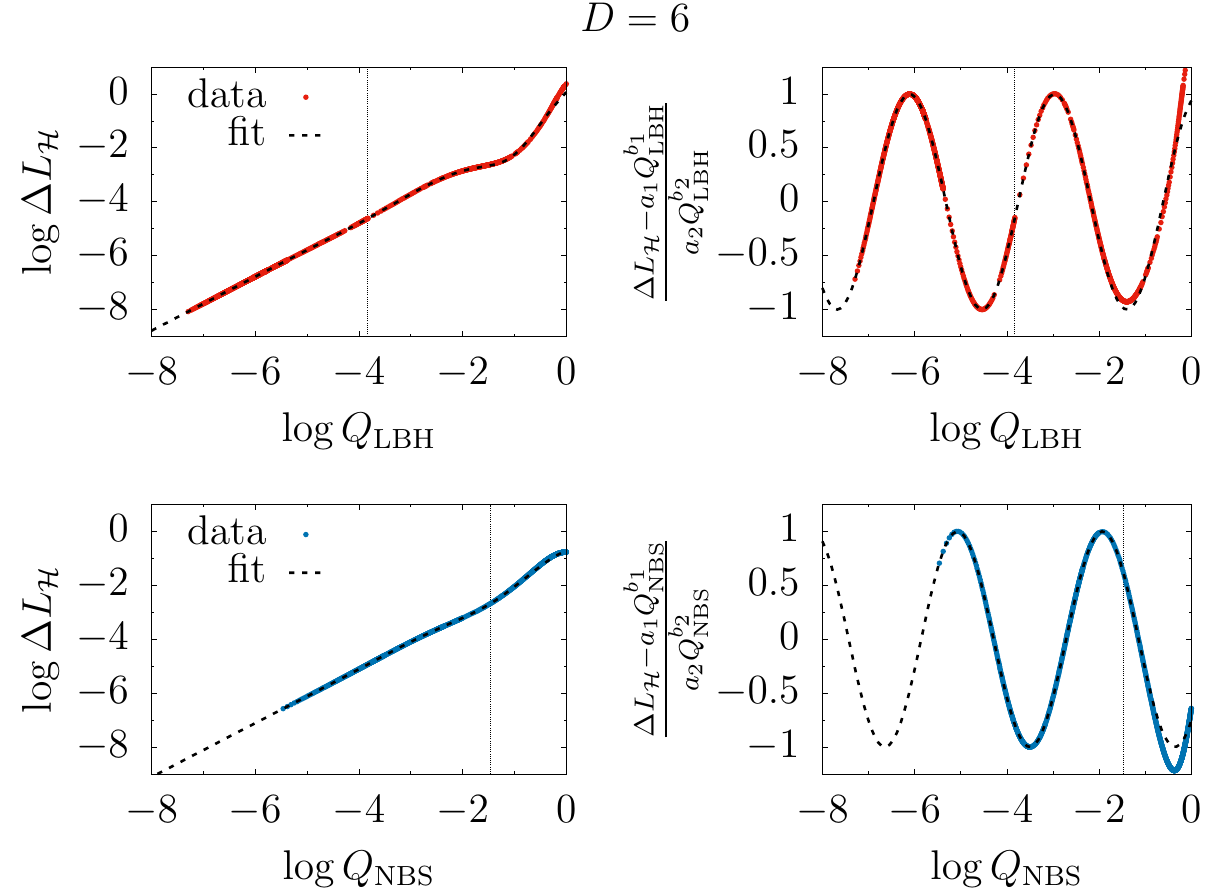}
	\caption{Data points (red dots for LBHs and blue dots for NBSs) and fit (blue solid lines) of the horizon length $L_\mathcal{H}$ as a function of  $Q_\text{LBH}$ or $Q_\text{NBS}$, respectively. The left column shows the explicit functional dependence, where both axes are log-scaled. We resolve the tiny subleading oscillations of the functions in the right column, where $\Delta L_\mathcal{H} = L_\text{c} - L_\mathcal{H}/L$. The first two rows correspond to $D=5$, while the last two rows correspond to $D=6$. In each plot the thin dotted vertical line indicates that all data points left to this line were used to produce the fit.}
	\label{fig:Scaling_HorizonLength}
\end{figure}

Finally, we observe from figure~\ref{fig:Geometry} that there is a physical quantity for which the ansatz~\eqref{eq:fit_ansatz_mass} is not suitable. As stated before, the proper horizon length $L_\mathcal H$ gradually increases in the critical regime and therefore we consider the fit ansatz 
\beq
	L_\mathcal{H} (Q) / L = L_\text{c} - a_1 Q^{b_1} + a_2 Q^{b_2} \cos \left( c_2 \log Q + d_2 \right) \, ,
	\label{eq:fit_ansatz_L}
\eeq 
which includes a non-oscillating leading term. Now, we have seven unknown parameters: $L_\text{c}$, $a_1$, $b_1$, $a_2$, $b_2$, $c_2$ and $d_2$. Obviously, $L_\text{c}$ is the horizon length (normalized with $L$) for $Q=0$, i.e.\ for the critical transit solution. It is apparent from equation~\eqref{eq:fit_ansatz_L} that $b_1 < b_2$ in order to really have a leading non-oscillating part. Figure~\ref{fig:Scaling_HorizonLength} compares data points and fit showing again great agreement for small $Q$.
 \\

We provide the obtained parameter values in the tables~\ref{tab:param_L5D} and~\ref{tab:param_L6D}. Once more, we observe that $b_2$ and $c_2$ are close to the values derived from the double-cone geometry. Interestingly, the leading exponent $b_1$ is approximately one in all cases. In other words, the horizon length is directly proportional to $Q$ at least to first order in the critical regime.
\begin{table}[h]
	\centering	
	\caption{Parameter values for the horizon length $L_\mathcal{H}$ when fitted with $L_\mathcal{H} (Q)/L = L_\text{c} - a_1 Q^{b_1} + a_2 Q^{b_2} \cos \left( c_2 \log Q + d_2 \right)$ in $D=5$. The upper line concerns the LBH case and the lower one the NBS case. }
	\label{tab:param_L5D}
	\begin{tabularx}{\textwidth}{cc *{6}{Y}}
		\toprule
					 	 	& $L_\text{c}$ 	& $a_1$		& $b_1$		& $a_2$		& $b_2$ 	& $c_2$ 	& $d_2$		\\
		\midrule
		LBH				 	& 1.428268		& 0.5548	& 1.0024	& 0.7976 	& 1.4976	& 1.9267	& 1.9116	\\
		NBS				 	& 1.428265		& 0.2441 	& 1.0041	& 0.2319 	& 1.5046 	& 1.9500	& 3.5614	\\
		\bottomrule
	\end{tabularx}
\end{table}
\begin{table}[h]
	\centering	
	\caption{Parameter values for the horizon length $L_\mathcal{H}$ when fitted with $L_\mathcal{H} (Q)/L = L_\text{c} - a_1 Q^{b_1} + a_2 Q^{b_2} \cos \left( c_2 \log Q + d_2 \right)$ in $D=6$. The upper line concerns the LBH case and the lower one the NBS case. }
	\label{tab:param_L6D}
	\begin{tabularx}{\textwidth}{cc *{6}{Y}}
		\toprule
					 	 	& $L_\text{c}$ 	& $a_1$		& $b_1$		& $a_2$		& $b_2$ 	& $c_2$ 	& $d_2$		\\
		\midrule
		LBH				 	& 1.464800		& 0.4564 	& 0.99999	& 0.6558 	& 1.9998	& 2.0001	& 2.7909	\\
		NBS				 	& 1.464801		& 0.3273 	& 0.99960	& 0.2143 	& 1.9898 	& 1.9985	& 3.8530	\\
		\bottomrule
	\end{tabularx}
\end{table}

  \chapter{Conclusions}
\label{chap:Conclusions}

In this thesis we studied static Kaluza-Klein (KK) black holes in five and six spacetime dimensions. To be more specific, we numerically constructed solutions of localized black holes (LBHs) and non-uniform black strings (NBSs). The former have a hyper-spherical horizon topology and are localized in the compact dimension. In contrast, the latter cover the entire compact dimension. Moreover, NBSs emanate from the Gregory-Laflamme (GL) instability of the analytically known uniform black strings (UBSs). We conclude this work by highlighting the main results and their physical implications in section~\ref{sec:Physical_implications} and, furthermore, we recapitulate in section~\ref{sec:Crucial_numerical_techniques} the crucial numerical techniques that provided us with sufficiently accurate solutions. Finally, in section~\ref{sec:Outlook}, we give an outlook and discuss possible future directions.

\section{Physical implications}
\label{sec:Physical_implications}

Due to the high-precision numerics we were able to complete the phase diagram of static KK black holes in $D=5$ and $D=6$. In particular, we investigated in detail the region in the phase diagram where the LBH and NBS branches meet. We found that in this critical regime thermodynamic quantities start to oscillate when approaching the transition. In the phase diagram of the microcanonical ensemble this leads to zigzag curves of the respective branches. In contrast, plotting other thermodynamic quantities against each other we observe typical spiral curves. While the singular transit solution itself is not attainable by our numerical implementation, we are able to resolve four turning points of the spiral in the LBH case and three turning points in the NBS case. In fact, with each turn of the spiral the two branches rapidly converge towards each other.

Moreover, we were able to describe the behavior near the transition qualitatively as well as quantitatively. Based on the analysis of perturbations of the double-cone metric~\cite{Kol:2002xz,Kol:2005vy} we fitted our numerical data of different physical quantities with an oscillating ansatz that has rapidly shrinking amplitude and wavelength when the transition is approached. Indeed, the obtained fit parameters that control amplitude and wavelength coincide remarkably well with the critical exponents derived from the double-cone metric. This, of course, gives compelling evidence in favor of the double-cone metric to be indeed a local model of the transit solution. Furthermore, we refer to this phenomenon as \textit{critical behavior}, since the critical exponents are universal for all thermodynamic quantities and both branches, but only depend on the number of spacetime dimensions. 

According to this critical behavior, we conclude that the spirals appearing in the thermodynamic diagrams discussed above are actually distorted logarithmic spirals, i.e.\ the extent of the spirals shrinks exponentially with each turn leading to an infinite number of turns before the endpoint is reached. Consequently, this gives rise to a \textit{discrete scaling symmetry} of thermodynamic quantities when the transition is approached.  Moreover, according to the so-called turning point method~\cite{Poincare:1885aa}, each turning point indicates the formation of an unstable mode, see also reference~\cite{Arcioni:2004ww}. Indeed, reference~\cite{Headrick:2009pv} found such a mode arising at the first turning point of the LBH branch. Consequently, our results suggest an infinite cascade of unstable modes close to the transition. Additionally, we found another interesting conclusion associated with the logarithmically spiraling behavior: Since mass and temperature have different phase shifts we identify infinitely many tiny regions within each branch where the corresponding object has positive specific heat.

Critical behavior or at least the appearance of a spiral curve seems to be a quite generic feature in the phase diagram of higher dimensional objects when there is a transition between two different branches with one branch emanating from the zero-mode of an instability. In reference~\cite{Bhattacharyya:2010yg} hairy black holes in AdS$_5\times \mathbb S^5$ were studied and the occurrence of critical exponents was shown in the soliton limit where regular and non-regular solitons meet, see also reference~\cite{Markeviciute:2016ivy}. The beginning of a spiral curve was observed in a similar context but in global AdS$_5$~\cite{Dias:2011tj}. In the two cases above the hairy black hole branch emanates from the superradiant instability of Reissner-Nordström black holes. Another interesting situation appears in asymptotically flat spacetime in $D\geq 6$, where references~\cite{Dias:2014cia,Emparan:2014pra} showed that the black ring branch approaches the so-called lumpy black hole branch. Again, the onset of an inspiral was found. Note that the lumpy black holes emanate from the ultraspinning instability of Myers-Perry black holes but in contrast to the Myers-Perry black holes they exhibit a pinched horizon.\footnote{We note that lumpy black holes are also referred to as bumpy black holes or pinched rotating black holes.} Interestingly, there are also black hole configurations in \textit{four} dimensions where a spiral curve in the phase diagram is present, for example in the context of hairy black holes~\cite{Herdeiro:2014goa,Herdeiro:2015gia}.

\section{Crucial numerical techniques}
\label{sec:Crucial_numerical_techniques}

To construct the corresponding LBH and NBS solutions we utilized a pseudo-spectral method. The basic ideas and techniques of this method are outlined in appendix~\ref{chap:Appendix}. However, we performed several important adaptions of the method that took into account the special behavior of the functions in the vicinity of the numerical boundaries. This led to two very different schemes for the two systems under study. The benefit of these rather complicated approaches is that our numerical solutions stand out due to their high accuracy even in the critical regime where the two branches approach the singular transit solution. Without these highly accurate results we would not be in a position to resolve the tiny oscillations of the physical quantities, at least not to the provided extent. 

We explained in detail the crucial adaptions of the method in the main text. However, some of them are of particular importance, for example the way we used appropriate domain decompositions and coordinate transformations in order to significantly increase the resolution near the region where the gradients become exceedingly high.\footnote{Recall the spacetime singularity of the transit solutions between LBHs and NBSs. For LBHs it forms at the axis that connects the poles while for NBSs it forms at the waist of the horizon. } Moreover, there are some further crucial adaptions that, to our knowledge, have not been utilized in the same way previously. Therefore, we list these adaptions below and summarize their benefits:
\begin{itemize}
	\item 	Domain decomposition in the asymptotic regions of LBHs and NBSs: \\
			We identified an asymptotic region in which, after compactifying infinity, we introduced several linearly connected subdomains. On the one hand, this takes into account the non-analytic behavior of the metric functions near infinity and thus ensures a rapid fall-off of the spectral coefficients with respect to the radial direction in each of the subdomains. On the other hand, this domain splitting allows us to use different resolutions, which is particularly important for the transverse direction, since a lower resolution is required when approaching infinity. Eventually, we end up with a significantly smaller total resolution, which leads to a dramatic saving of computational costs, i.e.\ memory capacity and computing time.  
	\item 	Decomposition of the metric functions in the asymptotic region of NBSs: \\
			We decomposed the metric functions into a part that only depends on the radial coordinate and a part that depends on the radial as well as the transverse coordinate. This allows us to directly extract the asymptotic coefficients from the functions that only depend on the radial coordinate. Thus, we obtain high accuracy for the physically important asymptotic charges.
	\item 	Exponential coordinate transformation in the asymptotic region of NBSs: \\
			In five spacetime dimensions, the metric functions show logarithmic behavior near infinity already at low orders. This is a problem for the modes that only depend on the radial coordinate, because their spectral coefficients decay slowly. We circumvented this via an exponential coordinate transformation that transforms the original logarithmic terms into infinitely smooth ones. Accordingly, this considerably improves the fall-off of the spectral coefficients, thus leading to higher accuracy with lower resolution. Again, this reduces the computational costs.     
	\item 	Use of a non-smooth reference metric for LBHs: \\
			The Einstein-DeTurck method requires an appropriately chosen prescribed reference metric. We considered two different ansätze that are suited to different boundaries of the integration domain. At the end we matched them at a contour lying within the integration domain. In particular, for simplicity we decided to match the different ansätze in a non-smooth way. To guarantee smoothness of the resulting functions of the target metric, we accordingly decomposed the domain of integration in order to have an inner boundary exactly on this special contour. At the end, the rather straight-forward ansatz for the reference metric together with simple coordinate transformations avoid lengthy and complicated expressions of the resulting field equations.  
\end{itemize}

\section{Outlook}
\label{sec:Outlook}

The most obvious extension of this work concerns the investigation of KK black holes in $D\geq 7$. In this case, numerical results for LBHs are rare apart from recent results in $D=10$~\cite{Dias:2017uyv}. The numerical scheme for the construction of LBH solutions described here should adapt straightforwardly to the higher-dimensional systems. However, when going to higher dimensions there arises a technical issue. In order to extract the asymptotic charges from the data we need to perform higher order numerical derivatives spoiling the accuracy of these observables. To solve this issue we could, in principle, incorporate a similar ansatz as for the NBSs by appropriately decomposing the metric functions near infinity, cf.\ section~\ref{sec:Construction_of_non-perturbative_solutions}. Alternatively, to circumvent the resulting complications one could completely refrain from doing so. Instead, for large $D$ one could use the first law $\delta M = T\, \delta S$ to obtain the mass and then Smarr's relation to obtain the relative tension. Of course, thereby we lose the opportunity to use the first law and Smarr's relation as consistency checks for the numerical results. 

Furthermore, it would be interesting to extend the existing results for NBSs for $D\geq 7$ in order to get closer to the critical transition. Note that numerical NBS solutions already exist up to $D=15$, see in particular reference~\cite{Figueras:2012xj}. Again, we emphasize that the methods used in this work should be capable to construct NBS solutions also for $D\geq 7$. In particular, the domain setup and the corresponding coordinate transformations are appropriate to enter the critical regime near the transition. As stated before, if one is not interested in highly accurate values of the asymptotic charges, one may refrain from doing the decomposition of the metric functions near infinity.

Following Kol's analysis of perturbations of the double-cone metric we expect the following picture for KK black holes in $D\geq 7$ to hold. For $7\leq D\leq 9$ the situation is qualitatively the same as in the cases $D=5$ and $D=6$ considered here, i.e.\ we have complex critical exponents leading to a damped oscillating behavior of the thermodynamic quantities when the transition is approached. The situation changes for $D\geq 10$ when the critical exponents become purely real and thus do not give rise to oscillations, at least to leading order. Nevertheless, the real critical exponents dictate the behavior of thermodynamic quantities near the transition, now giving rise to a continuous scaling symmetry (to leading order). According to the results of reference~\cite{Figueras:2012xj} there are further qualitative changes in the phase diagram of KK black holes for $D\geq 12$ and for $D\geq 14$, where in the former a part of and in the latter even the whole NBS branch becomes entropically favored over the UBSs. In any case, it would be extremely interesting to obtain the complete phase diagram and the exact location and behavior of the LBH/NBS transition therein. Moreover, we stress that the dimension $D=10$ is of particular interest for another reason. By using solution generating techniques one can relate the $D=10$ KK black hole solutions to type IIa supergravity solutions, which, by virtue of the AdS/CFT correspondence, are dual to certain thermal states of super-Yang-Mills theory on a circle~\cite{Dias:2017uyv}. More generally, KK black hole solutions in any dimension can be mapped to near-extremal branes on a circle~\cite{Harmark:2004ws}.

We expect that our results and the high-precision numerical methods are also relevant in other contexts. We note that GL instabilities towards non-uniform black objects and the competition between them and localized solutions are generic features of higher-dimensional gravity with compact extra dimensions. This includes generalizations where more than one extra dimension is compact or where the extended dimensions are subject to different asymptotic boundary conditions, such as in AdS spacetime. A related system appears for example in global AdS$_5 \times \mathbb S^5$ and has been under investigation recently~\cite{Dias:2015pda,Dias:2016eto}. The situation there is quite similar as in the context of static KK black holes, since there is a well-known static solution with horizon topology $\mathbb S^3\times \mathbb S^5$, which is subject to a GL instability caused by the different horizon and compactification length scale. From this instability emanates a new branch of solutions, which was numerically constructed in reference~\cite{Dias:2015pda}. Moreover, there are solutions that are localized on the $\mathbb S^5$ and have horizon topology $\mathbb S^8$, see reference~\cite{Dias:2016eto}. Again, these two branches are expected to meet at a topology changing singular solution, but the numerical data does not reach far enough to clarify this issue. An interesting question is whether the double-cone metric can serve as a local model of the transit solution in this context as well. Remarkably, if we consider the double-cone metric with arbitrary dimensionality of the two underlying (hyper-)spheres, it turns out that the physical implications only depend on the total number of spacetime dimensions $D$~\cite{Kol:2002xz}. 

The last considerations lead us to the final thoughts of this thesis and to the question: How generic is the double-cone metric as a possible local model for topology changing transit solutions? Reference~\cite{Emparan:2011ve} gives an explicit analytic example where two horizons merge to locally form a double-cone: a Kerr black hole in deSitter spacetime in $D\geq 6$ whose horizon touches the cosmological horizon at the equator. Moreover, in reference~\cite{Emparan:2011ve} further systems are discussed in which the double-cone metric is expected to describe the local geometry of a possible transition. One of them is the transition from black rings to lumpy black holes in $D\geq 6$, which we already mentioned above. Similar arguments hold for black saturns and circularly pinched lumpy black holes, also in $D\geq 6$. If the double-cone metric is indeed appropriate to describe the transitions in these situations, it is very likely that the critical behavior observed in the context of static KK black holes also occurs there. Clearly, high-precision numerics are needed to answer these questions. As useful tricks and techniques have been presented here, we thus think that this thesis can hold as a guideline for future work.

  \cleardoublepage %
  \pagestyle{scrheadings}
  \setcounter{page}{1}
  \pagenumbering{Roman}

  \appendix
  \chapter{Appendix}
\label{chap:Appendix}

Below, we provide supplementary material concerning the basic principles and techniques of the numerical scheme that was used to produce the results of this work.

\section{Pseudo-spectral method}
\label{sec:Pseudo-spectral_method}

In this section we review in detail the pseudo-spectral method as a tool to find numerical solutions to differential equations formulated as boundary value problems. The most striking advantage of pseudo-spectral methods compared to other numerical schemes is their ability to produce highly accurate results with a moderate consumption of computational resources, i.e.\ memory and time. Unfortunately, this advantage heavily relies on the smoothness of the underlying functions. Therefore, to unfold the full power of pseudo-spectral methods, one has to develop a deep understanding of the functions' behavior. Then, by utilizing appropriate coordinate transformations, function redefinitions and domain decompositions, one can design a scheme that is well adapted to the problem at hand and provides highly accurate results. This becomes most beneficial in situations where the mathematical structure becomes more involved, e.g.\ if a strongly pronounced peak appears that runs towards a singularity when a parameter is changed. This typically happens close to specifically interesting branch points or phase transitions, as it is the case in the localized black hole / non-uniform black string context. In these situations, where standard algorithms reach their limitations, pseudo-spectral methods are able to explore the critical regime, which may lead to the manifestation of unrevealed properties. 

The standard text books, references~\cite{Boyd:2001aa,Canuto:2007aa}, give a detailed description of the theoretical background of (pseudo-)spectral methods, while their applications to general relativity are discussed in references~\cite{Grandclement:2007sb,Meinel:2012aa}. Here, we will concentrate on the fundamental concepts and some technical aspects of the numerical scheme. After introducing the main ideas and concepts in subsection~\ref{subsec:Basic_ideas_and_concepts}, we will describe the overall numerical scheme in subsection~\ref{subsec:Solving_differential_equations}. Within this scheme a large linear system has to be solved. The efficient solution of this pseudo-spectral linear system is the subject of subsection~\ref{subsec:Solving_the_pseudo-spectral_linear_problem}.


\subsection{Basic ideas and concepts}
\label{subsec:Basic_ideas_and_concepts}

The idea of spectral methods is based on the expansion of a real-valued function $f(x)$ defined on a finite interval $x\in [a,b]$:
\beq
	f(x) = \sum _{k=0}^{\infty} c_k \Phi _k(x) \, ,
	\label{eq:SpectralExpansion}
\eeq
with spectral coefficients $c_k$ and a set of appropriate basis functions $\Phi _k(x)$. For example, in case of a periodic function $f(x)$ an appropriate basis is built of trigonometric functions, leading to a Fourier series representation. In practice, however, the calculation of infinitely many spectral coefficients $c_k$ is not feasible and thus we truncate
 the sum after $N$ terms, yielding a residual $\mathcal R_N(x)$: 
\beq
	f(x) = \sum _{k=0}^{N-1} c_k \Phi _k(x) + \mathcal R_N (x) \, .
	\label{eq:SpectralExpansionTruncated}
\eeq
In many cases the residual $\mathcal R_N(x)$ will decrease very rapidly with increasing $N$. Therefore, the finite sum over $c_k\Phi _k(x)$ is a good approximation of the original function $f(x)$ on $x\in [a,b]$ already for a moderate expansion order $N$. 

For non-periodic functions $f(x)$ another type of basis functions is commonly used: Chebyshev polynomials of the first kind 
\beq
	T_k(z) = \cos (k\arccos z) \, .
	\label{eq:ChebyshevPolynomials}
\eeq
Remarkably, there is a close relation of Chebyshev and Fourier series via the coordinate transformation $\tilde z = \arccos z$. Therefore many theorems concerning the Fourier series also apply to the Chebyshev expansion~\cite{Boyd:2001aa}.

\begin{figure}[ht]
	\centering
		\includegraphics[scale=1]{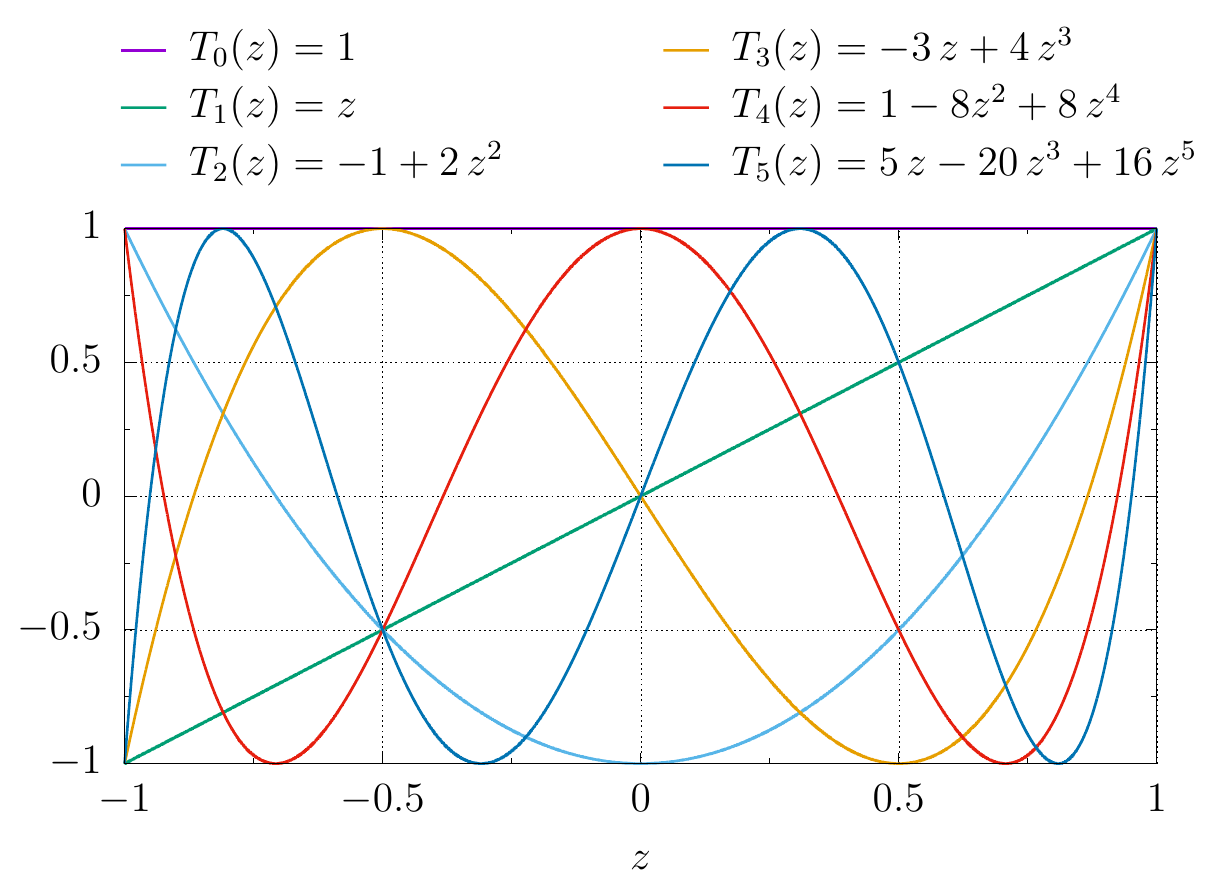}
	\caption{First six Chebyshev polynomials of the first kind $T_k(z)$. They are defined on $z\in [-1,1]$ and obey $-1\leq T_k(z)\leq 1$. The $k$th Chebyshev polynomial has $k$ zeros and (for $k\geq 1$) $k-1$ extremal points. Furthermore, Chebyshev polynomials with even $k$ are symmetric with respect to $z=0$, while those with uneven $k$ are antisymmetric.}
	\label{fig:ChebyshevPolynomials}
\end{figure}

The Chebyshev polynomials are defined on $z\in [-1,1]$, thus we have
\beq
	\Phi _k(x) = T_k \left( \frac{2x-b-a}{b-a} \right) \, ,
	\label{eq:ChebyshevBasis}
\eeq 
In the following, we concentrate on the use of Chebyshev polynomials as the basis functions for the spectral expansion. Figure~\ref{fig:ChebyshevPolynomials} illustrates a couple of Chebyshev polynomials and summarizes some of its properties. 

There are explicit integral formulas to calculate the spectral coefficients $c_k$ from the function $f(x)$. Nevertheless, there exists another way of obtaining the spectral coefficients, which is much more convenient for our purposes. 

\subsubsection{Collocation and interpolation}
\label{subsubsec:Collocation_and_interpolation}

In a numerical calculation we usually want to discretize the function $f(x)$ on certain grid points $x_k$. Throughout this work we use the so-called Lobatto grid points 
\beq
	x_k = \frac{b+a}{2} - \frac{b-a}{2} \cos \left(  \frac{\pi \, k}{N-1} \right)  \, , \quad k = 0,1, \ldots ,N-1 \, ,
	\label{eq:LobattoPoints}
\eeq
which represent the extremal points of $\Phi _{N-1}(x)$ together with the boundaries $a$ and $b$. Another common choice are so-called Gauss grid points, which originate from the zeros of the $N$th Chebyshev polynomial.

In this setup the spectral coefficients $c_k$ follow from the $N$ conditions 
\beq
	f_k := f(x_k) =  \sum _{l=0}^{N-1} c_l \Phi _l(x_k) \, , \quad k = 0,1, \ldots , N-1 \, ,
	\label{eq:ResidualCondition}
\eeq
i.e.\ we require that the residual $\mathcal R_N(x)$ vanishes at the grid points $x_k$. This yields an explicit expression for each spectral coefficient depending on the resolution $N$:
\beq
	c_k = (-1)^k \,\frac{2-\delta _{k,0}-\delta _{k,N-1}}{N-1} \left\{ \frac{1}{2}  \left[ f_0 + (-1)^k f_{N-1} \right] + \sum _{l=1}^{N-2} f_l \cos\left( \frac{\pi \, k l}{N-1} \right) \right\} \, .
	\label{eq:PseudospectralCoefficients}
\eeq 
There are different ways to calculate the spectral coefficients~\eqref{eq:PseudospectralCoefficients} efficiently. A straightforward approach is to implement the Clenshaw algorithm~\cite{Clenshaw:1955aa}, see also reference~\cite{Press:2007aa}. Beyond that, an even more efficient and sophisticated way is the use of a fast Fourier transformation algorithm, where one can exploit the close relation between the Chebyshev and the Fourier expansion to adapt this algorithm to the Chebyshev case. We recommend the FFTW library~\cite{Frigo:2005aa}, which is literally supposed to provide the fastest Fourier transformation algorithm in the West. 

Once the spectral coefficients are computed, we get an approximation of the function $f(x)$ at any point in $x\in [a,b]$ via
\beq
	f(x) \approx \sum _{k=0}^{N-1} c_k \Phi _k(x) \, , 
	\label{eq:SpectralExpansionTruncatedApprox}
\eeq
i.e.\ the pseudo-spectral method has a natural built-in technique of interpolation. Again, a straightforward calculation of the sum in formula~\eqref{eq:SpectralExpansionTruncatedApprox} is rather inefficient, but the Clenshaw algorithm does better. 

Obviously, the accuracy of the approximation~\eqref{eq:SpectralExpansionTruncatedApprox} depends on the resolution $N$. But even more crucial are the mathematical properties of the underlying function $f(x)$, as we will explain now. 

\subsubsection{Rates of convergence and error estimation}
\label{subsubsec:Rates_of_convergence_and_error_estimation}

\begin{figure}[!ht]
	\centering
		\includegraphics[scale=1]{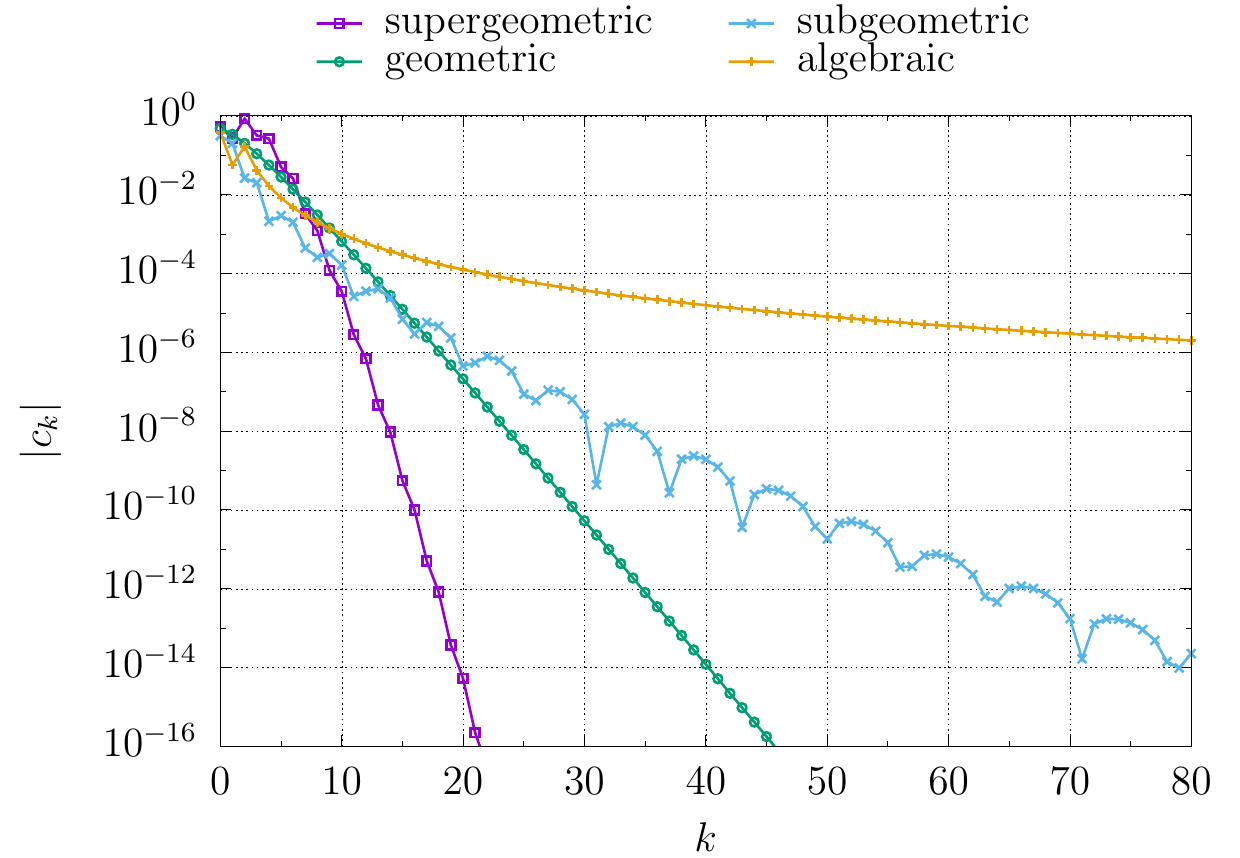} \\
		\includegraphics[scale=1]{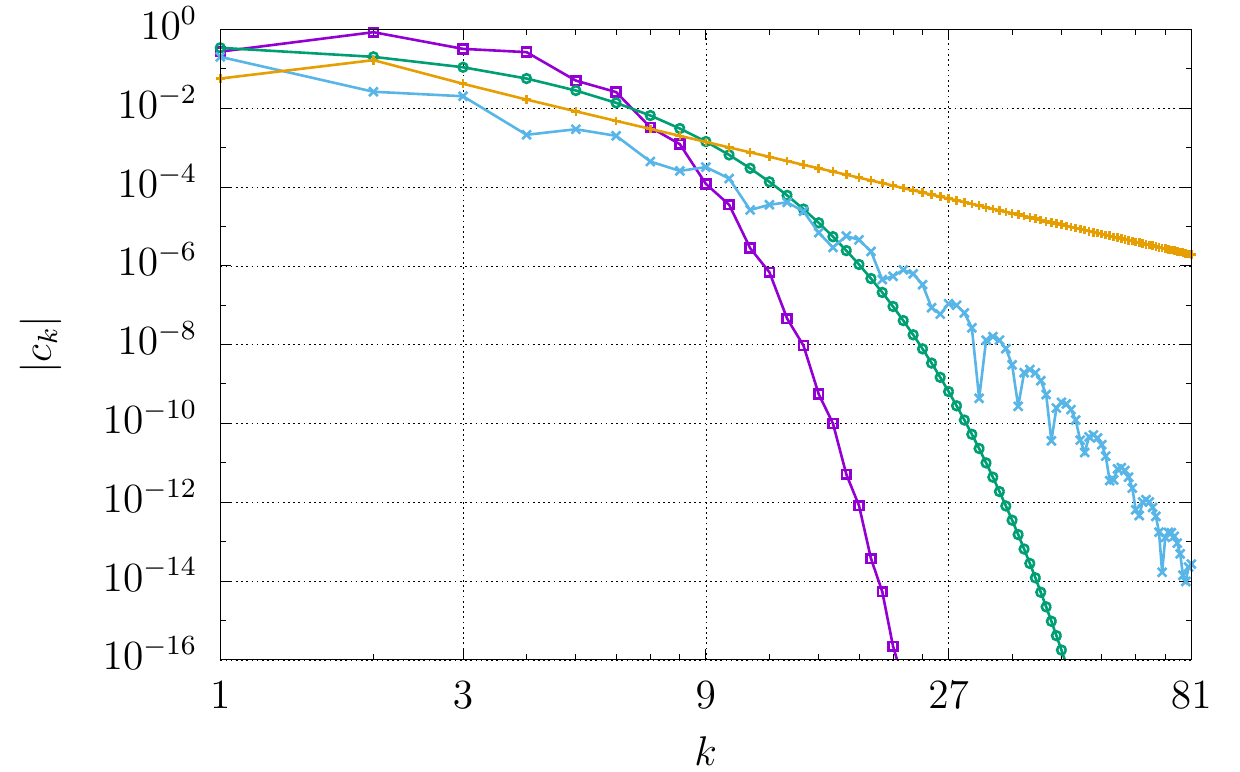} 
	\caption{Representative examples for different rates of convergence. We used the example functions given in the text for supergeometric~\eqref{eq:ExampleFunctionSupergeometricConvergence}, geometric~\eqref{eq:ExampleFunctionGeometricConvergence}, subgeometric~\eqref{eq:ExampleFunctionSubgeometricConvergence} and algebraic~\eqref{eq:ExampleFunctionAlgebraicConvergence} convergence. The upper plot incorporates a logarithmic rescaling of the ordinate (log-plot), while in the lower plot both the ordinate and abscissa are rescaled logarithmically (log-log-plot). }
	\label{fig:RatesOfConvergence}
\end{figure}

In the limit $N\to\infty$ the approximation~\eqref{eq:SpectralExpansionTruncatedApprox} converges towards the real continuous function $f(x)$, which means that the absolute values of the spectral coefficients $c_k$ decrease accordingly. If we consider $c_k$ as a sequence, then the question arises, what is the leading damping behavior in the limit $k\to\infty$. We follow Boyd~\cite{Boyd:2001aa} and classify different leading behaviors into four different rates of convergence:
\begin{itemize}
	\item 	Supergeometric convergence: \\
			The best case one can get is a supergeometric convergence, where the $c_k$ decay faster than any exponential $\exp (-\alpha k)$ with $\alpha >0$. Such an ideal convergence rate only occurs for entire functions, i.e.\ functions that, after analytic continuation into the complex plane, only have singularities at infinity. Examples for entire functions are polynomial, exponential, sine and cosine functions. Consider the example
			\beq
				f(x) = \cos (2\pi \, x + 0.5) 
				\label{eq:ExampleFunctionSupergeometricConvergence}
			\eeq
			on $x\in [0,1]$. Obviously, there is no singularity at finite distance from the interpolating interval $[0,1]$ even after analytic continuation $x\to x+\I y$.
	\item 	Geometric convergence: \\
			If the leading behavior of the spectral coefficients is $c_k\sim  \exp (-\alpha k)$, we call the rate of convergence geometric. This concerns functions that are not entire but analytic on the interval $[a,b]$, i.e.\ there exists a converging Taylor series in a neighborhood of every point $x\in [a,b]$. For example, the function
			\beq
				f(x) = \frac{1}{(1+5\, x)^2}
				\label{eq:ExampleFunctionGeometricConvergence}
			\eeq
			is analytic on $x\in [0,1]$ but the singularity at $x=-1/5$ spoils the supergeometric convergence of the spectral coefficients of $f(x)$ on $x\in [0,1]$. Note that the closer the singularities are to the interpolating interval the smaller is the parameter $\alpha$. Consequently, it is desirable to have singularities as far away as possible from the interpolating interval to get a rapid convergence. 
	\item 	Subgeometric convergence: \\
			The spectral coefficients are called to fall-off with a subgeometric rate of convergence if they decay more slowly than any exponential $\exp (-\alpha k)$ with $\alpha >0$ but faster than any inverse power $k^{-\beta}$ with $\beta >0$ of $k$. Usually, this behavior is present if the underlying function is not analytic but smooth on the interpolating interval, i.e.\ the function is infinitely many times differentiable but there is at least one point where no Taylor series converges in the neighborhood of this point. As an example consider
			\beq
				f(x) = \E ^{-\frac{1}{x}}
				\label{eq:ExampleFunctionSubgeometricConvergence}
			\eeq 
			on $x\in [0,1]$. Clearly, all derivatives of this function are finite on the interval $[0,1]$, but at $x=0$ all derivatives vanish. The corresponding Taylor series at this point, the zero function, does not converge towards $f(x)$ at any $x>0$. 
	\item 	Algebraic convergence: \\
			Finally, the worst case is an algebraic rate of convergence, which means that $c_k \sim k^{-\beta}$ with $\beta >0$. In other words, the $c_k$ decay with an inverse power law of order $\beta$. Naturally, algebraic convergence is present when only a finite number of derivatives of the underlying function exist. For instance, the function
			\beq
				f(x) = 	x \ln x \, ,
				\label{eq:ExampleFunctionAlgebraicConvergence}
			\eeq
			considered on $x\in [0,1]$, is continuous but already its first derivative diverges at $x=0$, which leads to a very slow convergence. Note that the more continuous derivatives exist, the higher is the order $\beta$ and the faster is the convergence. 			
\end{itemize}
We give an illustration of the different rates of convergence in figure~\ref{fig:RatesOfConvergence} by displaying the spectral coefficients, calculated via~\eqref{eq:PseudospectralCoefficients}, of the example functions~\eqref{eq:ExampleFunctionSupergeometricConvergence}, \eqref{eq:ExampleFunctionGeometricConvergence}, \eqref{eq:ExampleFunctionSubgeometricConvergence} and~\eqref{eq:ExampleFunctionAlgebraicConvergence} mentioned above. It is apparent from the definitions above that in the log-plot the geometric rate of convergence is represented by a straight line and similarly that in the log-log-plot the algebraic rate of convergence is represented by a straight line. This gives us the possibility to infer some properties of an unknown function from the decay of its spectral coefficients. However, we emphasize that, strictly speaking, these definitions only apply asymptotically, i.e.\ for large $k$. But in usual situations only a finite number of spectral coefficients are known. To make a statement about the rate of convergence in these cases, it is necessary to consider a wide range of $k$ that corresponds to spectral coefficients $c_k$ ranging over several orders of magnitude and going down to extremely tiny scales, like in figure~\ref{fig:RatesOfConvergence}. If a trend in the decay of the spectral coefficients is observed, it is very likely that this trend continues, since it is rather unlikely that the underlying function will change its behavior on these small scales. 

Most importantly, it follows from equation~\eqref{eq:SpectralExpansionTruncated} that for an increasing resolution $N$ the residual $\mathcal R_N(x)$ shows a similar behavior as the spectral coefficients~\cite{Boyd:2001aa}. Moreover, the last coefficient of a pseudo-spectral expansion $c_{N-1}$ gives a rough order of magnitude estimation for the maximal error of the approximation~\eqref{eq:SpectralExpansionTruncatedApprox}, i.e.\ the maximal absolute value of $\mathcal R_N(x)$. Nevertheless, this is not a reliable way of calculating the error of a spectral approximation of an unknown function. A better and necessary error estimation is to repeat the numerical algorithm for different resolutions $N$ and to compare results. 

From the discussion of the rates of convergence of spectral coefficients we conclude one of the striking advantages of pseudo-spectral methods: For many types of functions the spectral coefficients and the error decay faster than any inverse polynomial. This is often simply denoted as spectral or exponential convergence. In such situations the number of grid points $N$ to reach a certain accuracy usually stays moderate.

Sometimes one has to use some tricks to transform the function in such a way that its spectral coefficients fall-off appropriately fast. This is what chapters~\ref{chap:Numerical_construction_of_non-uniform_black_string_solutions} and~\ref{chap:Numerical_construction_of_localized_black_hole_solutions} deal with in the context of solving Einstein's vacuum field equations in order to obtain highly accurate non-uniform black string or localized black hole solutions.

\subsubsection{Differentiation and integration}
\label{subsubsec:Differentiation_and_integration}

For solving differential equations, like Einstein's field equations, it is necessary to compute the derivatives of functions. Fortunately, the pseudo-spectral approach provides a straightforward and rather simple way of doing this. Again, we want to expand the derivative of the function $f(x)$ into
\beq
	f'(x) \approx \sum _{k=0}^{N-1} c_k' \Phi _k(x) \, ,
	\label{eq:SpectralExpansionDerivative}
\eeq
where $c_k'$ are the spectral coefficients of the derivative $f'(x)$. Using various identities of the Chebyshev polynomials we obtain from the spectral coefficients $c_k$ of a function $f(x)$ the spectral coefficients of its derivative via
\beq
	c_{k-1}' = \left( 1-\frac{1}{2}\delta _{k,1} \right) \left( c_{k+1}' + 4 \, k c_k \right) \, , \quad k=N-1, N-2, \dots , 1 \, ,
	\label{eq:SpectralCoefficientsDerivative}
\eeq
with this recursive formula starting with $c_{N-1}' = c_N' = 0$. Note that by this construction the spectral expansion only has $N-1$ non-trivial coefficients, which becomes clear from the fact that by differentiating a polynomial of degree $N-1$ we get a polynomial of degree $N-2$. In addition, the coefficients of the derivative $c_k'$ are generically greater than the coefficients $c_k$ in terms of their absolute values, at least for large enough $k$. Consequently, the accuracy of the pseudo-spectral representation of the derivative $f'(x)$ will be slightly worse than that of $f(x)$. However, here we get back to the advantage of pseudo-spectral methods: Since we may be able to get a highly accurate approximation of $f(x)$ the approximation of its derivative $f'(x)$ is still very accurate.

Obviously, we get the spectral coefficients of the second derivative again from formula~\eqref{eq:SpectralCoefficientsDerivative} once the spectral coefficients of the first derivative are known. Furthermore, after rearranging equation~\eqref{eq:SpectralCoefficientsDerivative} we also obtain an expression for the spectral coefficients of the antiderivative of $f(x)$. Apart from that, the following formula gives an approximation of the definite integral over the whole interpolating interval $[a,b]$:
\beq
	\frac{1}{b-a}\int _a^b f(x)\, \D x \approx c_0 - \sum_{k=1}^{\lfloor{\frac{N-1}{2}}\rfloor} \frac{c_{2k}}{4\, k^2 -1} \, .
	\label{eq:SpectralIntegration}
\eeq

\subsection{Solving differential equations}
\label{subsec:Solving_differential_equations}

We now use the aforedescribed pseudo-spectral techniques of approximating functions and their derivatives to develop a numerical scheme that solves second order differential equations formulated as boundary value problems. For simplicity, we first consider an ordinary differential equation of the form
\beq
	F(f'',f',f;x) = 0 
	\label{eq:ODE}
\eeq
that is subject to the boundary conditions
\beq
	F_a(f(a),f'(a)) = 0 \quad \text{and} \quad F_b(f(b),f'(b)) = 0 \, .
	\label{eq:BoundaryConditions}
\eeq

\subsubsection{Discretization}
\label{subsubsection:Discretization}

We discretize the function $f(x)$ on Lobatto grid points $x_k$~\eqref{eq:LobattoPoints} yielding a set of \textit{a priori} unknown function values $f_k$. The vector
\beq
	\vec X = 	\begin{pmatrix}
					f_0 \\
					f_1 \\
					\vdots \\
					f_{N-1}			
				\end{pmatrix} 
	\label{eq:VectorX}
\eeq
collects all these unknowns sorted by their index. Accordingly, we define two further vectors, $\vec X'$ and $\vec X''$, that contain the values of the spectral derivatives of $f(x)$ at the Lobatto grid points $x_k$ obtained through equations~\eqref{eq:PseudospectralCoefficients}, \eqref{eq:SpectralCoefficientsDerivative} and~\eqref{eq:SpectralExpansionTruncatedApprox}. Then, we define yet another vector
\beq
	\vec F = 	\begin{pmatrix}
					F_0 \\
					F_1 \\
					\vdots \\
					F_{N-1}			
				\end{pmatrix} 
	\label{eq:VectorF}
\eeq
with 
\beq
	F_k = 	\begin{cases}
				F_a(f_0,f'_0) 			\, , &	\text{for } k=0			\, ,  \\  
				F(f''_k,f'_k,f_k;x_k)	\, , &	\text{for } 0<k<N-1 	\, ,  \\ 
				F_b(f_{N-1},f'_{N-1}) 	\, , &  \text{for } k=N-1		\, .  
			\end{cases}
	\label{eq:VectorFComponents}
\eeq
In other words, the vector $\vec F$ contains the discrete version of the differential equations and the corresponding boundary conditions. Eventually, we want to find a solution $\vec X$ to the set of equations
\beq
	\vec F(\vec X) = 0 \, .
	\label{eq:DiscreteDE}
\eeq

\subsubsection{Newton-Raphson scheme}
\label{subsubec:Newton-Raphson scheme}

We broke down the problem of solving a differential equation into finding the roots of a set of $N_\text{total}$ algebraic expressions that are combined in the vector $\vec F$. In the simple example of a single ordinary differential equation we have $N_\text{total} =N$. In order to find these roots we utilize a Newton-Raphson scheme that requires an initial guess $\vec X_0$ for the solution and then gradually improves this by
\beq
	\vec X_{m+1} = \vec X_m - \left[ \hat J\left( \vec X_m \right) \right] ^{-1} \vec F\left( \vec X_m \right) \, ,
	\label{eq:NewtonRaphson}
\eeq
where $\hat J$ represents the Jacobian matrix
\beq
	\hat J = \frac{\partial \vec F}{\partial \vec X} \quad \text{or equivalently} \quad J_{ij} = \frac{\partial F_i}{\partial X_j} \, , \quad i,j = 0,1,\dots ,N_\text{total}-1 \, .
	\label{eq:Jacobian}
\eeq
A simple and straightforward way to build up the Jacobian $\hat J$ numerically is to use a finite difference approximation
\beq
	J_{ij} = \frac{F_i\left( \vec X + \varepsilon \vec e_j\right) - F_i\left( \vec X - \varepsilon \vec e_j\right)}{2\,\varepsilon} \, ,
	\label{eq:JacobianComputationFD}
\eeq
where $\varepsilon$ is some small value and $\vec e_j$ is a unit vector pointing in the $j$th direction.

For non-linear problems the initial guess has to be rather close to the actual solution, otherwise the Newton-Raphson scheme will not converge.\footnote{For linear problems we do not have to utilize the Newton-Raphson scheme. We can rather solve them directly with a linear solver.} Mostly, it is not too difficult to construct an appropriate initial guess from known nearby functions or from some intuition about the mathematical structure of the solution. Then, the procedure terminates when there is no more significant improvement, i.e.\ when $\vec X_{m+1} - \vec X_m$ is small enough, and the vector $\vec F\left(\vec X_{m+1} \right)$ is reasonably close to zero. 

The biggest computational obstacle within the Newton-Raphson scheme is the solution of the linear system
\beq
	\hat J \, \delta \vec X = \vec F \, ,
	\label{eq:PseudoSpectralLinearProblem}
\eeq
yielding the correction $\delta \vec X$ to the vector $\vec X$ at each step. However, for a single ordinary differential equations the computational costs are manageable since the linear system only has dimension $N$ and, as explained above, a pseudo-spectral approximation gives accurate results already for moderate resolutions, typically for $N\lesssim 100$. The pseudo-spectral linear problem~\eqref{eq:PseudoSpectralLinearProblem}, arising within the Newton-Raphson scheme, can than be solved easily for example by using a standard LU-decomposition, e.g.\ see reference~\cite{Press:2007aa}. However, for partial differential equations the computational costs rise significantly.

\subsubsection{Extension to partial differential equations}
\label{subsubsec:Extension_to_partial_differential_equations}

Lets consider a situation where we have a system of second order partial differential equations for the functions $f(x,y)$, $g(x,y)$, $h(x,y)$, etc. The integration domain is a rectangle $(x,y)\in [a_x,b_x]\times [a_y,b_y]$. We discretize the functions on an $N_x\times N_y$ grid representing the Lobatto nodes $(x_k,y_l)$ with $k=0,1,\dots ,N_x-1$ and $l=0,1,\dots ,N_y-1$, see equation~\eqref{eq:LobattoPoints}. Then, we build up the vector $\vec X$ from the values of all functions at the Lobatto grid points. Furthermore, if the problem requires to determine some additional parameters, these are stored in $\vec X$ as well. Altogether, the  vector $\vec X$ contains $N_\text{total} = N_x N_y  N_f + N_\text{add}$ values, where $N_f$ denotes the total number of functions and $N_\text{add}$ the number of additional parameters.

As long as we are only interested in the functions' values on the grid lines, the one-dimensional spectral interpolation and differentiation apply straightforwardly to the two-dimensional case. For instance, we consider the function $f(x,y)$ at a certain grid line $y=y_l$ and regard it as a one-dimensional function $\tilde f(x) = f(x,y_l)$ for which we know how to calculate interpolations and derivatives.\footnote{However, if we are interested in an interpolation on points that do not lie exactly on the grid lines we employ the formula $f(x) = \sum _{k=1}^{N_x-1}\sum _{l=1}^{N_y-1} c_{kl}\Phi _k(x)\Phi _l(y)$. The two-dimensional coefficients $c_{kl}$ are obtained by utilizing equation~\eqref{eq:PseudospectralCoefficients} twice.} We repeat this for all grid lines in both directions and so the vector $\vec F$ arises similarly as in the one-dimensional case. It now contains the set of algebraic equations describing the discrete version of the system of partial differential equations on the grid points together with the boundary conditions at $x=a_x$, $x=b_x$, $y=a_y$ and $y=b_y$. In addition, if $N_\text{add}>0$ the vector $\vec F$ has to contain $N_\text{add}$ additional conditions that, together with the differential and boundary equations, fix the extra parameters. 

Finally, we are in place to apply the Newton-Raphson method~\eqref{eq:NewtonRaphson} to the algebraic system. In this case, however, the pseudo-spectral linear problem~\eqref{eq:PseudoSpectralLinearProblem} scales quadratically with the resolution (if we assume $N_x\approx N_y$). For example a problem arising from a single partial differential equations on a $100\times 100$ grid has $10^4$ unknowns leading to a dense Jacobian matrix with $10^8$ entries. The solution of this linear system becomes a serious task for standard LU solvers in terms of memory and time consumption.\footnote{Note that standard LU solvers have time complexity of $\mathcal O(N_\text{total}^3)$~\cite{Press:2007aa}. Consequently, in two dimensions it depends on the 6th power of the resolution $N$.\label{footnote:TimeComplexityLU}} Sometimes we also have even bigger linear problems to solve, as in the context of localized black holes and non-uniform black strings, where we reach dimensions of $N_\text{total} \sim 10^5$. Hence, we need more advanced algorithms to handle these situations as well.

\subsection{Solving the pseudo-spectral linear problem}
\label{subsec:Solving_the_pseudo-spectral_linear_problem}

Recall the central problem within the Newton-Raphson scheme: the solution of the linear problem $\hat J \, \delta \vec X = \vec F$, cf.\ equation~\eqref{eq:PseudoSpectralLinearProblem}, with the Jacobian $\hat J = \partial \vec F /\partial \vec X$, cf.\ equation~\eqref{eq:Jacobian}. Below, we describe another approach to solve this system in an efficient way.

\subsubsection{Computation of the Jacobian}
\label{subsubsec:Computation_of_the_Jacobian}

The given simple algorithm for building up the Jacobian~\eqref{eq:JacobianComputationFD} is rather expensive, since for the computation of a single matrix element $J_{ij}$ we have to calculate the whole vector $\vec F$ twice, i.e.\ for the entire Jacobian the vector $\vec F$ is computed $(2N_\text{total})^2$ times. We now describe a more elegant and efficient approach.

Usually, the explicit expression for $\vec F(\vec X)$ is given in the form $\vec F(\vec X'', \vec X' ,\vec X)$, where $\vec X'$ and $\vec X''$ represent the pseudo-spectral derivatives of $\vec X$ on the grid points. Since pseudo-spectral algorithms are linear operations we write $\vec X' = \hat D \vec X$ and $\vec X'' = \hat D \vec X' = \hat D^2 \vec X$ with a matrix $\hat D$. We use this fact to rewrite the differential of $\vec F$:
\begin{align}
	\D \vec F(\vec X'', \vec X' ,\vec X) & = \frac{\partial \vec F(\vec X'', \vec X' ,\vec X)}{\partial \vec X''} \, \D \vec X'' + \frac{\partial \vec F(\vec X'', \vec X' ,\vec X)}{\partial \vec X'} \, \D \vec X' + \frac{\partial \vec F(\vec X'', \vec X' ,\vec X)}{\partial \vec X} \, \D \vec X  \nonumber \\
		& = \left[ \frac{\partial \vec F(\vec X'', \vec X' ,\vec X)}{\partial \vec X''} \hat D^2 + \frac{\partial \vec F(\vec X'', \vec X' ,\vec X)}{\partial \vec X'} \hat D + \frac{\partial \vec F(\vec X'', \vec X' ,\vec X)}{\partial \vec X} \right] \, \D \vec X \nonumber \\
		& = \frac{\partial \vec F(\vec X)}{\partial \vec X} \, \D \vec X = \hat J(\vec X) \, \D \vec X \, .
	\label{eq:dF}
\end{align}
Now, we build up a vector $\vec D_F$ out of $\vec X$ and an arbitrary vector $\vec V$:
\beq
	\vec D_F(\vec X,\vec V) = \frac{\partial \vec F(\vec X'', \vec X' ,\vec X)}{\partial \vec X''} \, \vec V'' + \frac{\partial \vec F(\vec X'', \vec X' ,\vec X)}{\partial \vec X'} \, \vec V' + \frac{\partial \vec F(\vec X'', \vec X' ,\vec X)}{\partial \vec X} \, \vec V \, .
	\label{eq:D_F}
\eeq
Again, we note that $\vec X'$, $\vec X''$, $\vec V'$ and $\vec V'$ are obtained through the pseudo-spectral algorithms. Moreover, we emphasize that we get explicit expressions for the partial derivatives appearing in equation~\eqref{eq:D_F} from the system of differential equations, boundary and additional conditions. In consequence of the calculation~\eqref{eq:dF} we find
\beq
	\vec D_F (\vec X, \vec V ) = \hat J(\vec X ) \, \vec V \, , 
	\label{eq:JtimesV}
\eeq
i.e.\ the computation of the vector $\vec D_F(\vec X, \vec V)$ gives us the matrix-vector product of the Jacobian $\hat J(\vec X)$ and an arbitrary vector $\vec V$. Accordingly, we build up the Jacobian row by row through a successive computation of the vectors $\vec D_F(\vec X,\vec e_i)$ for $i=0,1,\dots ,N_\text{total}-1$. All in all, a vector $\vec D_F$ is computed $N_\text{total}$ times, which is a considerable reduction of operations compared to the straightforward construction of the Jacobian given by equation~\eqref{eq:JacobianComputationFD}. Moreover, an even greater benefit of equation~\eqref{eq:JtimesV} arises when iterative methods are used to solve the pseudo-spectral linear problem, like the one described below.

\subsubsection{BiCGSTAB method}
\label{subsubsec:BiCGSTAB method}

Our method of choice for the solution of the linear system $\hat J \, \delta \vec X = \vec F$ is the so-called BiCGSTAB (biconjugate gradient stabilized) method~\cite{vanderVorst:1992aa}. The implementation of this iterative method is rather simple, while its mathematical justification is more complex, see for example references~\cite{Barrett:1994aa,Press:2007aa}. Here, we concentrate on the discussion of the most crucial point within the BiCGSTAB method: the preconditioning. Since the BiCGSTAB method works best for well-conditioned matrices, we introduce a matrix $\hat J_\text{P}$, called the preconditioner, that has the following property
\beq
	\hat J_\text{P}^{-1} \, \hat J \approx \mathds{1} \, .
	\label{eq:Preconditioner}
\eeq
In other words, the preconditioner is an approximation of the Jacobian $\hat J$. Then, we rewrite the pseudo-spectral linear problem as
\beq
	\hat J_\text{P}^{-1} \, \hat J \, \delta \vec X = \hat J_\text{P}^{-1}  \, \vec F \, ,
	\label{eq:PreconditionedPseudoSpectralLinearProblem}
\eeq
where the matrix $\hat J_\text{P}^{-1} \, \hat J$ is close to the identity matrix by definition and is therefore well-conditioned. Then the BiCGSTAB method only solves a linear systems involving the preconditioner
\beq
	\hat J_\text{P} \, \vec V = \vec W
\eeq
within each BiCGSTAB iteration (with some vectors $\vec V$ and $\vec W$).\footnote{In fact, the LU-decomposition of the preconditioner only has to be performed once for each Newton-Raphson iteration.} Moreover, the Jacobian $\hat J$ itself is only needed to perform several matrix-vector multiplications. We do this efficiently by utilizing equation~\eqref{eq:JtimesV}. Therefore, we do not have to store the dense Jacobian matrix and consequently, we save a lot of memory. 

The iterative nature of the BiCGSTAB method entails that we merely get an improved approximation of the desired vector $\delta \vec X$ after each iteration. There are two factors that determine the runtime of the BiCGSTAB method: the number of iterations to reach a predetermined accuracy and the time each iteration lasts. Thus, the benefit of the BiCGSTAB method essentially depends on how good the preconditioner $\hat J_\text{P}$ approximates the Jacobian $\hat J$ (this reduces the number of iterations) and on how fast the linear system involving the preconditioner is solved (this reduces the time for each iteration). 

\subsubsection{Preconditioner}
\label{subsubsection:Preconditionier}

The idea to find an approximation of the Jacobian $\hat J$ is to use finite differencing rather than the pseudo-spectral algorithms to build up a Jacobian $\hat J_\text{P} = \hat J_\text{FD}$. In order to do so, we map the Lobatto grid points onto an equidistant grid through the coordinate transformation
\beq
	x(\xi ) = \frac{b+a}{2} - \frac{b-a}{2} \cos \left(  2\,\xi \right) \, ,
	\label{eq:CoordTrafoLobattoToEquidist}
\eeq
where $\xi \in [0,\pi /2]$. The grid points are
\beq
	\xi _k= \frac{\pi \, k}{2(N-1)} \, , \quad k = 0,1,\dots ,N-1 \, ,
	\label{eq:FDGridPoints}
\eeq
cf.\ equation~\eqref{eq:LobattoPoints}. We are now able to utilize the standard expressions of finite differencing to calculate the derivatives numerically. In the simplest case we use centered second order stencils
\beq
	\frac{\partial f_k}{\partial \xi} \approx 	\frac{f_{k+1} - f_{k-1}}{2\, h} \, , \quad \frac{\partial ^2 f_k}{\partial \xi^2} \approx 	\frac{f_{k+1} - 2\, f_k + f_{k-1}}{h^2} \, ,
	\label{eq:Stencils}
\eeq 
with spacing 
\beq
	h = \frac{\pi}{2(N-1)} \, .
	\label{eq:FDSpacing}
\eeq
A useful additional benefit of the coordinate transformation~\eqref{eq:CoordTrafoLobattoToEquidist} is that all functions of $\xi$ are symmetric to the boundaries $\xi =0$ and $\xi =\pi /2$. Thus, the centered stencils~\eqref{eq:Stencils} are applicable at the boundaries as well. At the end, in order to construct the finite difference Jacobian $\hat J_\text{FD}$, we convert the derivatives with respect to $\xi$ into derivatives with respect to $x$. We note that the extension to functions depending on several coordinates is straightforward. 

In contrast to the pseudo-spectral Jacobian $\hat J$ the finite difference Jacobian $\hat J_\text{FD}$ is sparse. This allows on the one hand for an efficient use of memory by only storing the non-zero elements. On the other hand there are very fast algorithms to solve sparse linear systems available. For instance, if we arrange the elements of the vectors $\vec X$ and $\vec F$ appropriately, the finite difference Jacobian $\hat J_\text{FD}$ will exhibit a band structure. There exist simple adaptions to the standard LU-decomposition algorithm that take advantage of the band structure and considerably reduce the time complexity, e.g.\ see reference~\cite{Press:2007aa}.\footnote{In the best case we get a time complexity of roughly $\mathcal O(N^2_\text{total})$, which, in the two-dimensional case, goes as $\mathcal O(N^4)$ with respect to the resolution $N$, cf.\ footnote~\ref{footnote:TimeComplexityLU}.} We also implemented a more sophisticated approach by utilizing the high-level SuperLU library~\cite{superlu99,superlu_ug99} leading to considerably lower runtimes.

\subsection{Further common techniques}
\label{subsec:Further_common_techniques}

In this subsection, we outline two important techniques that are of particular importance for the present work: the multi-domain method, which is used very commonly, and the analytic mesh refinement, which proved to be crucial for our purposes.

\subsubsection{Multi-domain method}
\label{subsubsect:Multi-domain_method}

There are several reasons to perform a decomposition of the domain of integration into several subdomains:
\begin{itemize}
	\item 	The desired function shows a problematic behavior, for example a strongly pronounced peak. Then, an appropriate domain decomposition allows us to enhance the resolution near the critical region while keeping the resolution moderate in the other regions. 
	\item	The domain of integration has more than four edges. To avoid singular coordinate transformations a decomposition into several subdomains with only four edges is needed. 
	\item 	The desired function or some of its derivatives is expected to be discontinuous. Therefore, the spectral representation of this function will exhibit only an algebraic rate of convergence. If the subdomains can be arranged in such a way that this discontinuity locates exactly on the boundary between two subdomains, the function will be smooth within each subdomain leading to a spectral convergence. 
\end{itemize}
Note that we find examples for all of these points in the work at hand. 

\begin{figure}[!ht]
	\centering
		\includegraphics[scale=1]{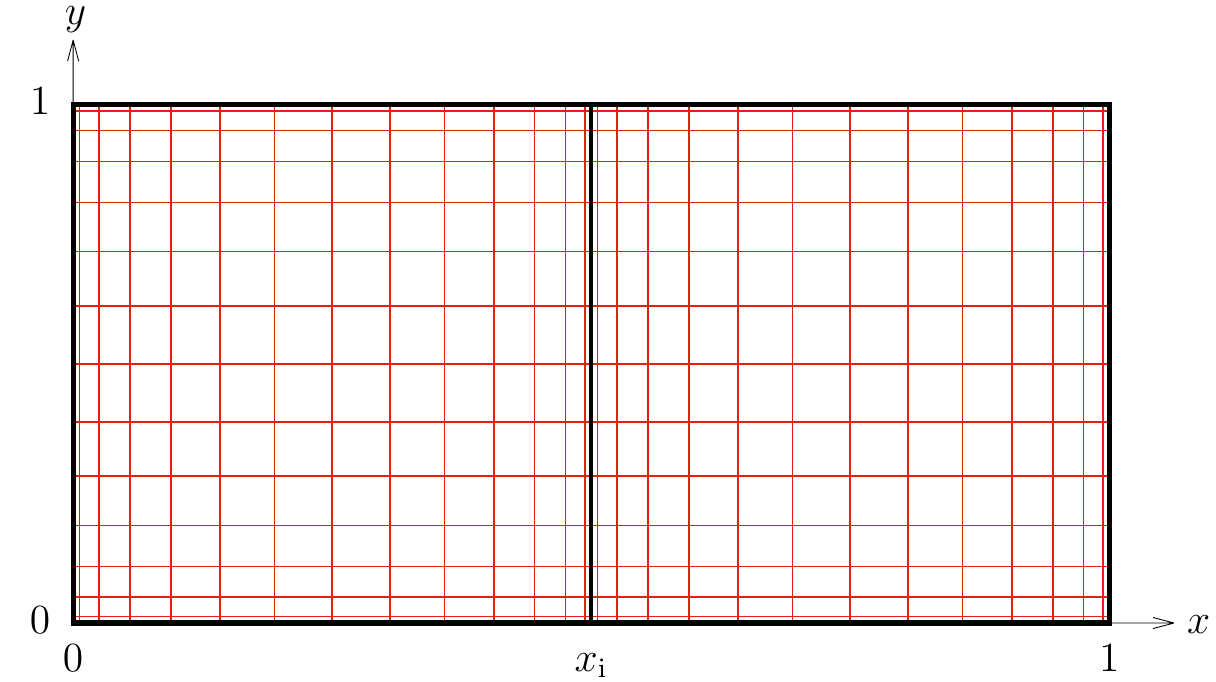} 
	\caption{Simple example of a domain decomposition. The domain of integration $(x,y)\in [0,1]\times [0,1]$ is split into two subdomains: $(x,y)\in [0,x_\text{i}]\times [0,1]$ and $(x,y)\in [x_\text{i} ,1]\times [0,1]$. In each subdomain we consider a two-dimensional Lobatto grid. Consequently, the inner boundary at $x=x_\text{i}$ is represented twice. New boundary conditions are required at the inner boundary $x=x_\text{i}$ in each subdomain. For a second order differential equation we demand equality of the function values and the values of the corresponding normal derivatives at $x=x_\text{i}$. Obviously, in this simple case the normal derivative is the derivative in $x$ direction.  }
	\label{fig:MultiDomain}
\end{figure}

Let us explain the multi-domain method in case of a two-dimensional problem.\footnote{For a one-dimensional problem the situation is even simpler.} In figure~\ref{fig:MultiDomain} we illustrate the simple case of a rectangular domain of integration that is decomposed into two rectangular subdomains. In such a situation the original coordinates are suitable to parametrize both subdomains. Nevertheless, we usually have to deal with more complicated cases and have to find appropriate coordinate transformations in order to parametrize non-rectangular subdomains. Normally, it is not too difficult to find such coordinate transformations. In addition, there is a way to construct a coordinate transformation that maps a non-rectangular to a rectangular domain only from the knowledge of the boundary curves of the domain, see reference~\cite{Dias:2015nua}.

It is straightforward to incorporate the multi-domain method into the pseudo-spectral scheme for solving differential equations as described above. Instead of having a two-dimensional Lobatto grid covering the entire domain of integration, we now have several two-dimensional Lobatto grids covering each of the subdomains. Nevertheless, we have to work out new boundary conditions at the inner boundaries between adjacent subdomains. Note that an inner boundary is represented twice in the numerical grid as a boundary of each subdomain. At the corresponding grid points we demand continuity of the desired functions and their normal derivatives, i.e.\ normal with respect to the inner boundary. This is appropriate for the most important case of second order differential equations. If the grid points of two touching subdomains do not match at the common inner boundary, for example due to a different resolution, we have to apply interpolation techniques. But as explained in subsection~\ref{subsec:Basic_ideas_and_concepts} the standard pseudo-spectral algorithms do this very accurately and efficiently. 

\newpage
\subsubsection{Analytic mesh refinement}
\label{subsubsec:Analytic_mesh_refinement}

\begin{figure}[!ht]
	\centering
		\includegraphics[scale=1]{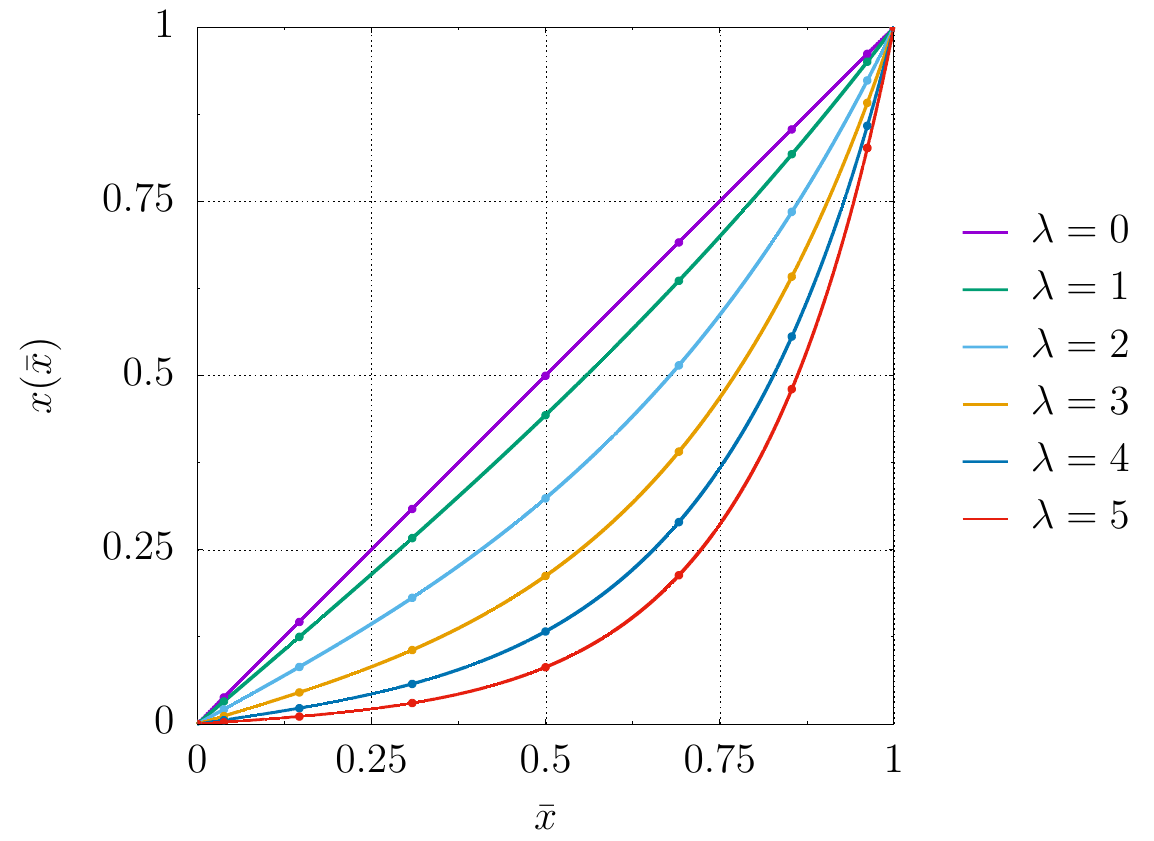} 
	\caption{Illustration of the function $x(\bar x) = \sinh (\lambda \bar x) / \sinh \lambda$ for different $\lambda$. The small circles on top of each graph represent Lobatto grid points in $\bar x$ mapped to $x$. We see that for increasing $\lambda$ the grid points become more densely distributed around $x=0$. Therefore, the analytic mesh refinement enhances the resolution near $x=0$ while it reduces the resolution around the other edge, $x=1$. }
	\label{fig:meshrefinement}
\end{figure}

Often we encounter problems in which the underlying functions exhibit steep gradients. Following references~\cite{Meinel:2012aa,Macedo:2014bfa} one way to treat such a behavior is an analytic mesh refinement. Consider a function $f(x)$ on the interval $x\in [0,1]$ with a peak at $x=0$. The coordinate transformation
\beq
	x(\bar x) = \frac{\sinh (\lambda \bar x) }{\sinh \lambda} 
	\label{eq:meshrefinement}
\eeq
maps $x\in [0,1]$ to $\bar x\in [0,1]$, see figure~\ref{fig:meshrefinement} for illustration. For $\lambda >0$ the gradients of $f(x)$ become flatter when considered as $f(x(\bar x))$. In a numerical implementation this corresponds to an increase of the density of grid points around $x=0$ while the density decreases close to the other edge $x=1$, cf.\ figure~\ref{fig:meshrefinement}. Accordingly, representing $f(x)$ and $f(x(\bar x))$ on a Lobatto grid with equal resolution, the peak at $x=0=\bar x$ is better resolved if $f$ is considered as a function of $\bar x$. Note that the limit $\lambda\to 0$ yields the identity transformation $x(\bar x) = \bar x$. 

\begin{figure}[!ht]
	\centering
		\includegraphics[scale=1]{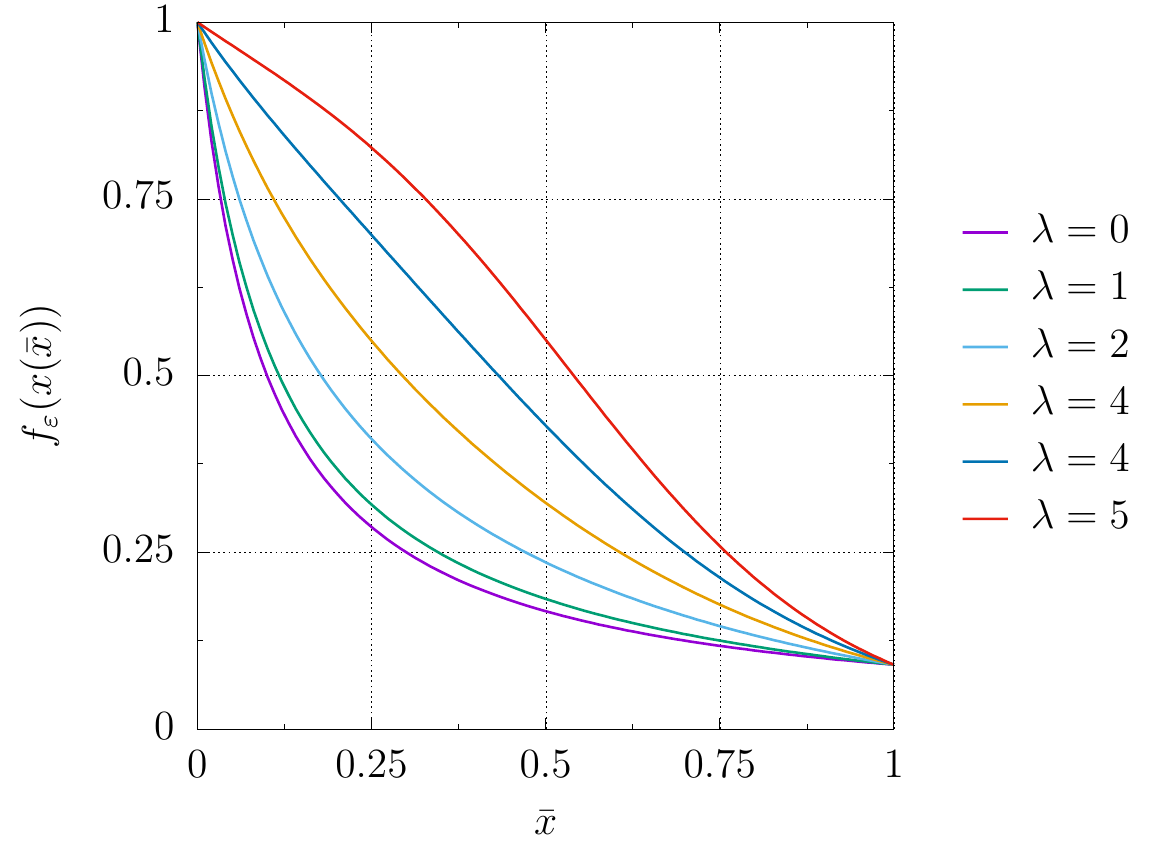} 
	\caption{Illustration of the function $f_{\varepsilon}(x(\bar x)) = \varepsilon / [\varepsilon + \sinh (\lambda \bar x) / \sinh \lambda ]$ with $\varepsilon =0.1$ for different $\lambda$. For increasing $\lambda$ the gradients around $\bar x=0$ flatten out.  }
	\label{fig:meshrefinementfunction}
\end{figure}

As an example we consider the function
\beq
	f_{\varepsilon}(x) = \frac{\varepsilon}{\varepsilon +x} \, ,
	\label{eq:functionformeshrefinement}
\eeq
see also references~\cite{Meinel:2012aa,Macedo:2014bfa}. For $\varepsilon \ll 1$ this function has a clearly pronounced peak at $x=0$. In figure~\ref{fig:meshrefinementfunction} we display the change of the function $f_{\varepsilon}(x(\bar x))$ for different $\lambda$ in the case $\varepsilon =0.1$.

\begin{figure}[!ht]
	\centering
		\includegraphics[scale=1]{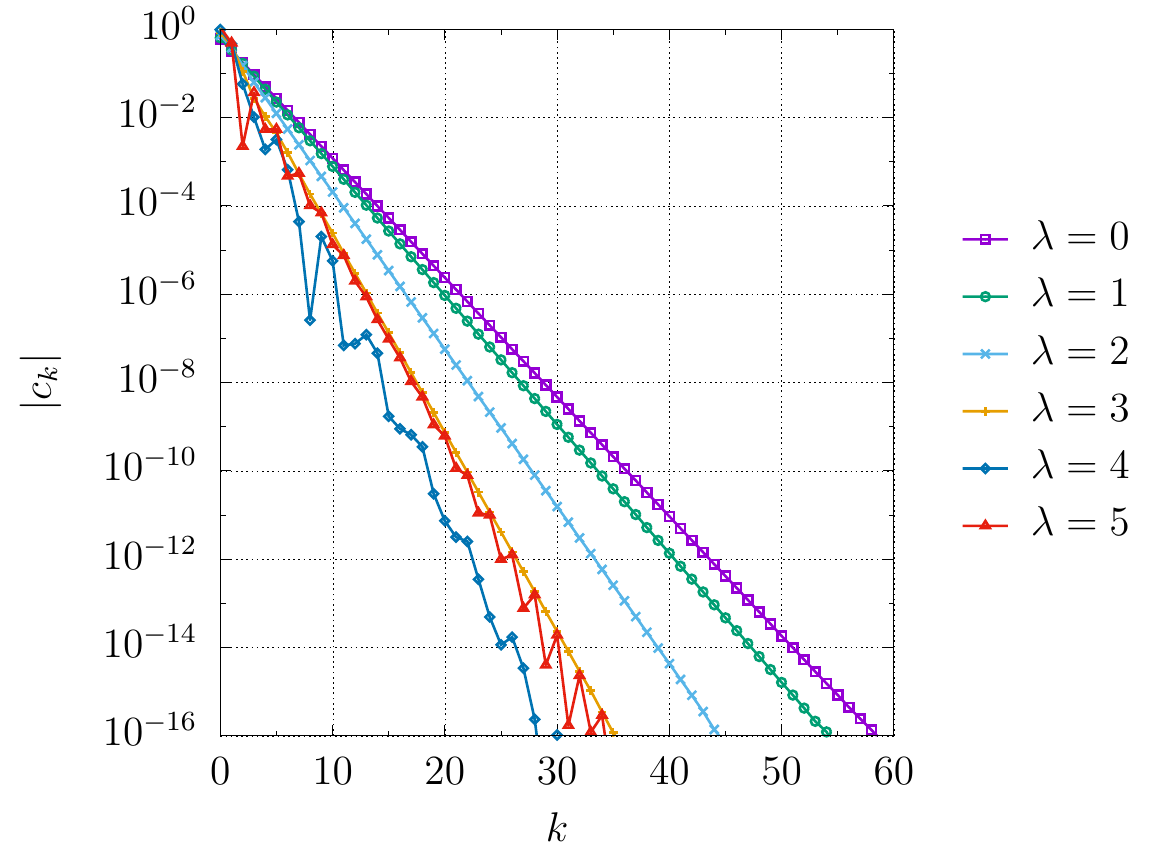} 
	\caption{Spectral coefficients of the function $f_{\varepsilon}(x(\bar x)) = \varepsilon / [\varepsilon + \sinh (\lambda \bar x) / \sinh \lambda ]$ with $\varepsilon =0.1$ for different $\lambda$. For increasing $\lambda$ the convergence improves until $\lambda =4$. For greater values the convergence worsens again. Note that in the optimal case, $\lambda =4$, we need only half of the spectral coefficients to reach a certain accuracy compared with the case without analytic mesh refinement, $\lambda =0$.}
	\label{fig:meshrefinementconvergence}
\end{figure}

Now we show the advantage of the analytic mesh refinement~\eqref{eq:meshrefinement} in a pseudo-spectral scheme. Again, we consider the example function $f_\varepsilon(x(\bar x))$ and show the decay of its spectral coefficients for different $\lambda$ in figure~\ref{fig:meshrefinementconvergence}, where $\varepsilon =0.1$ was set. The benefit of the analytic mesh refinement is apparent since for $\lambda =4$ we only need half of the spectral coefficients to reach a certain accuracy compared to $\lambda =0$. For smaller $\varepsilon$ this becomes even more crucial. Note that there is an optimal $\lambda$ since for higher $\lambda$ the density of grid points around the other edge, $\bar x=1$, becomes too low. In other words, for high $\lambda$ the gradients near $\bar x=1$ become steep as well, cf.\ figure~\ref{fig:meshrefinementfunction}.


    {
      \emergencystretch=2.5em
      \hbadness=10000
      \printbibliography
    }


  \chapter{List of Abbreviations}

\begin{list}{}{
    \setlength{\labelwidth}{3cm}
    \setlength{\labelsep}{0.3cm}
    \setlength{\leftmargin}{\labelwidth+\labelsep}
}
	\item[AdS] 	anti-deSitter
	\item[CFT]	conformal field theory
	\item[GR]	general relativity
	\item[GL]	Gregory-Laflamme	
  	\item[KK]  	Kaluza-Klein
  	\item[LBH] 	localized black hole
  	\item[NBS] 	non-uniform black string
  	\item[ST]	Schwarzschild-Tangherlini
  	\item[UBS] 	uniform black string
\end{list}

\chapter{Acknowledgments}

I am deeply grateful to my former supervisor, Marcus Ansorg, for everything he taught me. At any time I could rely on his help and on his ideas, especially when I was close to desperation.

I thank Martin Ammon, who gave me the opportunity to complete this thesis and who supervised me during the last months. Furthermore, I am grateful to Sebastian Möckel for the fruitful collaboration and for the access to his APDES (automatic PDE solver) code. It is a pleasure to thank Burkhard Kleihaus, Jutta Kunz and Eugen Radu for inspiring this work and for many discussions. Moreover, I thank all my colleagues for the pleasant and stimulating working atmosphere and all my friends for support. In particular, I am grateful to Alexander Blinne and David Schinkel for IT support and to Christian Kohlfürst, Julian Leiber, Sebastian Möckel and David Schinkel for their comments on this manuscript. 

Furthermore, I would like to thank the referees of this dissertation, Martin Ammon, Jutta Kunz and Toby Wiseman, for working through this thesis and for writing their referee reports. I particularly appreciate Toby Wiseman's comments and suggestions of improvement.  

Finally, I acknowledge financial support by the Deutsche Forschungsgemeinschaft (DFG) graduate school GRK 1523/2 and by the University of Jena.


\end{document}